%% file: pensions.tex
\documentclass{article}
\usepackage{amsmath}
\usepackage{amssymb}
\usepackage{amsthm}
\usepackage{bbm,mathtools}
\usepackage{booktabs}
\usepackage{url}
\usepackage{etoolbox}
\usepackage{lscape}
\usepackage[table]{xcolor}
\usepackage{multirow}
\usepackage{xfrac}
\usepackage{makecell}

\newtoggle{bulletpoints}
\togglefalse{bulletpoints}

\def\reals{\mathbb{R}}

\def\comp{\raise 1pt \hbox{$\scriptstyle\circ$}}

\def\logit{\mathop{\rm logit}\limits}

\def\upto{{\raise 1pt \hbox{$\scriptstyle \,\nearrow\,$}}}
\def\downto{{\raise 1pt \hbox{$\scriptstyle \,\searrow\,$}}}

\def\L{{\cal L}}

\def\YTM{{\rm YTM}}
\def\AWE{{\rm AWE}}
\def\CPI{{\rm CPI}}
\def\GDP{{\rm GDP}}
\def\LTIE{{\rm LTIE}}
\def\STOCK{{\rm Stock Index}}

\theoremstyle{definition}

\theoremstyle{empty}

\newcommand{\glrdelim}[4]
{\ifcase#1\left#2\or#2\or\bigl#2\or\Bigl#2\or\biggl#2\or\Biggl#2\fi%
#4\ifcase#1\right#3\or#3\or\bigr#3\or\Bigr#3\or\biggr#3\or\Biggr#3\fi}

\newcommand{\lrp}[2][0]{\glrdelim{#1}{(}{)}{#2}} % left-right parentheses 
\newcommand{\lrs}[2][0]{\glrdelim{#1}{[}{]}{#2}} % left-right square brackets
\begin{document}
%\title{Stochastic modeling of pension assets and liabilities}
%\title{Stochastic modeling for longevity risk management}
\title{Stochastic modeling of assets and liabilities with mortality risk
%\footnote{The work of the first author was partially supported by the Pensions Policy Institute and Professor Sir Richad Trainor PhD-scholarship. The authors are grateful to Matt Hamilton-Glover for his help on longevity modelling and to Chris Curry and Timothy Pike of PPI for helpful discussions throughout the project.}
}

%\author{Sergio Alvares Maffra \and John Armstrong \and Teemu Pennanen}
%
%\maketitle

\author{Sergio Alvares Maffra~\footnote{Research partially supported by the Pensions Policy Institute, {\tt sergio.maffra@kcl.ac.uk}} \and John Armstrong \and Teemu Pennanen\footnote{Department of Mathematics, King's College London, {\tt teemu.pennanen@kcl.ac.uk}}}

\maketitle

\begin{abstract}
This paper describes a general approach for stochastic modeling of assets returns and liability cash-flows of a typical pensions insurer. On the asset side, we model the investment returns on equities and various classes of fixed-income instruments including short- and long-maturity fixed-rate bonds as well as index-linked and corporate bonds. On the liability side, the risks are driven by future mortality developments as well as price and wage inflation. All the risk factors are modeled as a multivariate stochastic process that captures the dynamics and the dependencies across different risk factors. The model is easy to interpret and to calibrate to both historical data and to forecasts or expert views concerning the future. The simple structure of the model allows for efficient computations. The construction of a million scenarios takes only a few minutes on a personal computer. The approach is illustrated with an asset-liability analysis of a defined benefit pension fund.
\end{abstract}

{\bf Subject classifications:}  Investment, Insurance, Stochastic modelling

%\tableofcontents

\input{doc_introduction}

\input{doc_risk_factors}

\input{doc_stochastic_model}

\input{doc_model_calibration}

\input{doc_model_uk}

\input{doc_model_uk_simulation}

%\input{examples}

%\input{conclusion}

%\section{Acknowledgments}

%\begin{itemize}
%    \item King's for providing the Prof Richard Trainor scholarship
%    \item Pension Policy Institute
%    \item Mathew Glover, for sharing his Python code for the estimation of mortality risk factors
%\end{itemize}

\bibliographystyle{plain}
%\bibliography{sp}
%\bibliographystyle{amsplain}
%\addcontentsline{toc}{section}{References}
\bibliography{doc_references}

\appendix

\end{document}

%% file: doc_introduction.tex
\section{Introduction}\label{introduction}

Stochastic simulations have important applications in life and pension insurance. They are used in risk management, asset-allocation, pricing and hedging of longevity-related products, and in the valuation of pension liabilities. 
In the context of pensions, probably the best known stochastic models are the CAP:LINK model built by Tillinghast Towers Perrin (now Towers Watson)~\cite{mulvey1996generating,mulvey1998towers} and the Wilkie model~\cite{wilkie1984stochastic} for actuarial use in the UK. While the above are concerned with economic and financial risk factors, there now exist many stochastic models for longevity, another major risk in pension and life insurance; see e.g.~\cite{leecarter, CAIRNS2011355,hunt2014general, aro2014stochastic}.

This paper presents a general framework for the construction of stochastic simulation models for the assets and longevity-linked liabilities of a typical pension or life insurer. The framework accommodates longevity risk as well as the main asset classes used by pension funds: equities, short-term and long-term fixed-rate bonds, index-linked bonds, and corporate bonds. Our models incorporate the often neglected dependencies between longevity and asset returns which is important in valuation and hedging of long-dated longevity related products and liabilities; see e.g.~\cite{aro2014stochastic, glover.mortality}.

After suitable nonlinear transformations, all the relevant risk factors are modelled by vector autoregressive (VAR) models much as in~\cite{koivu2005cointegration, koivu2005modeling, aro2014stochastic, wilkieyet5}. VAR models were recommended also in \cite{wilkieyet5} as an alternative to the cascading structures adopted in~\cite{wilkie1984stochastic, wilkie1995more, mulvey1996generating, mulvey1998towers}. The VAR structure allows for more general dependencies across different risk factors. Moreover, VAR models lend themselves to efficient vectorized computations which allow the simulation of millions of scenarios within minutes on a personal computer. Another advantage of the VAR structure is that it allows for easy incorporation of short-term forecasts as well as long-term views. Such subjective modifications were also proposed in \cite{wilkieyet2} and \cite{wilkie1984stochastic}. 

We use the general approach of \cite{aro2011user}, where longevity/mortality-risk is described by population wide interpretable {\em systemic risk factors} as well as by {\em idiosyncratic} binomial risk factors describing yearly deaths. While the idiosyncratic risk can be diversified away by increasing the size of a fund, the systemic risk represents undiversifiable longevity risk. The explicit inclusion of the idiosyncratic risk makes the model ideal for studying the benefits of e.g.\ mergers or consolidations of multiple pension funds. The systemic risk factors allow for natural interpretations which facilitates the calibration and analysis of the model. This is a significant advantage over e.g.\ the famous Lee-Carter~\cite{leecarter} model where the risk factors are computed through a singular value decomposition which changes every time new data points are introduced.

As an illustration of the general approach, we build a stochastic simulation model for UK pension funds. The modelled asset classes cover the majority of pension investments in the UK, namely, equities, short- and long-term fixed rate bonds, inflation-linked bonds and corporate bonds which, according to the ``Purple Book''~\cite{purplebook2017short}, constitute approximately 90\% of the UK pension assets. On the liability side, the model includes longevity risk as well as risks coming from the indexation of future pension benefits. The UK model is illustrated with a series of simulation experiments where we analyse the dynamics of population sizes, asset returns and macroeconomic risk factors that are statistically connected to both assets and liabilities. The simulation model was developed and implemented in Python~3.7 and run on Intel(R) Core(TM) i7-7700HQ laptop with 32 GB of RAM. Computation of one million random scenarios took 147 seconds.

The simulations show that idiosyncratic mortality risk is diversified away as cohort sizes are increased but it remains significant even in fairly large populations. We analyze the risks involved with defined benefit liabilities and consider the special case USS scheme in the UK. The risk contributions from longevity and indexation of the benefits vary with age and cohort sizes but over longer periods, both risks are significant. We also find that a strong connection between economic growth and old age mortality, confirming the findings of \cite{aro2014stochastic} and establishing a link between longevity-linked liabilities and investment returns. We end the analysis with an asset-liability management study where we simulate DB-pension schemes forward in time until run-off. We compare the performance of the average allocations PPF-eligible schemes in 2008 and 2019. While the significant reduction in stock investments results in lower uncertainty in future net wealth, the reduction is mainly on the upside while the downside risk remains almost the same.

The rest of the paper is organized as follows. Section~\ref{sec:risk_factors} describes the different types of risk factors that can be incorporated into our models. Section~\ref{stochastic_model}, describes in general terms how one can specify time series models for such risk factors, and in Section~\ref{calibration} their calibration to historical data and to views. After the theoretical part, we describe the UK model and its calibration in Section~\ref{model_uk}. Analysis of the simulation experiments can be found in Section~\ref{simulation_results}.

%% file: doc_risk_factors.tex
\section{Risk factors}\label{sec:risk_factors}

The first step in the creation of any stochastic model is the identification of the most relevant \emph{risk factors} that affect the quantities of interest. In the context of pension fund management, the interesting quantities are the investment returns in different asset classes and the pension expenditure. On the liability side, the most important sources of uncertainty are longevity, and price and wage inflation that are often used in indexation of defined benefit liabilities. There are also some other macroeconomic risk factors, such as the gross domestic product, that may affect investment returns or liabilities indirectly. Indeed, it was found in \cite{aro2014stochastic} that gross domestic product (GDP) has an effect on old age mortality over longer periods of time. In~\cite{glover.mortality}, a similar link was found between old age mortality and average weekly earnings. GDP affects pension liabilities also through inflation which is often used in indexation of the benefits. GDP has statistically significant connections not only with inflation but also many other factors that affect investment returns.

In a typical pension fund, one can easily identify thousands of risk factors that affect the assets and liabilities. Fortunately, it is often possible to come up with significant reductions in the number of risk factors while still capturing the main distributional properties of asset returns and liability payments. This section reviews some of the most relevant risk factors of a typical pension fund as well as some useful reductions for describing longevity and bond investments. A more detailed case study can be found in Section~\ref{model_uk}.

\subsection{Longevity}\label{longevity}

Longevity risk is perhaps the most important source of risk faced by a defined benefit pension fund. There has been an increasing interest in stochastic modelling of longevity risk; see e.g.\ \cite{CAIRNS2011355,hunt2014general}. Common to all these models is the aim to describe the longevity risk across all relevant cohorts by a small number of risk factors whose future development is then modelled stochastically. We describe here the flexible approach from \cite{aro2011user}, which is easy to implement and allows for interpretable risk factors, unlike the famous Lee-Carter model~\cite{leecarter}.

We will denote the size of cohort aged $a$ in year $t$ by $E_{a,t}$. It is natural to assume that the future population sizes follow Binomial distributions with
%\begin{block}{Population size dynamics}
\begin{equation}
E_{a+1,t+1} \sim \text{Bin}\lrp{E_{a,t}; p_{a,t}},
\end{equation}
where $p_{a,t}$ is the {\em survival probability} of an individual aged $a$ at the beginning of year $t$. For each year $t$, we assume that
\begin{equation}
 \logit p_{a,t} := \ln\left(\frac{p_{a,t}}{1 - p_{a,t}}\right) 
 = \sum_{i=1}^n v^i_t\phi^i(a),\label{eq_logit_risk_factor}
\end{equation}
where $\phi^i$ are a given collection of {\em basis functions} and $v_t^i$ are the corresponding risk factors that may vary over time and across different scenarios. Thus, instead of considering the survival probability in each age cohort as a separate risk factor, we describe the probabilities of all ages by the $n$ risk factors $v^i$ with $n$ much less than the number of ages considered. This reduction facilitates both the stochastic modelling of the future survival probabilities as well as their numerical simulation; see Section~\ref{computation_times}. The interpretation of the risk factors $v^i$ depends on the choice of the corresponding basis function $\phi^i$. With the choices made in \cite{aro2014stochastic}, each factor corresponds to the logistic survival probability of an age group; see also Section~\ref{longevity_uk} below.

Historical values of the risk factors $v_t^i$ are obtained by maximizing, year by year, the likelihood function
\begin{equation*}
{\mathcal L}_t(v_t) := \sum_{a\in A}\left[\lrp{E_{a,t} - D_{a,t}}\sum_{i=1}^n v_t^i\phi^{i}\left(a\right) - E_{a,t}\ln \left(1+\text{exp}\lrp{\sum_{i=1}^n v_t^i\phi^{i}\left(a\right)}\right)\right],
\end{equation*}
where $D_{a,t}$ is the number of deaths of individuals aged $[a,a+1)$ in year $t$. The function $\L_t$ is concave which greatly facilitates its maximization; see \cite[Proposition~3]{aro2011user}. 

We model the future development of the risk factors $v$ stochastically in a joint model with other relevant risk factors. While the uncertainty concerning the development of $v$ may be interpreted as {\em systemic} risk, the uncertainty concerning the binomial cohort sizes is an {\em idiosyncratic} risk that may be reduced with diversification. Such diversification effects are the key factors when considering mergers and consolidations of pension funds and schemes.

\subsection{Investments}\label{asset_returns} %Asset returns

On the investment side, our aim is to describe the future returns in typical asset classes of pension funds. As we are concerned with long-term strategic investments, we will group investments in broader classes as is often done is strategic asset-liability management. Assets can be grouped in many different ways but for purposes of asset allocation, the most useful is done according to the statistical properties of the returns.

In the present work, we concentrate on the asset classes most relevant to UK pension funds. According to Table~\ref{tab:purplebookallocation}, extracted from the ``Purple Book''~\cite{purplebook2017short}, nearly 90\% of the UK pension funds are allocated in equities, bonds, property and cash and deposits.

\input{table_purple_book_allocations}

Liquid equity investments are described, as usual, by a total return index $P$ that tracks the changes in value due to price movements as well as dividend payments. Yearly equity returns are then given simply by $R_t=P_t/P_{t-1}$.

Bond investments can be grouped according e.g.\ to maturities, credit ratings and the possible underlying indices. Table~\ref{tab:purplebookallocationbonds} gives a breakdown of the UK pension fund bond investments; see \cite{purplebook2017short}. 

\input{table_purple_book_allocations_bonds}

For bond investments, we will use the approximation from \cite{Koivu78} that describes investment returns on bond portfolios using only two risk factors, the {\em yield to maturity} and the underlying index. Consider a bond or a portfolio of bonds with outstanding payments at times $t_1<t_2<\cdots<t_N$. The portfolio's {\em yield to maturity} (YTM) $Y_t$ at time $t<t_1$ is defined as the solution of the equation
\begin{equation}\label{y}
P_t = \sum_{n=1}^N e^{-Y_t(t_n-t)}I_tc_n,
\end{equation}
where $P_t$ is the portfolio's market price, $c_n$ are the outstanding payments, and $I_t$ is an index. 
Fixed rate bonds correspond to a constant $I$ while in the case of inflation-linked bonds, $I$ is the {\em consumer/retail price index}. The index can also be used to describe {\em default risk} in corporate bonds. In that case, $I$ is the {\em recovery rate} that describes the reduction of the outstanding payments due to defaults of bond issuers.

As shown in \cite{Koivu78}, the first order Taylor-approximation of the logarithmic price with respect to time and the YTM gives
\begin{equation}\label{bondreturn}
\Delta\ln P \approx Y_t\Delta t - D_t\Delta Y + \Delta\ln I,
\end{equation}
where $\Delta Y:=Y_s-Y_t$, $\Delta Y =Y_s-Y_t$, $\Delta\ln I=\ln I_s-\ln I_t$ and
\[
D_t := \frac{1}{B_t}\frac{\partial B_t}{\partial Y} = \frac{1}{B_t}\sum_{n=1}^N (t_n-t)e^{-Y_t(t_n-t)}c_nI_t 
\]
is the {\em duration} of the portfolio. Thus, the portfolio return is given by
\[
\frac{P_s}{P_t} \approx \exp\left(Y_t\Delta t - D_t\Delta Y + \Delta\ln I\right).
\]
This simple formula expresses the return on a bond portfolio in terms of only two risk factors, the YTM and the underlying index. The possibly very complicated structure of the outstanding payments is captured by a single characteristic, the duration. Statistical analysis of \cite{Koivu78} on bond data from six different countries shows that formula \eqref{bondreturn} explains consistently more than 99\% of monthly return variations of log-returns on bond portfolios.

Similarly, the second order Taylor's approximation gives
\begin{equation}\label{bondreturn_order2}
\Delta\ln P \approx Y_s\Delta t - D_t\Delta Y +\frac{1}{2}(C_t-D_t^2)(\Delta Y)^2 + \Delta\ln I,
\end{equation}
where 
\[
C_t := \frac{1}{B_t}\frac{\partial^2B_t}{\partial Y^2} =  \frac{1}{B_t}\sum_{n=1}^N(t_n-t)^2e^{-Y_t(t_n-t)}I_tc_n,
\]
is the {\em convexity} of the bond portfolio. The statistical analysis of \cite{Koivu78} shows, however, that the second order terms does not add much precision as the first order formula is already nearly perfect.

%% file: table_purple_book_allocations.tex
\begin{table}[ht]
\begin{small}
%\begin{center}
\resizebox{\textwidth}{!}{
\begin{tabular}{ccccccccc}
\toprule
Year & Equities & Bonds & Property & \thead{Cash and\\deposits} & \thead{Insurance\\policies} & Hedge funds & Annuities & Misc. \\
%     &          &       &          & deposits & policies   &             &  \\
\midrule
%2006 & 61.1\% & 28.3\% & 4.3\% &  2.3\% & 0.9\% & n/a & n/a & 3.1\% \\
%\rowcolor{black!5}2007 & 59.5\% & 29.6\% & 5.2\% &  2.3\% & 0.8\% & n/a & n/a & 2.5\% \\
2008 & 53.6\% & 32.9\% & 5.6\% &  3.0\% & 1.1\% & n/a & n/a & 3.8\% \\
\rowcolor{black!5}2009 & 46.4\% & 37.1\% & 5.2\% &  3.9\% & 1.4\% & 1.5\% & n/a & 4.5\% \\
2010 & 42.0\% & 40.4\% & 4.6\% &  3.9\% & 1.4\% & 2.2\% & n/a & 5.4\% \\
\rowcolor{black!5}2011 & 41.1\% & 40.1\% & 4.4\% &  4.1\% & 1.6\% & 2.4\% & n/a & 6.3\% \\
2012 & 38.5\% & 43.2\% & 4.9\% &  5.1\% & 0.2\% & 4.5\% & n/a & 3.6\% \\
\rowcolor{black!5}2013 & 35.1\% & 44.8\% & 4.7\% &  6.7\% & 0.1\% & 5.2\% & n/a & 3.5\% \\
2014 & 35.0\% & 44.1\% & 4.6\% &  6.1\% & 0.1\% & 5.8\% & n/a & 4.3\% \\
\rowcolor{black!5}2015 & 33.0\% & 47.7\% & 4.9\% &  3.5\% & 0.1\% & 6.1\% & n/a & 4.7\% \\
2016 & 30.3\% & 51.3\% & 4.8\% &  3.0\% & 0.1\% & 6.6\% & 2.1\% & 1.7\% \\
\rowcolor{black!5}2017 & 29.0\% & 55.7\% & 5.3\% & -0.9\% & 0.1\% & 6.7\% & 3.3\% & 0.8\% \\
2018 & 27.0\% & 59.0\% & 4.8\% & -2.5\% & 0.1\% & 7.0\% & 3.4\% & 1.2\% \\
\rowcolor{black!5}2019 & 24.0\% & 62.8\% & 5.0\% & -4.4\% & 0.3\% & 7.4\% & 4.0\% & 1.0\% \\
\bottomrule
\end{tabular}
}
%\end{center}
\end{small}
\caption{Asset allocation in total assets for DB pension schemes in the Purple Book data set (weighted averages)}\label{tab:purplebookallocation}
\end{table}

%% file: table_purple_book_allocations_bonds.tex
\begin{table}[ht]
\begin{small}
\begin{center}
%\resizebox{\textwidth}{!}{
\begin{tabular}{cccc}
\toprule
Year & \thead{Government\\fixed interest} & \thead{Corporate\\fixed interest} & Index-linked \\
\midrule
2008 &  33.2\%  &  32.6\%  &  33.9\%  \\
\rowcolor{black!5}2009 &  29.0\%  &  38.3\%  &  32.6\%  \\
2010 &  24.6\%  &  42.2\%  &  33.1\%  \\
\rowcolor{black!5}2011 &  19.6\%  &  44.3\%  &  36.1\%  \\
2012 &  17.7\%  &  44.8\%  &  37.5\%  \\
\rowcolor{black!5}2013 &  18.5\%  &  40.6\%  &  40.9\%  \\
2014 &  18.6\%  &  40.3\%  &  41.1\%  \\
\rowcolor{black!5}2015 &  20.3\%  &  37.7\%  &  42.0\%  \\
2016 &  21.9\%  &  33.7\%  &  44.4\%  \\
\rowcolor{black!5}2017 &  24.1\%  &  31.4\%  &  44.5\%  \\
2018 &  24.1\%  &  28.8\%  &  47.1\%  \\
\rowcolor{black!5}2019 &  25.4\%  &  28.4\%  &  46.2\%  \\
\bottomrule
\end{tabular}
%}
\end{center}
\end{small}
\caption{Proportions of bond investment by DB pension schemes in the Purple Book data set (weighted averages)}\label{tab:purplebookallocationbonds}
\end{table}

%% file: doc_stochastic_model.tex
\section{The stochastic model}\label{stochastic_model}

Our aim is to construct a stochastic model that provides a reasonable description of the future development of the relevant risk factors and thus, of the asset returns and liability cash-flows. What is reasonable depends, to a large extent, on subjective views and knowledge concerning the involved risk factors. We will develop a simple model that allows for an easy calibration to both the user's views and to historical data.

In order to capture certain natural features of the risk factors, we start by applying invertible transformations to the original ones. For example, it is common to model the logarithms of various price processes when they are known to stay positive in all scenarios. A general yet simple approach is described in the classic paper of Box and Cox~\cite{box1964analysis}. With appropriate transformations, one can often get away with Gaussian processes in modeling the transformed risk factors; see Section~\ref{model_uk} for a case study with UK data.

We will model the future development of the transformed risk factors $x_t$, by a multivariate stochastic difference equation of the form
\begin{equation}\label{ivar}
\Delta x_t = A x_{t-1} + a_t + \varepsilon_t,
\end{equation}
where $\Delta x_t := x_t-x_{t-1}$, $A$ is a square matrix, $a=(a_t)_{t\ge 0}$ is a given sequence and $\varepsilon_t$ are zero-mean random (innovation) vectors all of appropriate dimensions. This can be written as an inhomogeneous {\em vector autoregressive} time series model
\[
x_t = (A+I)x_{t-1} + a_t + \varepsilon_t.
\]
Vector autoregressive models have been extensively studied in the literature; see e.g.\ \cite{sims1980macroeconomics,hamilton}. Applications to macroeconomic modelling can be found in \cite{canovavar, hildevar, koivu2005cointegration, koivu2005modeling, aro2014stochastic} and Wilkie and \c{S}ahin~\cite{wilkieyet5}. One may view \eqref{ivar} also as a time-discretization of the linear inhomogeneous stochastic differential equation
\[
dx_t = (\tilde Ax_t+\tilde a_t)dt + dW_t,
\]
where $W$ is a martingale whose increments correspond to $\varepsilon$ in \eqref{ivar}. Continuous time stochastic differential equations have been used e.g.\ in the stochastic asset model CAP:Link built by Tillinghast Towers Perrin (now Towers Watson); see \cite{mulvey1996generating, mulvey1998towers}.

Already in the scalar case, \eqref{ivar} subsumes many well known stochastic 
process models as special cases. If $A=0$ we obtain a discrete-time version of the classical 
Brownian motion with drift $a=(a_t)_{t\ge 0}$. If $x$ is the logarithm of the stock price, we then recover the classical geometric Brownian motion. The {\em mean reversion} model
\[
\Delta x_t = -\alpha(x_{t-1}-\bar x) + \varepsilon_t
\]
is another instance of \eqref{ivar}. If $x$ is the logarithm of an interest rate, we recover the Black--Karasinski interest rate model~\cite{blackkarasinski}. 

The multivariate VAR-format gives a natural way of describing interactions between different risk factors without having to specify directional causalities as in the classic Wilkie model~\cite{wilkie1984stochastic}. Note that dependencies can come from the autoregression matrix $A$ or through the dependencies between the components of the innovation vectors $\varepsilon_t$. Imaginary eigenvalues of the matrix $A$ result in oscillatory behaviour of $x$ which is often observed in the long-term analysis of various macroeconomic variables~\cite{hamilton}. A single instance of \eqref{ivar} can describe stationary, mean-reverting and even cointegrated risk factors. Indeed, the {\em vector error correction model} (see e.g.\ \cite{engle1987co})
\[
\Delta x_t = \tilde A x_{t-1} + B(Cx_{t-1}-d) + \varepsilon_t,
\]
is also an instance of \eqref{ivar}. This format was used in the context of pensions in \cite{koivu2005modeling}. 

Through the term $a=(a_t)_{t\ge 0}$, one can incorporate forecasts and expert views concerning the future; see Section~\ref{ssec:views} below. This is a particularly important feature in long-term applications where the historical data may be a poor description of the future. For example, developments in monetary policy, medicine and lifestyle have changed the way we now view interest rates, inflation and mortality developments. When specifying views concerning specific risk factors, it is important that the risk factors have natural interpretations. Such interpretations may be lost in approaches based on modelling principal components or singular values as in the Lee-Carter model of mortality~\cite{leecarter}.

%% file: doc_model_calibration.tex
\section{Model calibration}\label{calibration}

In order to get a reasonable description of the future, we will calibrate \eqref{ivar} to both historical data as well as expert views/forecasts concerning the future. User views are particularly important when the historical values of the risk factors don't correspond to what one expects to see in the future. For example, due to improvements in healthcare and monetary policy, most of us expect mortality and interest rates to be lower in the future than 50 years ago. The historical values may, nevertheless, still exhibit dependencies that one expects to prevail in the future as well. Accordingly, we calibrate the parameters of the model in two steps: we first estimate the parameter matrix $A$ as well as the distribution of the innovations $\varepsilon_t$ using historical data and, in the second step, we incorporate user's views by an appropriate specification of the sequence $(a_t)_{t\ge 1}$.

\subsection{Calibration to historical data}

The elements of the autoregression matrix $A$ are estimated by simple linear regression. We aim for a parsimonious model where we retain only the regressors that are statistically significant 
and have a clear economic interpretation. In addition, we perform a simple robustness test to validate the regressions, by varying the estimation window. Similar tests have been suggested in~\cite{huber1997review, wilkieyet1}.
All the elements of $A$ should remain statistically significant and have the sign consistent with economic theory. As an additional consistency check, we compute the eigenvalues of $A$. If we know, for example, that all the risk factors $x$ are all stationary, the eigenvalues of $(A+I)$ should all lie strictly inside the unit circle in the complex plane. Nonstationary risk factors $x^i$ whose increments $\Delta x^i$ are stationary correspond to eigenvalues on the unit circle. Imaginary eigenvalues result in oscillations may be associated with economic cycles.

Once the autoregression matrix $A$ has been estimated, we use the regression residuals to estimate the distribution of the innovations $\varepsilon_t$. In the simplest case, a multivariate Gaussian distribution is used. Alternatives include multivariate t-distribution and copula models. One could also use more complicated stochastic volatility models such as multivariate GARCH models as e.g.\ in \cite{hilli2011optimal, van2002go}.

\subsection{Incorporating views and forecasts}\label{ssec:views}

Asset managers base their investment decisions on their best knowledge and views concerning the future investment returns. The returns are highly uncertain but with proper financial/statistical analysis, the best asset managers tend to guess the median returns better than the average person. The importance of views and educated guesses is pronounced in pension funds that aim to cover the uncertain pension expenditure in the future. Such views also have a strong effect on the valuation of pension liabilities and the pricing of life insurance products.

In the inhomogeneous VAR model\eqref{ivar}, one can incorporate views on the future median values of $x$ by an appropriate specification of the sequence $(a_t)_{t\ge 1}$. Indeed, if $x=(x_t)_{t\ge 0}$ follows \eqref{ivar} with given $x_0$, we have
\[
x_t = \bar x_t + e_t
\]
where $\bar x=(\bar x_t)_{t\ge 0}$ follows the deterministic difference equation
\[
\Delta\bar x_t = A\bar x_{t-1} + a_t
\]
with $\bar x_0=x_0$ and $e$ is the zero-mean process following the homogeneous stochastic difference equation
\begin{equation}\label{vare}
\Delta e_t = Ae_{t-1} + \varepsilon_t
\end{equation}
with $e_0=0$. Conversely, if we are given future median values $\bar x$ of $x$, and we set
\begin{equation}\label{a}
a_t := \Delta\bar x_t - A\bar x_{t-1}
\end{equation}
in \eqref{ivar}, we obtain a stochastic process $x$ with median $\bar x$. Indeed, if $x$ follows \eqref{ivar} with this specification of $a$, we have that $e=x-\bar x$ follows \eqref{vare} so it has zero median. To incorporate the future views into the inhomogeneous VAR \eqref{ivar} one can thus first specify the median values $\bar x$ and then define $a$ by \eqref{a}.

One rarely has a forecast for the whole trajectory $\bar x=(\bar x_t)_{t\ge 1}$. Instead, forecasts are usually available only for some of the risk factors and only for a few years into the future. In addition, one may have views on long-term mean/median values of e.g.\ interest rates, inflation and various investment returns. We will construct a median trajectory $\bar x$ which is consistent with such information as well as with the estimated parameter matrix $A$. To this end, we assume that the risk factors can be decomposed as $x=(x^0,x^1)$ such that $x^0$ and $\Delta x^1$ are stationary. Typically, $x^0$ contains risk factors such as interest rates, yields and inflations while $x^1$ contains stock indices and other risk factors with a positive drift. Taking expectations and limits in \eqref{ivar}, then gives
\begin{equation}\label{lime}
\lim_{t\to\infty}E\Delta x_t = A^0\lim_{t\to\infty}Ex^0_t + \lim_{t\to\infty}EA^1x^1_t + \lim_{t\to\infty}a_t,
\end{equation}
where the decomposition $A=[A^0,A^1]$ corresponds to that of $x$.

We assume that the {\em long term views} are given in the form
\begin{align}
\lim_{t\to\infty}Ex^0_t&=\bar x^0,\label{views1}\\
\lim_{t\to\infty}E\Delta x^1_t&=d^1,\label{views2}\\
\lim_{t\to\infty}EA^1x^1_t&=c.\label{views3}
\end{align}
The first equation can be used to specify mean reversion levels of e.g.\ interest rates, yields and inflations while the second can be used to specify long-term median stock returns. The third equation specifies long-term means of cointegration vectors. Thus, the long-term views are specified by a total of $n+m$ parameters, where $n$ is the dimension of $x$ and $m$ is the number of cointegration vectors. For the sequence $(a_t)_{t\ge 1}$ to be consistent with \eqref{views1}--\eqref{views3}, equation \eqref{lime} implies that we must have
\begin{equation}\label{along}
a:=\lim_{t\to\infty}a_t = (0,d^1) - A^0\bar x^0 - c.
\end{equation}

In summary, we assume that the long-terms views are given in terms of
\begin{itemize}
    \item 
    $\bar x^0\in\reals^{n_0}$: the long-term median values of the stationary risk factors,
    \item
    $d^1\in\reals^{n_1}$: the long-term median drifts of the nonstationary risk factors,
    \item
    $c\in\reals^m$: the long-term median values of the cointegration vectors.
\end{itemize}
Note that we prefer to specify medians instead of means. When the innovations are Gaussian (or more general symmetrically distributed random variables), there is no difference between the two but the components of $x$ are often obtained through nonlinear monotone transformations of economic risk factors. Medians pass directly through the transformations but means do not. More specifically, if $x^i=g^i(X^i)$ and $X^i$ has median $\bar X^i$, then $x^i$ has median $g^i(\bar X^i)$.

For \eqref{views2} to be consistent with \eqref{views3}, the specifications must satisfy
\[
A^1d^1=0.
\]
Also, if a row of $A^1$ has only zero elements, the corresponding component of $c$ has to be zero as well. To combine these with possible short-term forecasts, we define the median trajectory recursively by $\bar x_0:=x_0$ (the current values of the risk factors) and
\[
\bar x_t :=\bar x_{t-1}+A\bar x_{t-1}+a\quad t=1,2,\ldots
\]
where $a$ is given by \eqref{along} and, at each step, we overwrite the values of those components of $\bar x_t$ for which forecasts are available.

%% file: doc_model_uk.tex
\section{An implementation for the UK}\label{model_uk}

Following the approach presented above, we now develop a stochastic model for UK pension insurers. On the asset side, the model incorporates the most important asset classes used by pension funds in the UK, as described in the Purple Book: cash and deposits (short-term fixed income instruments), equities, government bonds, inflation-linked bonds, and corporate bonds. Additional asset classes can be easily included, if necessary. On the liabilities side, the model incorporates mortality risk as described in Section~\ref{longevity}. In addition, the model incorporates the average weekly earnings (AWE), GDP growth and inflation. Earlier works on stochastic simulation models for UK pensions include the Wilkie model~\cite{wilkie1984stochastic} and the PPI-model~\cite{levelcontributionppi}, but neither of them considered mortality risk. 

\subsection{Historical data}\label{sec:historical_data}

Finding appropriate historical data for the calibration of high-dimensional models with various economic as well as longevity risk factors can be challenging. First, one must deal with the different sampling frequencies found in time series data. Financial data, such as stock indices and bond yields, is often available on a daily basis while macroeconomic indicators such as inflation and GDP are usually available only monthly or quarterly. On a national level, demographic data is only available on a yearly basis. In addition, the available time series usually span different periods of time. Here, we deal with the first problem by adopting the lowest sampling frequency found in the available data sets, and with the second problem by splicing closely related time series. 

We will use the database ``A Millennium of Macroeconomic Data''~\cite{boe1k}, an extensive collection of spliced time series data on the UK economy that has been organized in the Bank of England. This database was our main source of data, as it supplied the historical yields used to extend the bond indices collected from the ICE Index Platform~\cite{iceindexplatform}, and the macroeconomic indicators used in the project: the gross domestic product (GDP), the consumer price index (CPI), and the average weekly earnings (AWE). For the estimation of survival probabilities, we have used demographics data for the UK from the Human Mortality Database~\cite{hmd}. In the following sections, we describe in mode detail the historical data used in the estimation of parameters for the risk factors in our model.

\input{plots_input_data}

\subsubsection{Survival probabilities}\label{longevity_uk}

\begin{figure}[ht]
\begin{center}
\includegraphics[scale=0.4]{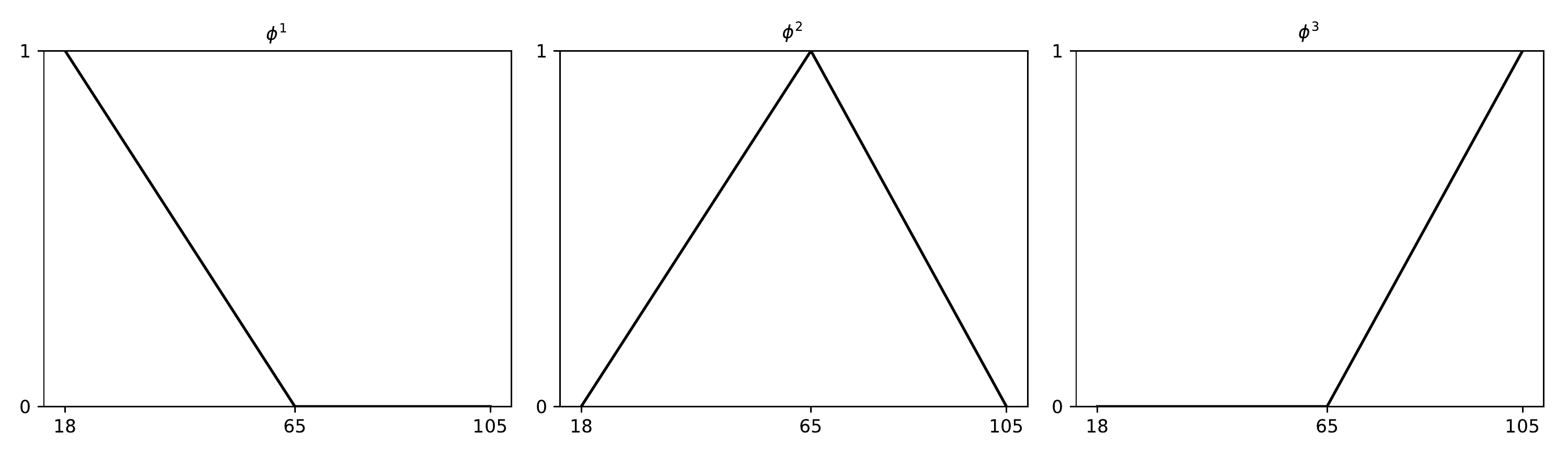}
\caption{Set of basis functions used in the estimation of the historical mortality risk factors. With this particular choice, the risk factors $v^1$, $v^2$, and $v^3$ correspond to logistic survival probabilities of cohorts with age 18, 65, and 105.}\label{fig:basis_functions}
\end{center}
\end{figure}

Once a set of basis functions is specified, survival probabilities can be easily estimated from demographics data by maximizing the likelihood function shown in Section~\ref{longevity}. Following~\cite{aro2011user}, we adopt three risk factors for each gender, and the piecewise linear basis functions
{\small
\begin{equation*}
 \phi^1(x) = 
  \begin{cases} 
   \frac{65 - x}{47} &  x \leq 65 \\
   0                 &  x \geq 65
  \end{cases},\quad
 \phi^2(x) = 
  \begin{cases} 
   \frac{x-18}{47}    &  x\leq 65 \\
   \frac{105-x}{40}   &  x \geq 65
  \end{cases},\quad
 \phi^3(x) = 
  \begin{cases} 
   0               &  x\leq 65 \\
   \frac{x-65}{40} &  x \geq 65
  \end{cases};
\end{equation*}}
see Figure~\ref{fig:basis_functions}. Equation~\eqref{eq_logit_risk_factor} implies that
\[
v^{1}_{t}=\text{logit }p_{18,t},\quad 
v^{2}_{t}=\text{logit }p_{65,t},\quad 
v^{3}_{t}=\text{logit }p_{105,t},  
\]
i.e, $v^1_t$, $v^2_t$ and $v^3_t$ are the logistic survival probabilities of the 18, 65 and 105 year old individuals. Such simple interpretations of the risk factors facilitates the development of stochastic models as it helps assessing the sensibility of parameter estimates and, in particular, the incorporation of views concerning the future development of the factors.

Using historical population sizes and numbers of deaths for the UK from the Human Mortality Database~\cite{hmd}, we estimated the yearly historical longevity risk factors in the period between the years 1922 and 2016. The yearly values are plotted in Figure~\ref{fig:historical_survival_risk_factors}. As one would expect, the logistic survival probabilities decrease with age, and, in general, females have a larger chance of survival than males of similar age. There is a strong positive trend in all ages but the younger cohorts seem to have already reached a saturation level. Similar phenomena con be observed in most developed countries~\cite{aro2014stochastic}. One can also observe a decrease in the survival probabilities of young males during the Second World War. Figure~\ref{fig:hist_surv_prob} plots the historical survival ratios $E_{a+1,t+1}/E_{a,t}$ and the estimated survival probabilities for comparison. The overall shapes are similar but, as expected, the probability surfaces are smoother as they do not incorporate the idiosyncratic binomial risk.

\input{plots_mortality_risk_factors_uk}

\subsubsection{Short-term bonds}

Returns for the portfolio of short-term bonds in our model are based on the YTMs of the ``ICE BofA 1-4 Year UK Gilt Index'', found in the ICE Index Platform with the code GFL0. Given the short length of this time series, we extend it with the ``Bank of England/Official Interest Rates in the UK'' yields for the period between 1948 and 1998, found in the Millennium database.
The values recorded in the time series match quite well, as illustrated in Figure~\ref{fig:ytm_match}, where the daily values of the YTM for the ICE index is plotted against the the annual average values found in the Millennium database.
%For the duration of the portfolio, we assume that it is annually adjusted and kept at the constant value of 2.3 by its portfolio manager.

\subsubsection{Long-term bonds}\label{sec:govt_bonds}

For the portfolio of long-term government bonds, we adopt the ``ICE BofA 5-10 Year UK Gilt'' fixed income index, with code G6L0 in the ICE Index Platform. As for short-term bonds, we obtain the YTM of the portfolio by splicing two time series. In this case, the G6L0 index and the ``Yield on 10 year/medium-term British Government Securities'' historical yields (1948-1991) from the Millennium database. In the top-right corner of Figure~\ref{fig:ytm_match}, one can notice how well both time series match. %Finally, this portfolio is also considered to be annually adjusted to maintain a constant duration value of 6.0. 

\subsubsection{Inflation-linked bonds}

For the YTM of our portfolio of inflation-linked bonds (ILBs), we adopt the difference between the yields on long-term government bonds and long-term inflation expectations (LTIE), both found in the Millennium database. This proxy matches the YTM of the ``ICE BofA 5-15 Year UK Inflation-Linked Gilt Index'' (GWLI) fixed income index, as illustrated in Figure~\ref{fig:ytm_match}. %Based on the historical data of the GWLI index, we assume the constant value of 7.0 for the duration of the portfolio.

\subsubsection{Corporate bonds}

The YTMs of the portfolio of corporate bonds are obtained by extending the ``ICE BofA Sterling Industrial Index'' (UI00) index with the historical yields found in the Millennium database for the years between 1948 and 1996. These correspond to a spliced series of ``Yield on debentures, loan stocks and other corporate bonds 1945-2005'' that have been collected from several sources and the ``Sterling Corporate bond yields on industrials rated AAA-BBB'' fixed income index from BofA Merrill Lynch Global Research. As shown in Figure~\ref{fig:ytm_match}, the YTM time series match quite well. %For the duration of the portfolio, we assume the constant value of 9.2, based on the historical data of the UI00 index. 

\input{plots_bond_returns_proxies_uk}

\subsubsection{Stock index}

The stock price index in our model is a spliced time series taken from the Millennium database. It is composed of the ``Actuaries' Investment index of ordinary industrial shares'',  from various issues of the Monthly Digest of Statistics, and the ``FTSE All-share'' index, scaled to 100 in the $10^{th}$ of April, 1962.

\subsection{Data transformations}\label{data_transformations}

Power transformations, such as those introduced in~\cite{box1964analysis} and~\cite{yeo2000new}, are commonly used in regression models to correct for skewness and non-gaussianity found in regression residuals. Here, we express most risk factors in real terms (discounting inflation) and use special cases of the Box-Cox transform to correct for the skewness and lack of gaussianity found in historical data before the estimation of parameters for our time series model, which is assumed to contain only gaussian risk factors.

The simplest transform used in our model is the natural logarithm, used to map positive values into the whole real line. This transform is commonly used with stock prices and here is used for the stock price index and AWE, which gives us the real earnings $E = \ln\lrp{\sfrac{\AWE}{\CPI}}$ with the discounting of inflation. Including a small shift, we obtain a transformation that allows for slightly negative values and, therefore, is particularly useful when dealing with the credit spread and real yields. For the latter we use the transformation  
\[
Y_t = \ln\lrp{\frac{\YTM_t}{\sfrac{\CPI_t}{\CPI_{t-1}}} + \mu },
\]
where $\YTM_t$ stands for the yield-to-maturity and $\mu\ge 0$ for the aforementioned shift. 

It is interesting to note that real yields are modeled as non-negative in~\cite{wilkie1995more}. We allow negative values to reflect current market conditions. As described in a series of articles by the Financial Times~\cite{ftnegativeyields}, negative yields have appeared in Japan, Sweden, Switzerland, Denmark and the euro zone. Also in the UK, as reported in~\cite{justsaynonegativeyields,impactnegativeyields}.

Similarly to~\cite{wilkie1984stochastic}, we model inflation using the log-growth transform, $\Delta \ln x_t := \ln \lrp{ \sfrac{x_t}{x_{t-1}}}$, of the CPI, %Finally, discounting inflation, we obtain 
and the real GDP-growth as $G=\Delta\ln\lrp{\sfrac{\GDP}{\CPI}}$. Long-term inflation expectations (LTIE) are modeled through their spread over inflation. Mortality risk factors are modeled as such. For a summary of all risk factors, transforms and their parameters, please refer to Table~\ref{risk_factors_uk}. The historical data is plotted in Figure~\ref{fig:historical_data}.

\begin{table}[ht]
\begin{center}
\resizebox{\textwidth}{!}{
\begin{tabular}{lrcc}
\toprule
 Risk factor & Description & Transform & Parameters \\
\midrule
$I$ & CPI log-growth (inflation) & $\Delta\ln\lrp{\CPI}$ & -- \\
\rowcolor{black!5} $\hat{I}$ & Inflation expectation spread & $\LTIE - I$ & -- \\
$G$ & Real GDP log-growth  & $\Delta\ln\lrp{\sfrac{\GDP}{\CPI}}$  & -- \\
\rowcolor{black!5} $E$ & Log real AWE  & $\ln\lrp{\sfrac{\AWE}{\CPI}}$ & -- \\
$S$ & Log stock index  & $\ln\lrp{\STOCK}$ & -- \\ 
\rowcolor{black!5} $Y^{s}$ & Short-term bonds real YTM & $\ln\lrp{\frac{\YTM^{s}}{\sfrac{\CPI_t}{\CPI_{t-1}}} + \mu }$ & $\mu=0.05$ \\
$Y^{l}$ & Long-term bonds real YTM  & $\ln\lrp{\frac{\YTM^{l}}{\sfrac{\CPI_t}{\CPI_{t-1}}} + \mu }$ & $\mu=0.05$ \\
\rowcolor{black!5} $C$ & Log credit spread  & $\ln\lrp{\YTM^{corp} - \YTM^{l} + \mu}$ & $\mu=0.01$ \\
$v^{i,m}$, $v^{i,f}$ & Mortality risk factors, $i=1,2,3$  & --  & -- \\ 
\bottomrule
\end{tabular}
}
\end{center}
\caption{Risk factors for the UK model}\label{risk_factors_uk}
\end{table}

In summary, with the application of the invertible transformations described above, the random vector in our time series model is 
\[
x_t = \left[I_t, \hat{I}_t, G_t, E_t, S_t, Y_t^{s}, Y_t^{l}, C_t, v^{1,m}_{t}, v^{2,m}_{t}, v^{3,m}_{t}, v^{1,f}_{t},v^{2,f}_{t},v^{3,f}_{t}\right]^T.
\]

\subsection{Model calibration}

\subsubsection{Calibration to historical data}\label{calibration_historical_data}

The estimated autoregressive matrix $A$ is given in Table~\ref{tab:A}. All the remaining parameters have p-values less than $2.2\%$. The negative diagonal elements correspond to mean reversion in the corresponding risk factors. 
As in~\cite{wilkie1984stochastic}, the real yields in the model are mean-reverting.
The only nonstationary risk factors are the stock index, average weekly earnings and the longevity risk factors $v^{2,m}$ and $v^{2,f}$. For the first two, the nonstationarity is quite natural as the two indices are expected to have a positive drift. The nonstationarity of the longevity risk factors is more questionable in the long run but looking at the historical data, it is not surprising that the mean-reversion is not statistically significant. 
As to the nondiagonal elements, the GDP-growth affects the short- and long-term yields positively while for the credit-spread the effect is negative. This is quite natural as during strong economic growth, one expects the overall level of interest rates to increase due to monetary policy decisions, while the credit spreads tend to widen during economic crisis. The short- and long-term real yields are also affected by inflation which may seem puzzling at first. The explanation is that the GDP risk factor is in fact the {\em real} GDP-growth so adding a positive multiple of the inflation makes the regressor become closer to the nominal GDP-growth. The average weekly earnings affect the old-age longevity factors $v^{3,m}$ and $v^{3,f}$ positively. This is quite natural as higher income levels have been found to have a positive effect on the health of the population; see e.g. \cite{aro2014stochastic,glover.mortality} and the references therein. 

The correlation matrix and variance vector for the risk factors of the time series model are presented in Table~\ref{tab:sigma}. No historical data before 1985 was used in the estimation of the empirical covariance matrix to avoid periods when inflation was too high. Such periods are unlikely to occur again, as the UK has adopted an inflation targeting policy in 1992~\cite{allsopp2006united}. 

% notes to comment on covariance matrix:
% significant correlations:
% GDP, INF: -0.33 (explained by inflation control?)
% AWE, INF: 0.62 (increase in wages increase the consumption and then inflation)
% Y^s, AWE: 0.33 (increase in wages increases the demand for short-term loans as people buy durable goods)
% Y^l, Y^s: 0.5
% C, GDP: -0.41
% C, Y^l: -0.55
% \hat{I}, INF: -0.84
% \hat{I}, GDP: 0.53

\input{table_matrix_autoregression}

\input{table_matrix_correlation}

\subsubsection{Duration of bond portfolios}

We start the calibration for the returns on bond portfolios by validating the choice of the first-order model~\eqref{bondreturn}. Using monthly historical data for the ICE BofA indices mentioned in Section~\ref{sec:historical_data}, we compute the approximation errors for the models~\eqref{bondreturn} and~\eqref{bondreturn_order2}. The results in Table~\ref{bonds_r2} validate the choice of the first-order model, as no significant gains were obtained with the inclusion of the convexity term in the second-order model. The $R^2$ values for portfolios of fixed rate bonds (GFL0 and G6L0) are, respectively, 98.7\% and 99.6\%. The smaller value obtained for corporate bonds is due to the the credit risk which is not captured by formula \eqref{bondreturn}. We will describe the credit risk by modeling $\Delta\ln I$ as a nonpositive random variable; see Section~\ref{calibration_corporate_bonds}.

Our bond portfolios are assumed to be annually adjusted by a portfolio manager to maintain constant duration values. Those chosen for our portfolios can be found in Table~\ref{bonds_r2}, and annual values for their historical duration are illustrated in Figure~\ref{fig:bond_durations}.

\begin{table}[h!]
\begin{center}
%\makebox[\textwidth][c]{
\begin{small}
%\resizebox{\textwidth}{!}{
\begin{tabular}{llll}
\toprule
Portfolio & $R^2$ (no convexity)& $R^2$ (with convexity) & Duration\\
\midrule
Short-term bonds &  0.987 & 0.987 & 2.30\\
\rowcolor{black!5} Long-term bonds &  0.996 & 0.996 & 6.00\\
Inflation-linked bonds &  0.954 & 0.933 & 7.00\\
\rowcolor{black!5} Corporate bonds & 0.964 & 0.964 & 9.20\\
\bottomrule
\end{tabular}
%}
\end{small}
\end{center}
\caption{$R^2$ of bond return approximations and duration values}\label{bonds_r2}
\end{table}

\input{plots_bond_durations}

\subsubsection{Corporate bonds}\label{calibration_corporate_bonds}

In the case of corporate bonds, the effect of default losses is represented by the term $\Delta\ln I$ in~\eqref{bondreturn}. Historical values for the recovery rate can be computed by  
\[
\Delta\ln I \approx \Delta\ln P - Y_t \Delta t + D_t \Delta Y
\]
from historical prices, yields and durations, as shown in Section~\ref{asset_returns}; see Figure~\ref{fig:corporate}. The values for the historical recovery rate are mostly negative. We will model the recovery rates as iid random variables with
\[
\log\lrp{-\Delta\ln I + \delta} \sim N\lrp{\mu, \sigma^2}.
\]
The qq-plot in Figure~\ref{fig:corporate} shows the adherence of the recovery rates to the log-normal model. With a shift $\delta = 0.1$, the maximum likelihood estimators for the mean $\mu$ and variance $\sigma^2$  are respectively equal to $-2.29$ and $7.47e-4$.

\input{plots_corporate_bonds}

\subsubsection{User views}\label{views}

Short-term forecasts and views on the long-term developments of the median values of the risk factors can be incorporated into the model, as described in Section~\ref{ssec:views}. For the experiments in Section~\ref{simulation_results}, we choose not to include short-term forecasts, and adopt the long-term medians present in Table~\ref{table:views}. The chosen values could be questioned and should only be taken as an illustration. The values chosen for the real YTM on short and long-term bonds reflect a belief that those will increase in the long run, ending the current period of low interest rates. For inflation and inflation expectation spreads, the chosen values consider the well-known inflation targeting policy adopted by the Bank of England~\cite{inflationtargeting2}. 
%Given its existence, a non-zero spread for inflation expectations seems unrealistic. 

\begin{table}
\begin{center}
\begin{small}
\begin{tabular}{llc}
    \toprule
    Risk factor  & Value\\
    \midrule
    CPI log-growth (inflation) & 0.02 \\
    \rowcolor{black!5}Inflation expectation spread & 0.00 \\
    Real GDP log-growth & 0.02 \\
    \rowcolor{black!5}Short-term bonds real YTM & 0.02 \\
    Long-term bonds real YTM & 0.04 \\
    \rowcolor{black!5}Credit spread & 0.02 \\
    \bottomrule
\end{tabular}
\caption{Views on long-term medians used in the simulation of the risk factors}\label{table:views}
\end{small}
\end{center}
\end{table}

%% file: plots_input_data.tex
\begin{figure}[ht]
\begin{center}
\makebox[\textwidth][c]{
\includegraphics[scale=0.5]{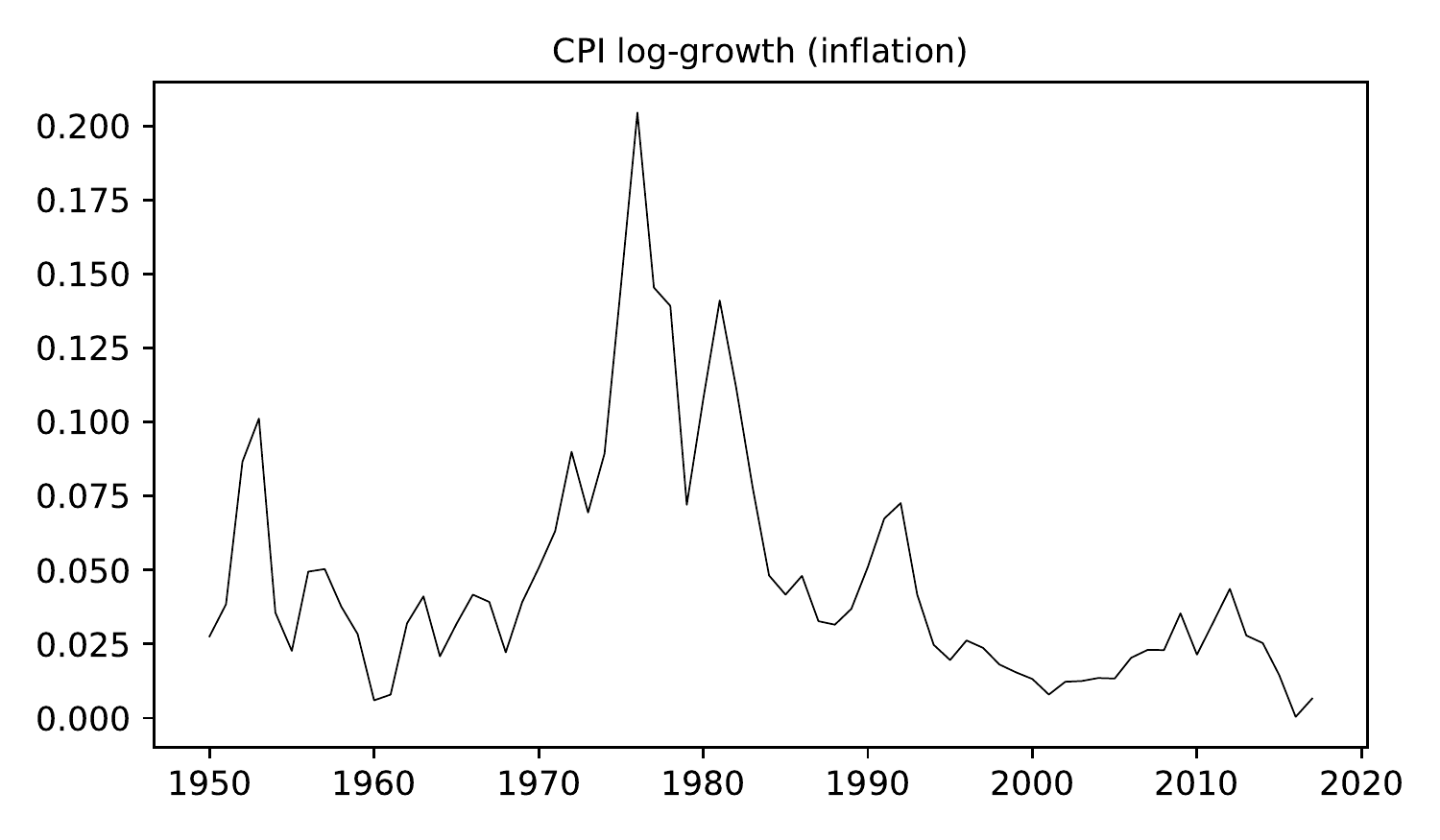}\includegraphics[scale=0.5]{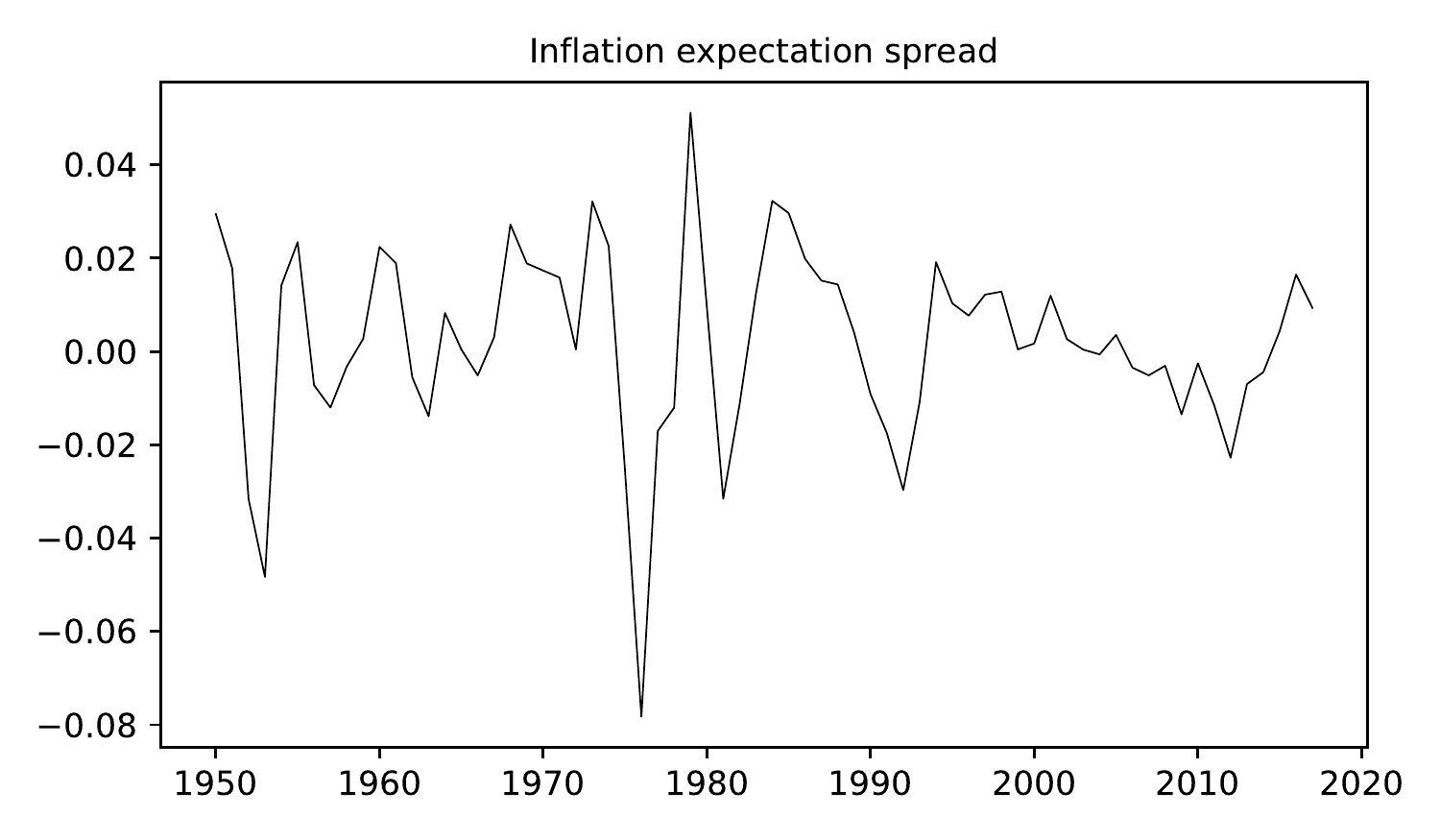} 
}
\makebox[\textwidth][c]{
\includegraphics[scale=0.5]{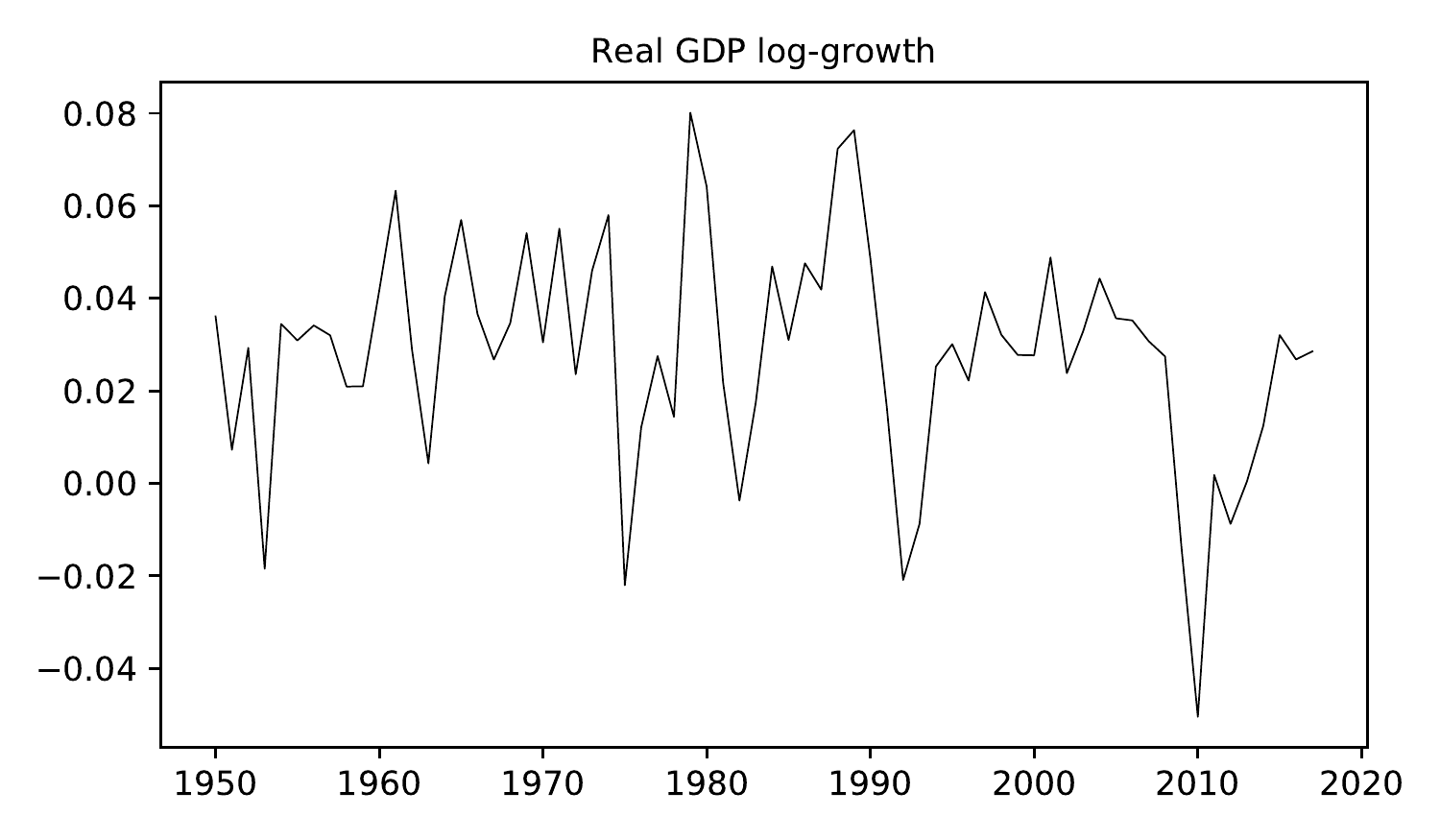}\includegraphics[scale=0.5]{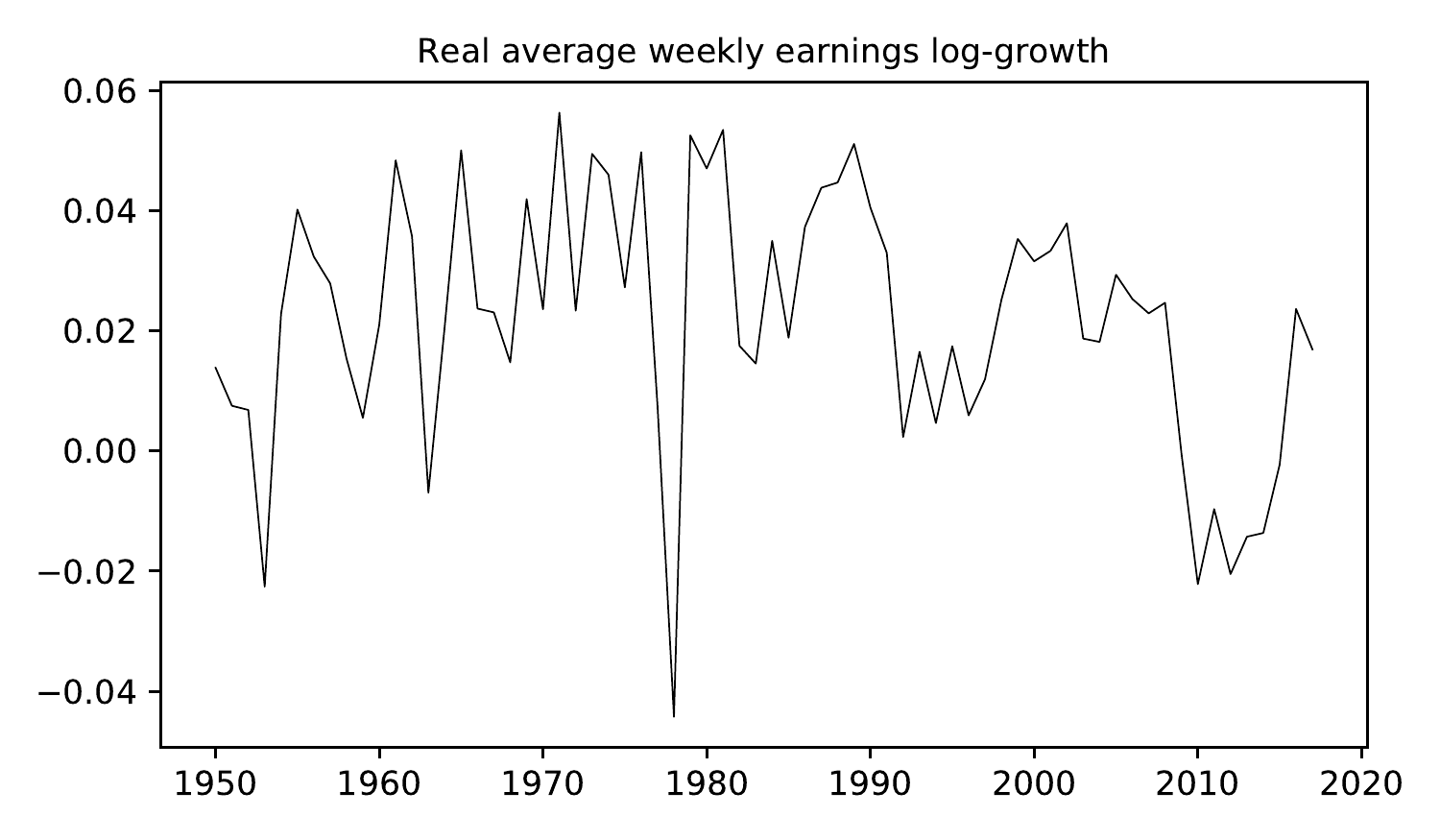} 
}
\makebox[\textwidth][c]{
\includegraphics[scale=0.5]{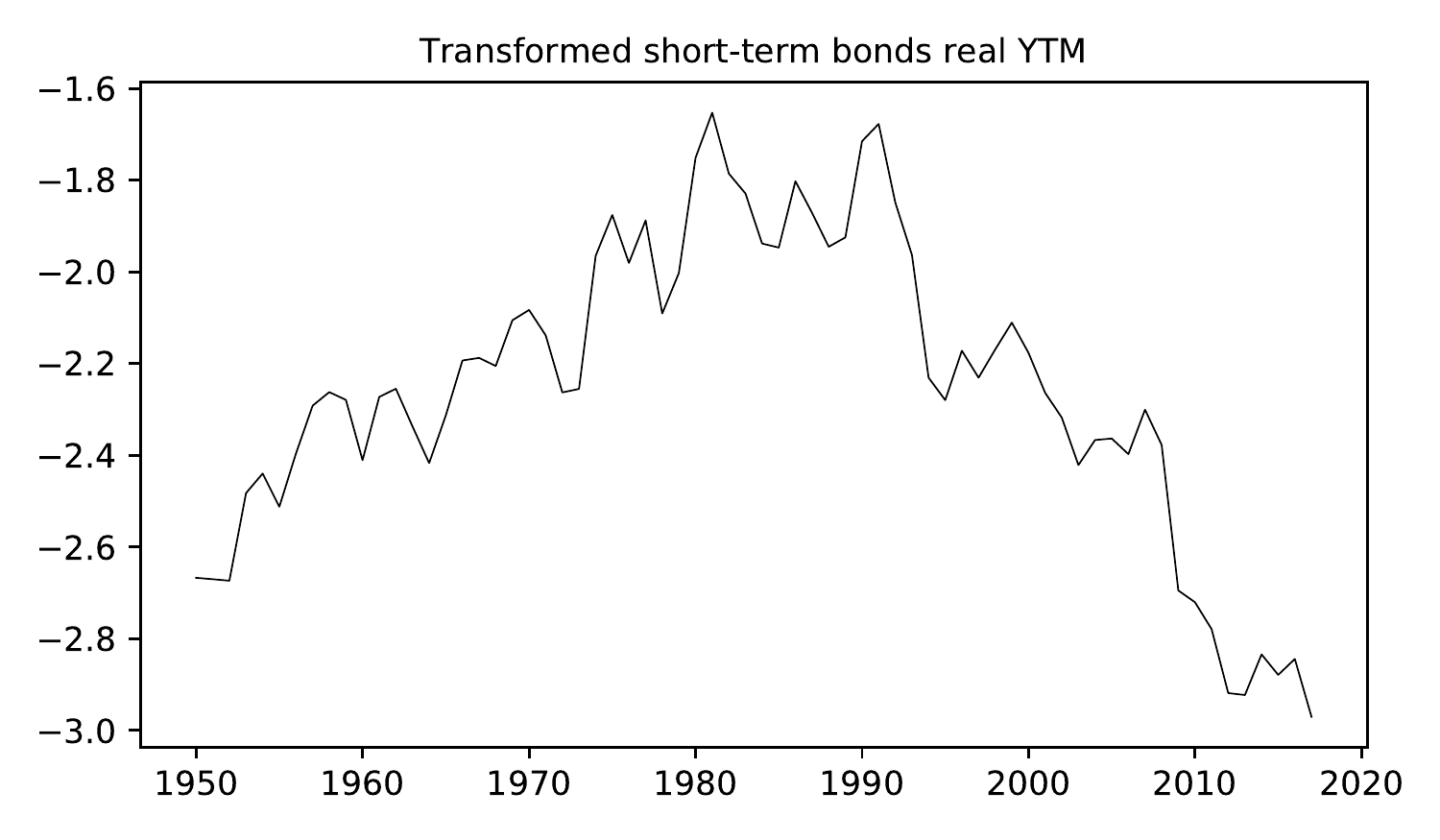}\includegraphics[scale=0.5]{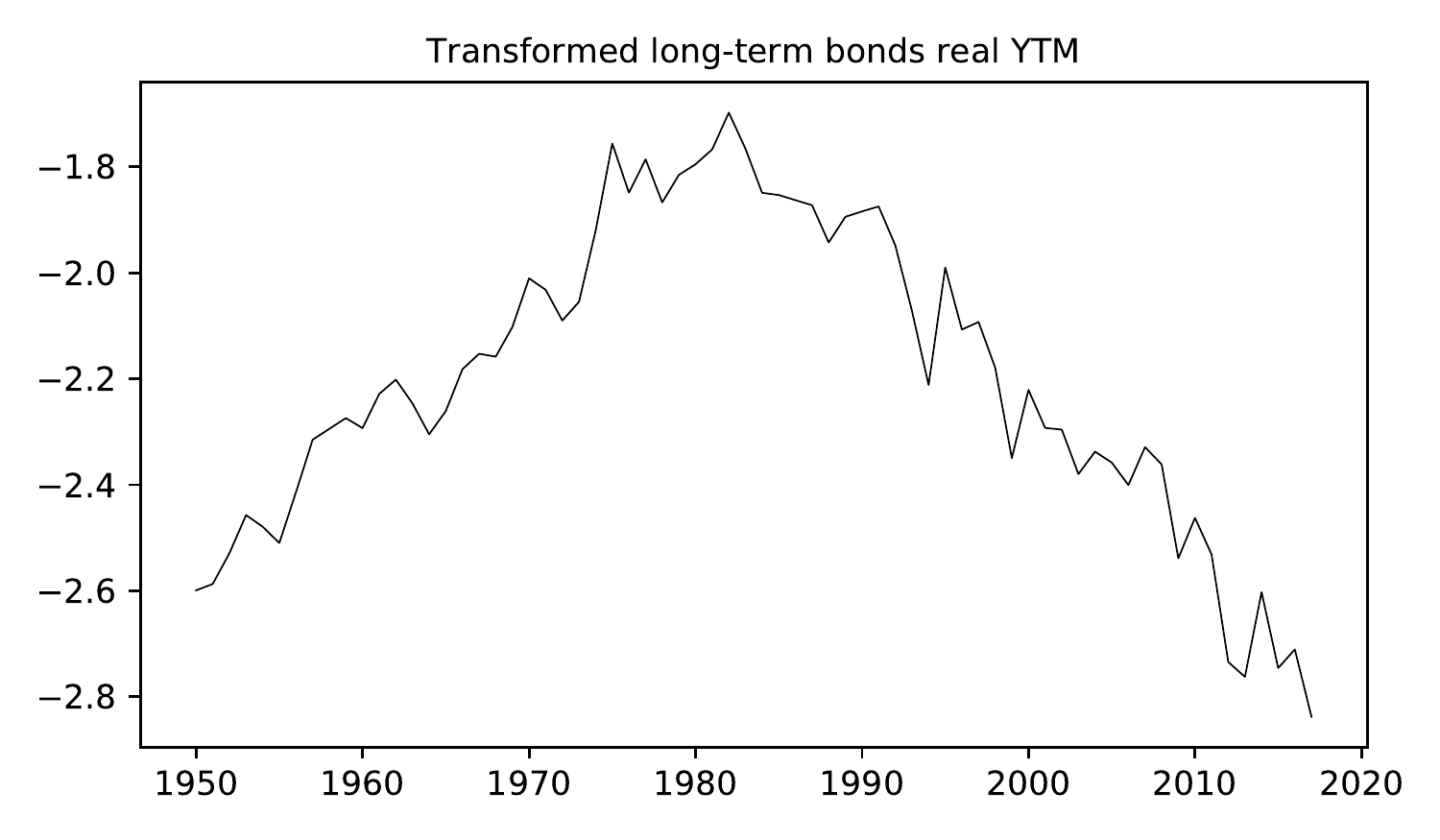} 
}
\makebox[\textwidth][c]{
\includegraphics[scale=0.5]{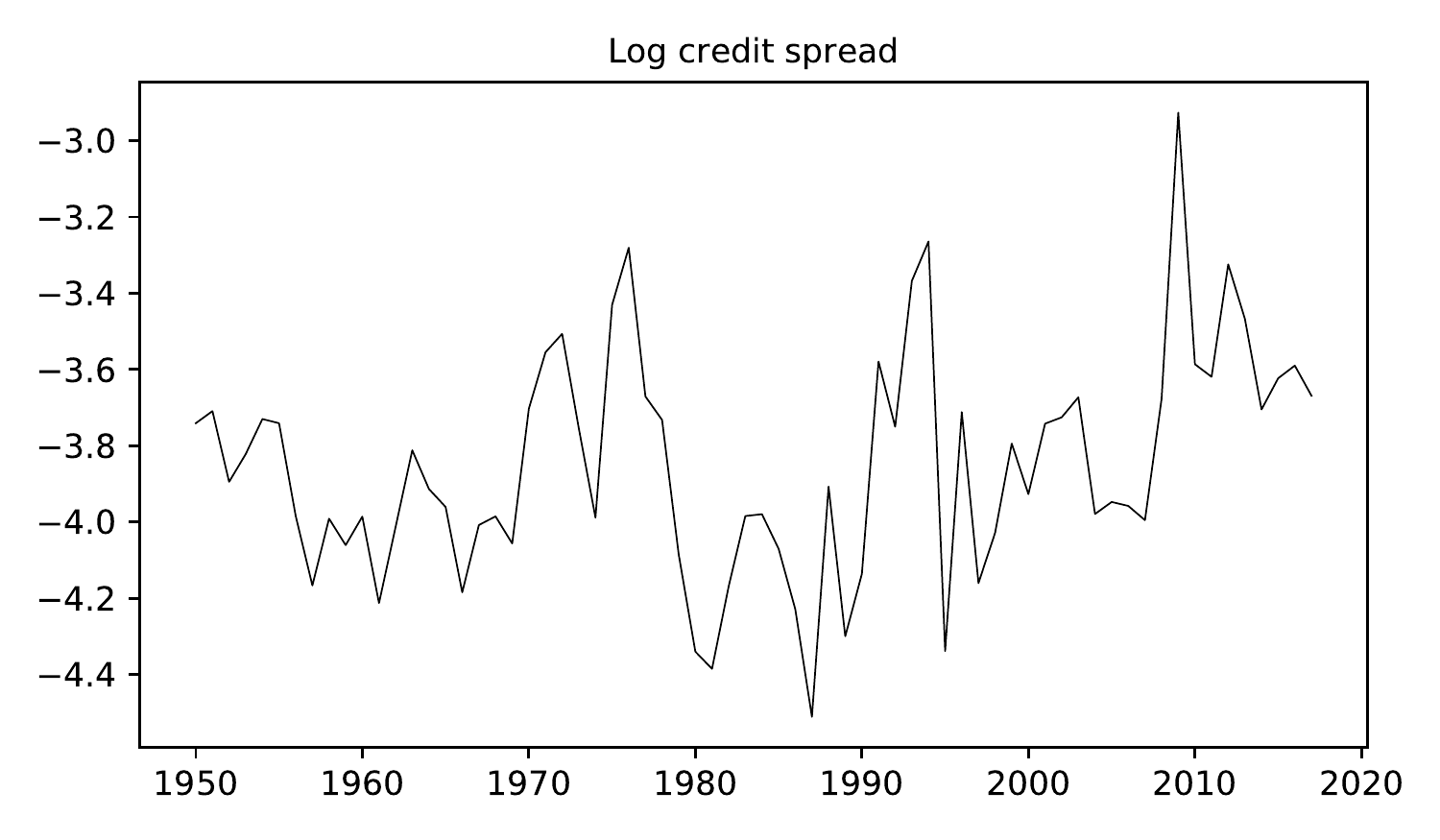}\includegraphics[scale=0.5]{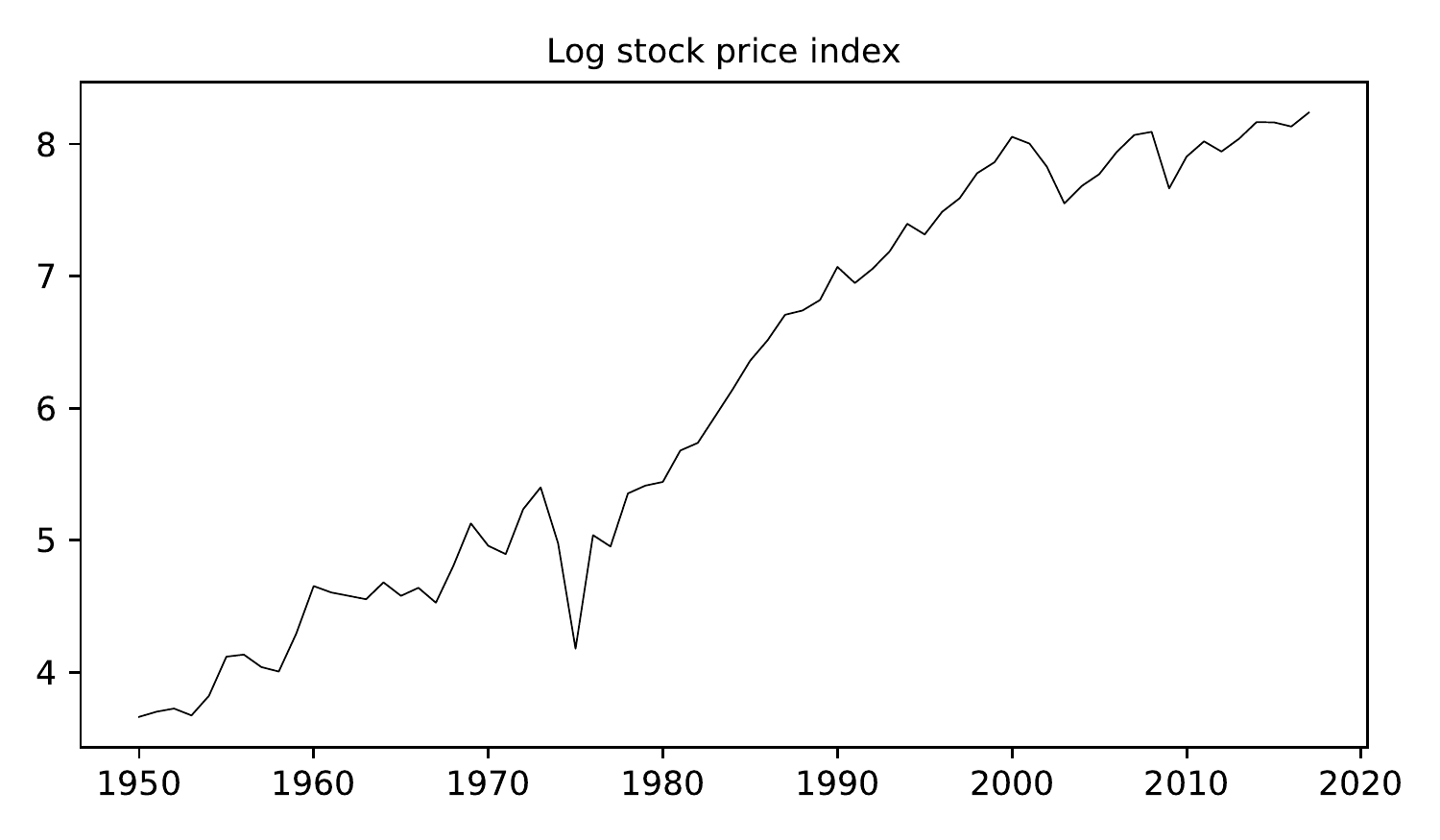} 
}
\end{center}
\caption{Historical data used in the calibration of the time series model.}\label{fig:historical_data}
\end{figure}

%% file: plots_mortality_risk_factors_uk.tex
\begin{figure}[ht]
\begin{center}
\includegraphics[scale=0.6]{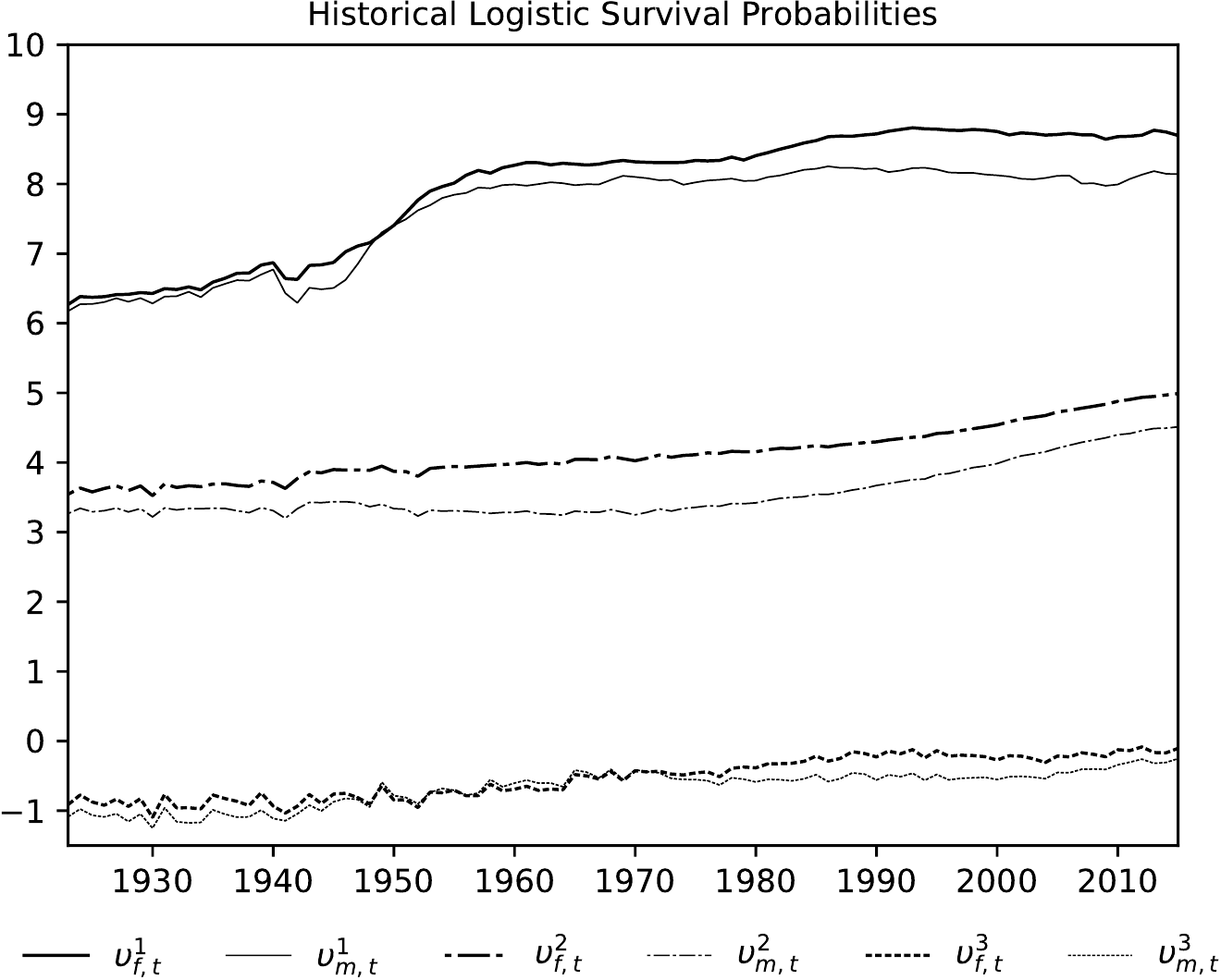}
\end{center}
\caption{Historical values for the mortality risk factors in the UK. The comparison of the risk factors in the plot shows that survival probabilities decrease with increasing age ($v^{1}>v^{2}>v^{3}$) and that, in general, females are more likely to survive than males of similar age ($v^{f} > v^{m}$). The plots also show an interesting feature that is captured by the model: a decrease in the survival probabilities of young males during the Second World War.}\label{fig:historical_survival_risk_factors}
\end{figure}

\begin{figure}[ht]
\begin{center}
\includegraphics[scale=0.45]{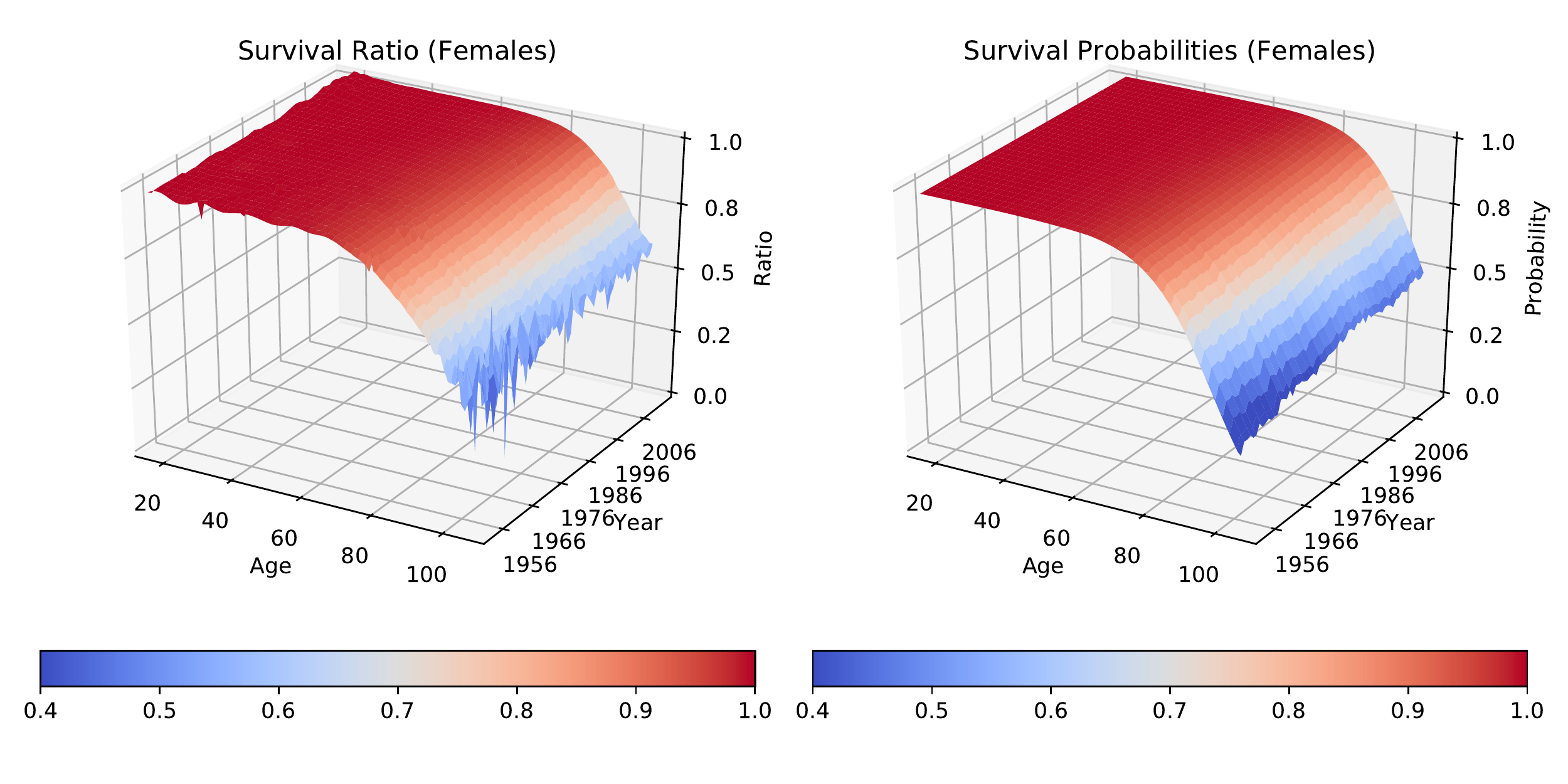} \\
\end{center}
\caption{Comparison of historical survival ratios ($\sfrac{E_{a+1, t+1}}{E_{a, t}}$) with the corresponding estimated survival probabilities for females in the UK.}\label{fig:hist_surv_prob}

\end{figure}

%% file: plots_bond_returns_proxies_uk.tex
\begin{figure}[ht]
\begin{center}
\makebox[\textwidth][c]{
\includegraphics[scale=0.5]{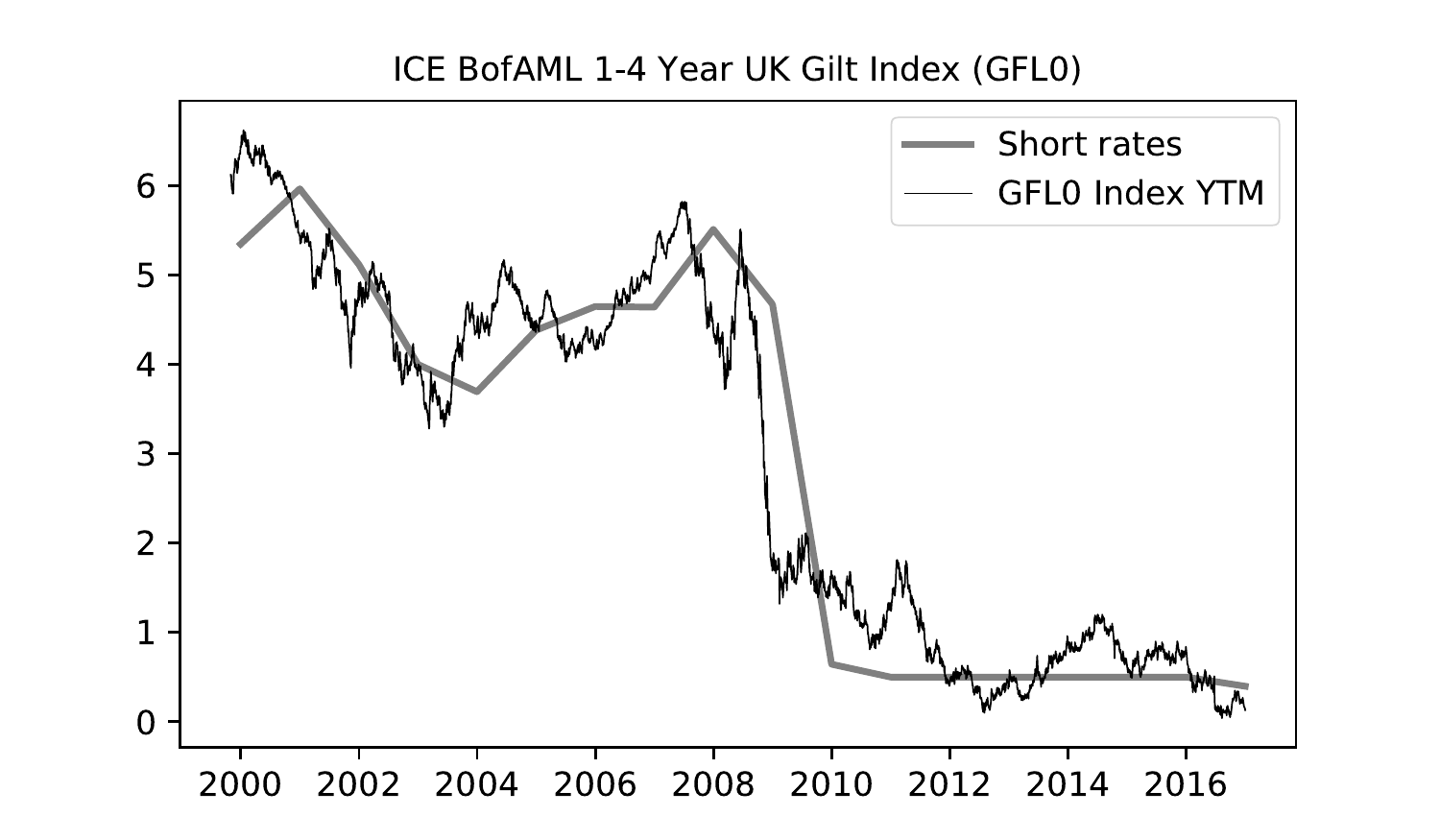}\includegraphics[scale=0.5]{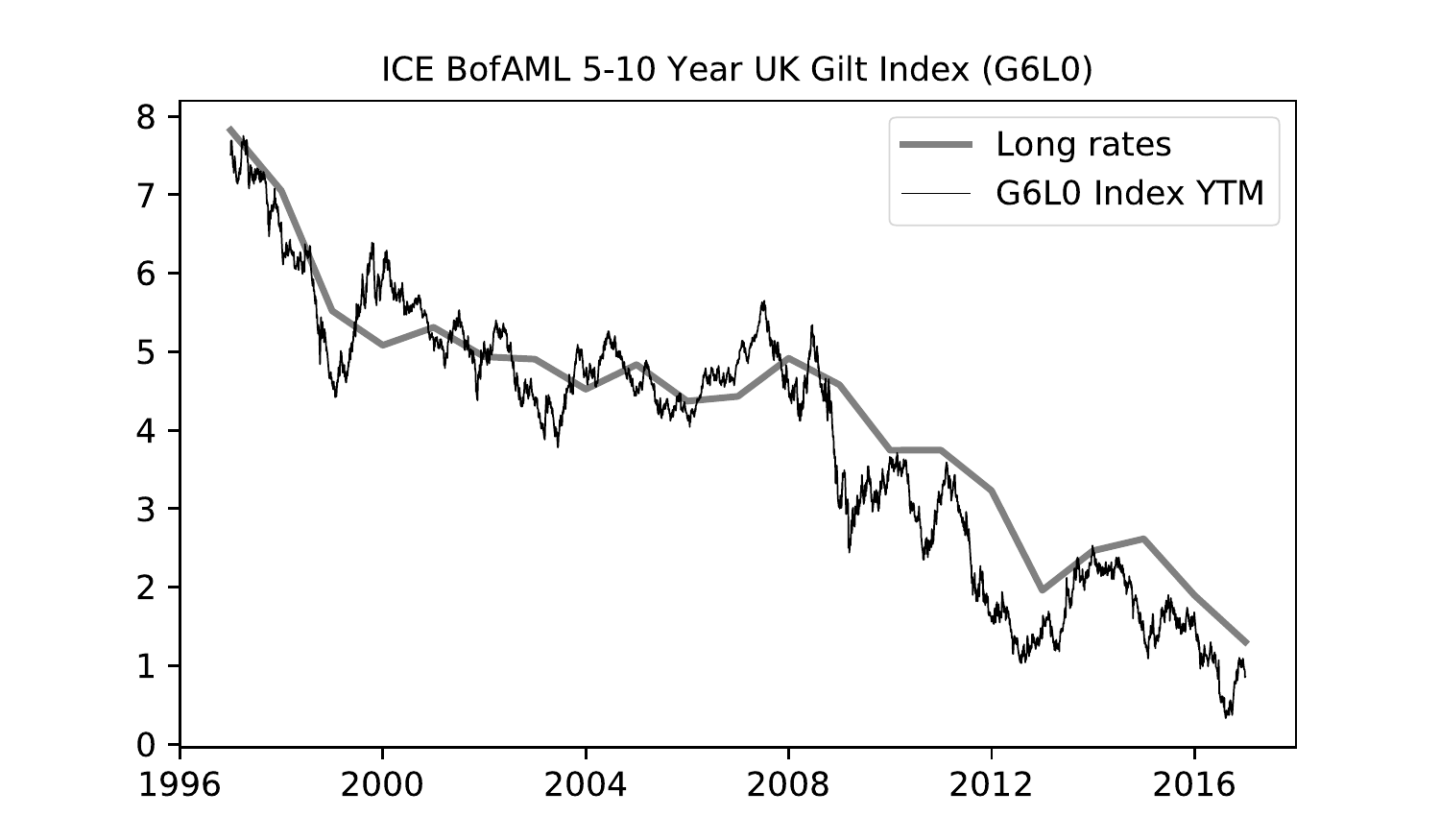}
}
\makebox[\textwidth][c]{
\includegraphics[scale=0.5]{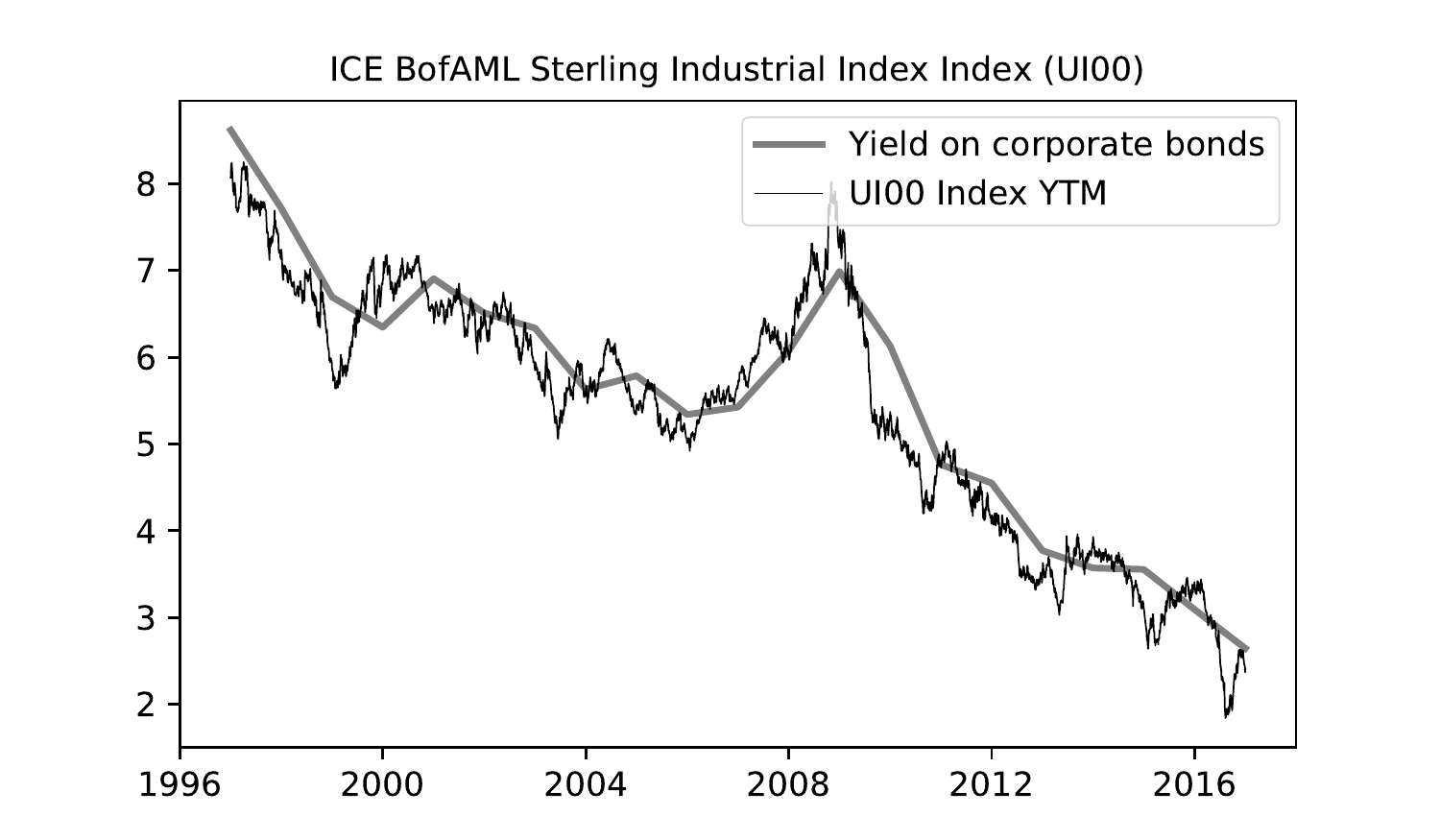}\includegraphics[scale=0.5]{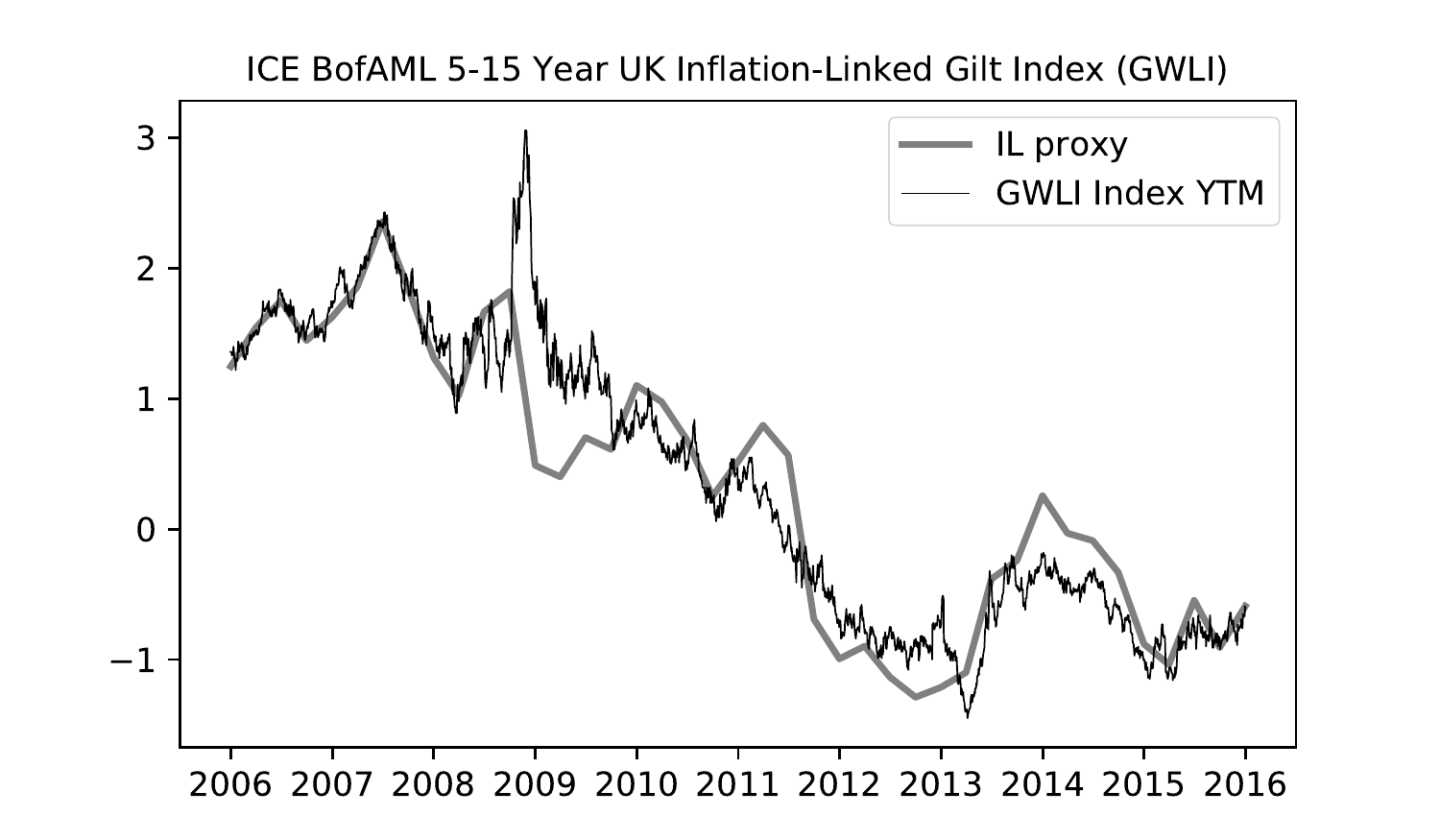}
}
\end{center}
\caption{The pictures illustrate the matching of historical yields found in the Millennium database and in the ICE Index Platform. Yearly values from the latter were spliced with values from the first to provide the long time series necessary for the calibration of the time series model.}\label{fig:ytm_match}
\end{figure}

%% file: table_matrix_autoregression.tex
\begin{center}
\begin{table}[ht]
\resizebox{\textwidth}{!}{
%\begin{scriptsize}
%\begin{tabular}{>{\columncolor{olive!20}}lrrrrrrrrrrrrrr}
\begin{tabular}{lrrrrrrrrrrrrrr}
%\toprule
& $I$ & $G$ & $E$ & $S$ & $Y^{s}$ & $Y^{l}$ & $C$ & $\hat{I}$ & $v^{1,m}$ & $v^{2,m}$ & $v^{3,m}$ & $v^{1,f}$ & $v^{2,f}$ & $v^{3,f}$ \\ 
\midrule
 $I$&-0.16&&&&&&&&&&&&& \\
 &(0.02)&&&&&&&&&&&&& \\
\rowcolor{black!5} $G$&&-0.59&&&&&&&&&&&& \\
\rowcolor{black!5} &&(0.00)&&&&&&&&&&&& \\
 $E$&&0.43&&&&&&&&&&&& \\
 &&(0.00)&&&&&&&&&&&& \\
\rowcolor{black!5} $S$&&&&&&&&&&&&&& \\
\rowcolor{black!5} &&&&&&&&&&&&&& \\
 $Y^{s}$&1.02&2.88&&&-0.16&&&&&&&&& \\
 &(0.00)&(0.00)&&&(0.00)&&&&&&&&& \\
\rowcolor{black!5} $Y^{l}$&0.90&1.76&&&&-0.13&&&&&&&& \\
\rowcolor{black!5} &(0.00)&(0.00)&&&&(0.00)&&&&&&&& \\
 $C$&&-3.64&&&&&-0.66&&&&&&& \\
 &&(0.02)&&&&&(0.00)&&&&&&& \\
\rowcolor{black!5} $\hat{I}$&&&&&&&&-0.61&&&&&& \\
 &&&&&&&&(0.00)&&&&&& \\
 $v^{1,m}$&&&&&&&&&-0.16&&&&& \\
 &&&&&&&&&(0.00)&&&&& \\
\rowcolor{black!5} $v^{2,m}$&&&&&&&&&&&&&& \\
\rowcolor{black!5} &&&&&&&&&&&&&& \\
 $v^{3,m}$&&&0.06&&&&&&&&-0.32&&& \\
 &&&(0.02)&&&&&&&&(0.00)&&& \\
\rowcolor{black!5} $v^{1,f}$&&&&&&&&&&&&-0.10&& \\
\rowcolor{black!5} &&&&&&&&&&&&(0.00)&& \\
 $v^{2,f}$&&&&&&&&&&&&&& \\
 &&&&&&&&&&&&&& \\
\rowcolor{black!5} $v^{3,f}$&&&0.16&&&&&&&&&&&-0.41 \\
\rowcolor{black!5} &&&(0.00)&&&&&&&&&&&(0.00) \\
\bottomrule
\end{tabular}
%\end{scriptsize}
}
\caption{Coefficients of the autoregression matrix $A$ and their corresponding p-values (in parenthesis). Zero elements were omitted for clarity.}\label{tab:A}
\end{table}
\end{center}

%% file: table_matrix_correlation.tex
\begin{center}
\begin{table}[ht]
\resizebox{\textwidth}{!}{
%\begin{tabular}{>{\columncolor{olive!20}}lrrrrrrrrrrrrrr}
\begin{tabular}{lrrrrrrrrrrrrrr}
%\toprule
& $I$ & $G$ & $E$ & $S$ & $Y^{s}$ & $Y^{l}$ & $C$ & $\hat{I}$ & $v^{1,m}$ & $v^{2,m}$ & $v^{3,m}$ & $v^{1,f}$ & $v^{2,f}$ & $v^{3,f}$ \\ 
\midrule
 $I$&1.0&-0.33&-0.05&-0.11&0.04&0.03&0.26&-0.84&-0.03&-0.01&-0.09&0.09&-8.45e-03&-0.03 \\
 &&(0.00)&(0.68)&(0.40)&(0.77)&(0.82)&(0.04)&(0.00)&(0.80)&(0.91)&(0.48)&(0.49)&(0.95)&(0.79) \\
\rowcolor{black!5} $G$&&1.0&0.62&0.22&0.28&0.09&-0.41&0.53&-0.06&-0.20&-0.08&-0.22&-0.26&-0.05 \\
\rowcolor{black!5} &&&(0.00)&(0.08)&(0.02)&(0.46)&(0.00)&(0.00)&(0.64)&(0.10)&(0.52)&(0.08)&(0.04)&(0.68) \\
 $E$&&&1.0&0.14&0.33&0.18&-0.18&0.21&-0.03&-0.15&-0.14&-0.19&-0.20&-0.02 \\
 &&&&(0.27)&(0.00)&(0.14)&(0.14)&(0.09)&(0.78)&(0.24)&(0.24)&(0.13)&(0.11)&(0.90) \\
\rowcolor{black!5} $S$&&&&1.0&-0.01&-0.13&-0.29&0.04&0.30&-0.01&-0.10&0.14&-0.02&0.02 \\
\rowcolor{black!5} &&&&&(0.91)&(0.29)&(0.02)&(0.77)&(0.01)&(0.91)&(0.40)&(0.27)&(0.86)&(0.88) \\
 $Y^{s}$&&&&&1.0&0.50&-0.41&0.02&0.06&-0.09&-0.13&0.18&-0.08&-0.04 \\
 &&&&&&(0.00)&(0.00)&(0.88)&(0.62)&(0.47)&(0.28)&(0.15)&(0.52)&(0.77) \\
\rowcolor{black!5} $Y^{l}$&&&&&&1.0&-0.55&0.05&-0.04&-0.11&-0.04&0.07&-0.08&-0.02 \\
\rowcolor{black!5} &&&&&&&(0.00)&(0.68)&(0.73)&(0.37)&(0.72)&(0.55)&(0.50)&(0.90) \\
 $C$&&&&&&&1.0&-0.24&-0.15&0.07&0.08&-0.21&0.09&-0.02 \\
 &&&&&&&&(0.05)&(0.22)&(0.55)&(0.54)&(0.09)&(0.47)&(0.86) \\
\rowcolor{black!5} $\hat{I}$&&&&&&&&1.0&-0.08&-0.10&0.04&-0.20&-0.16&0.06 \\
\rowcolor{black!5} &&&&&&&&&(0.53)&(0.44)&(0.72)&(0.10)&(0.20)&(0.62) \\
 $v^{1,m}$&&&&&&&&&1.0&-0.12&-0.16&0.46&-0.11&-0.06 \\
 &&&&&&&&&&(0.35)&(0.20)&(0.00)&(0.36)&(0.65) \\
\rowcolor{black!5} $v^{2,m}$&&&&&&&&&&1.0&0.34&0.13&0.89&0.47 \\
\rowcolor{black!5} &&&&&&&&&&&(0.00)&(0.29)&(0.00)&(0.00) \\
 $v^{3,m}$&&&&&&&&&&&1.0&-0.34&0.49&0.82 \\
 &&&&&&&&&&&&(0.00)&(0.00)&(0.00) \\
\rowcolor{black!5} $v^{1,f}$&&&&&&&&&&&&1.0&0.02&-0.16 \\
\rowcolor{black!5} &&&&&&&&&&&&&(0.88)&(0.20) \\
 $v^{2,f}$&&&&&&&&&&&&&1.0&0.49 \\
 &&&&&&&&&&&&&&(0.00) \\
\rowcolor{black!5} $v^{3,f}$&&&&&&&&&&&&&&1.0 \\
\rowcolor{black!5} &&&&&&&&&&&&&& \\
\bottomrule
Variance 
& 5.00e-04
& 4.84e-04
& 3.43e-04
& 0.05
& 8.89e-03
& 5.78e-03
& 0.06
& 3.40e-04
& 1.13e-03
& 8.66e-04
& 4.47e-03
& 1.21e-03
& 6.54e-04
& 4.50e-03
\\
\bottomrule
\end{tabular}
}
\caption{Correlation coefficients and corresponding p-values (in parenthesis) for the risk factors in the time series model for UK pensions insurers. Variances can be found in the bottom row of the table.}\label{tab:sigma}
\end{table}
\end{center}

%% file: plots_bond_durations.tex
\begin{figure}[ht]
\begin{center}
\makebox[\textwidth][c]{
\includegraphics[scale=0.5]{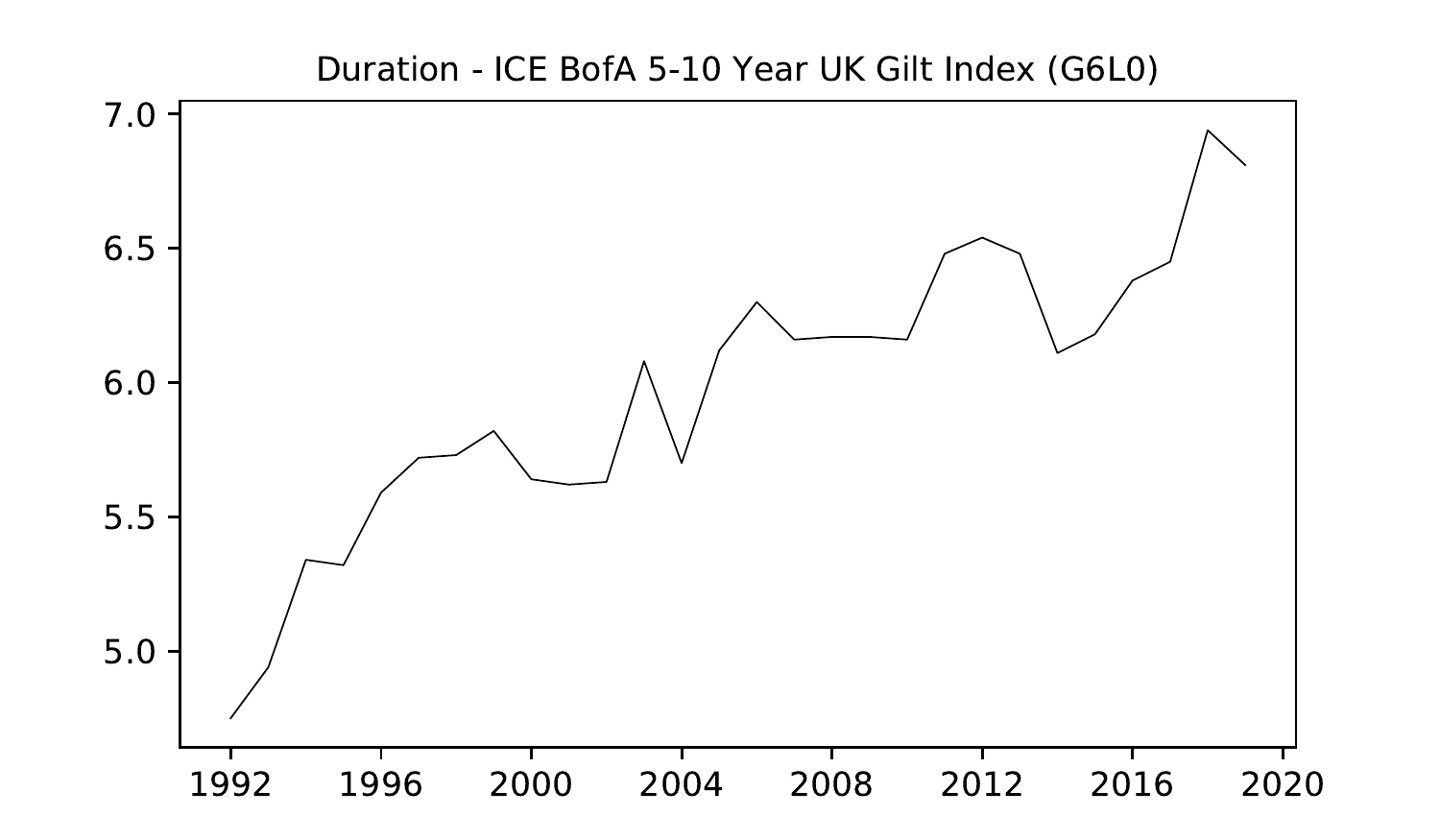}\includegraphics[scale=0.5]{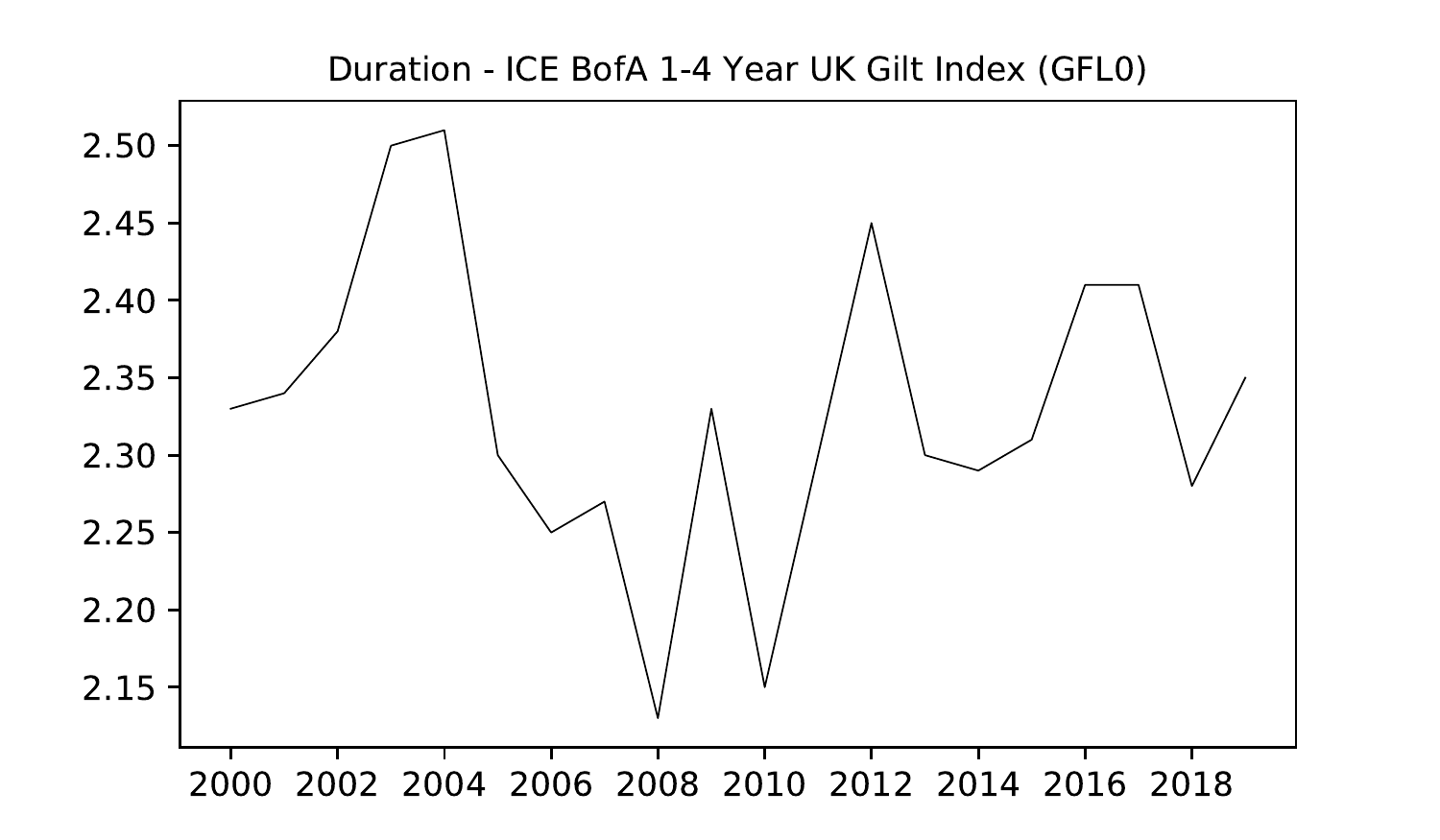}
}
\makebox[\textwidth][c]{
\includegraphics[scale=0.5]{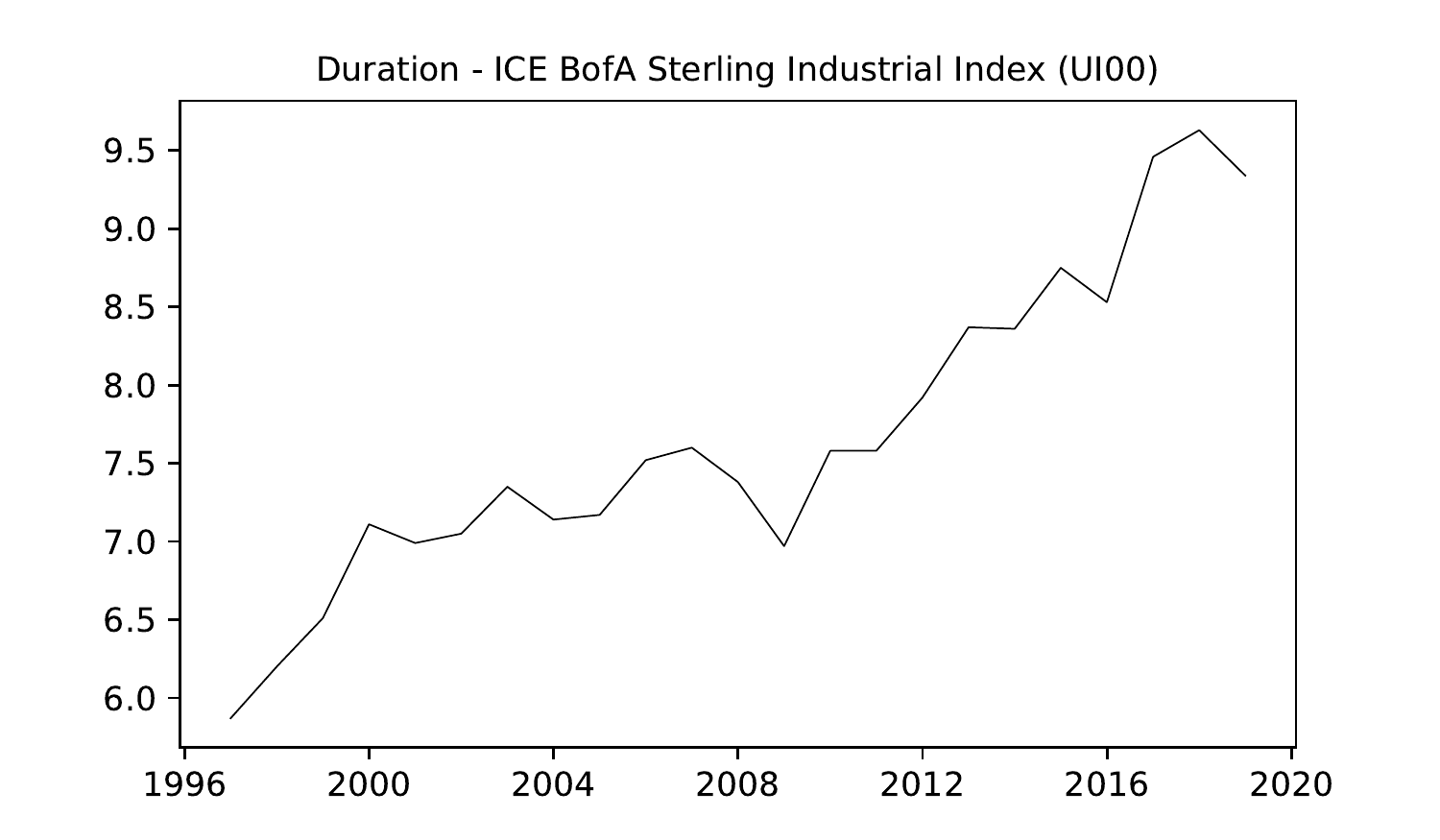}\includegraphics[scale=0.5]{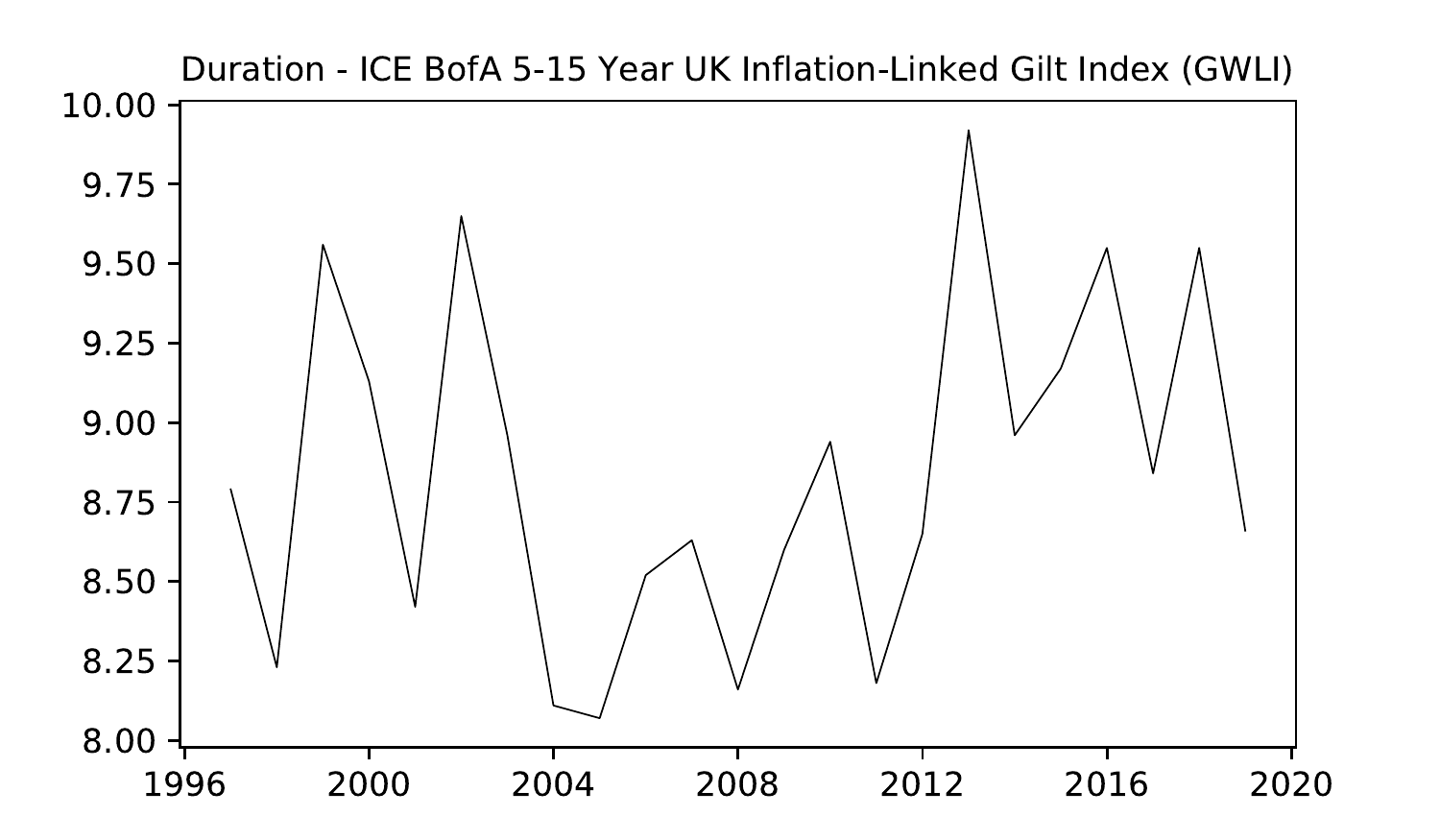}
}
\end{center}
\caption{Historical durations of the bond portfolios used in the UK model.}\label{fig:bond_durations}
\end{figure}

%% file: plots_corporate_bonds.tex
\begin{figure}[ht]
\begin{center}
\makebox[\textwidth][c]{
\includegraphics[scale=0.5]{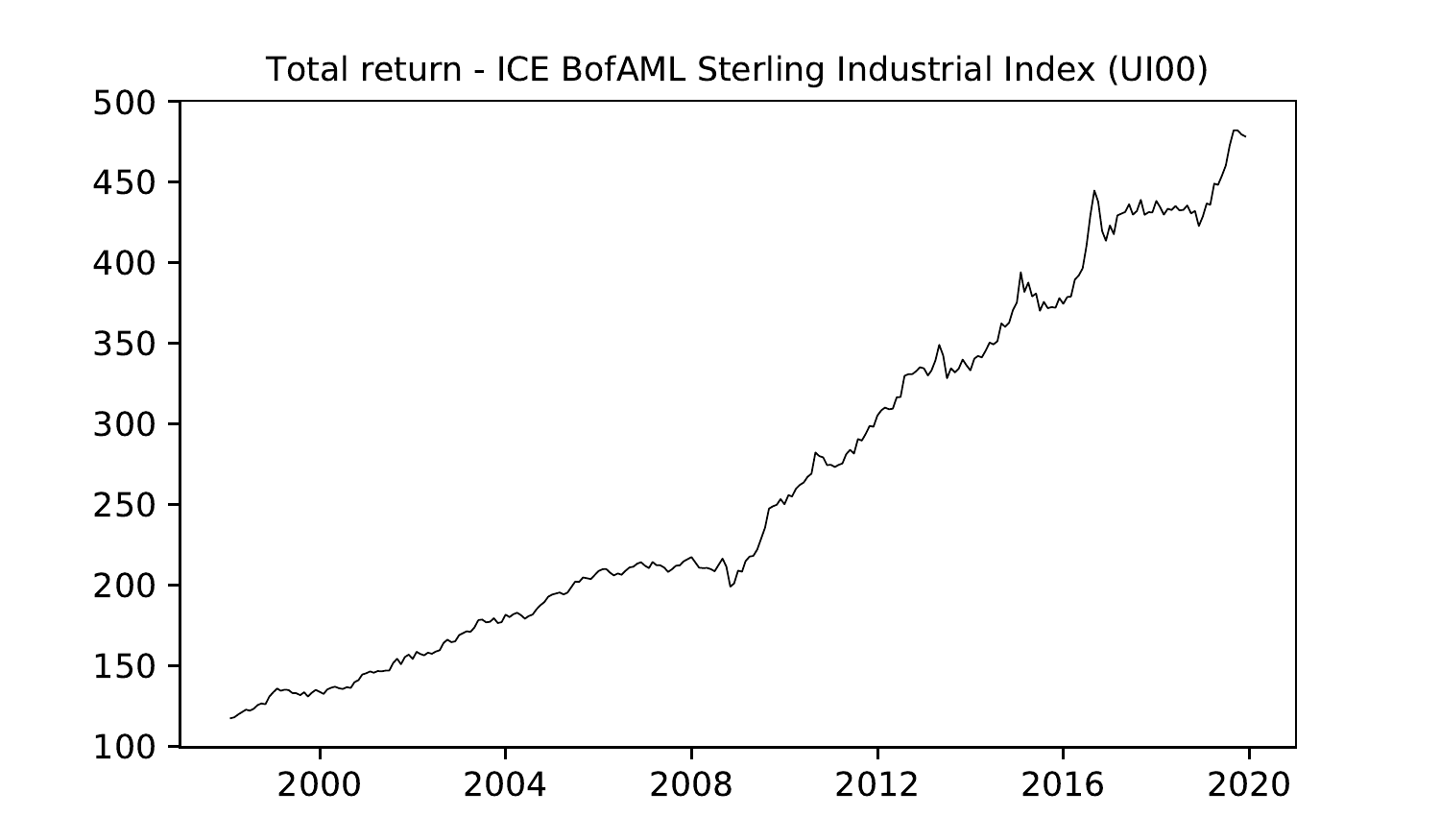}\includegraphics[scale=0.5]{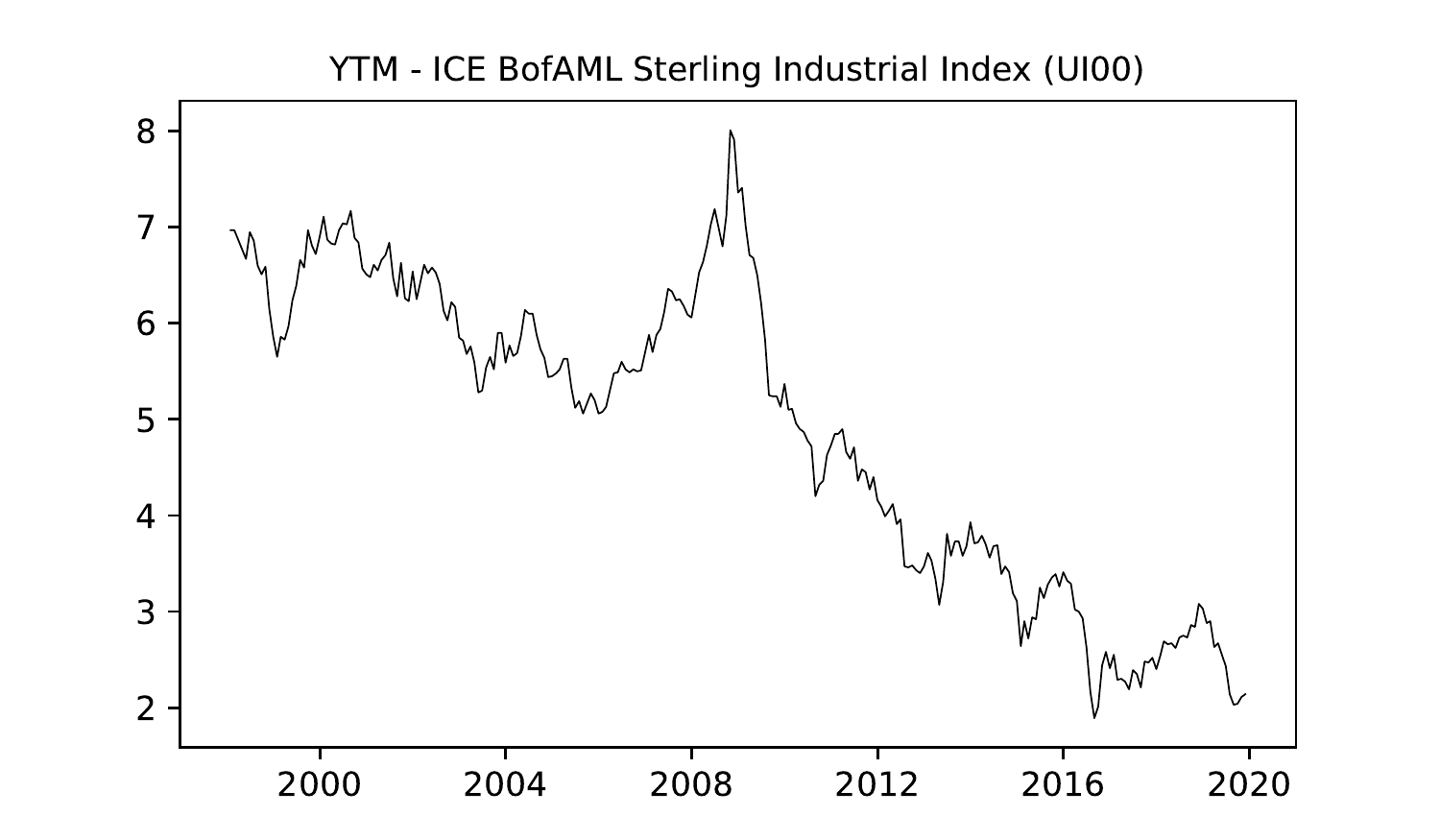}
}
\makebox[\textwidth][c]{
\includegraphics[scale=0.5]{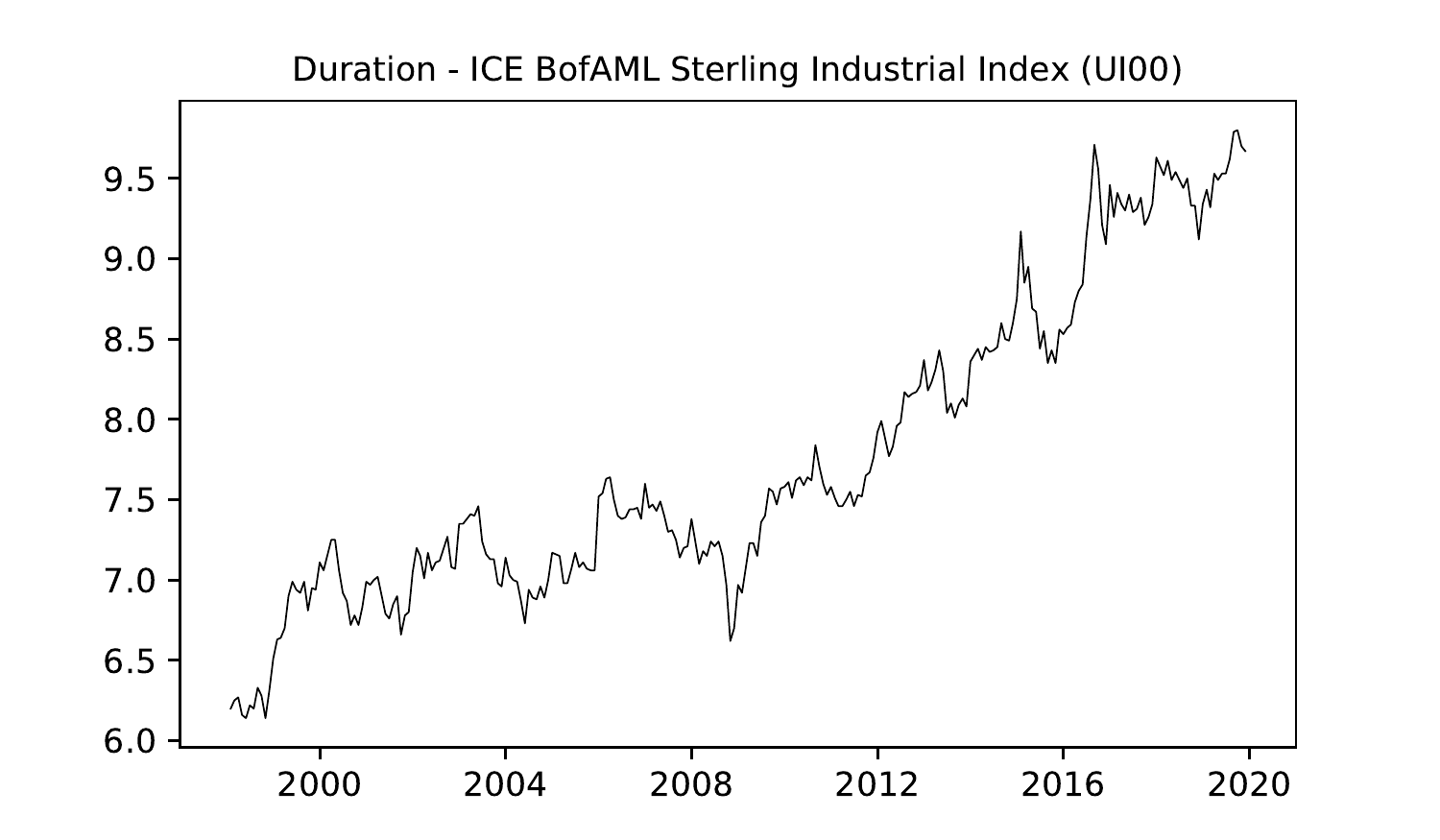}\includegraphics[scale=0.5]{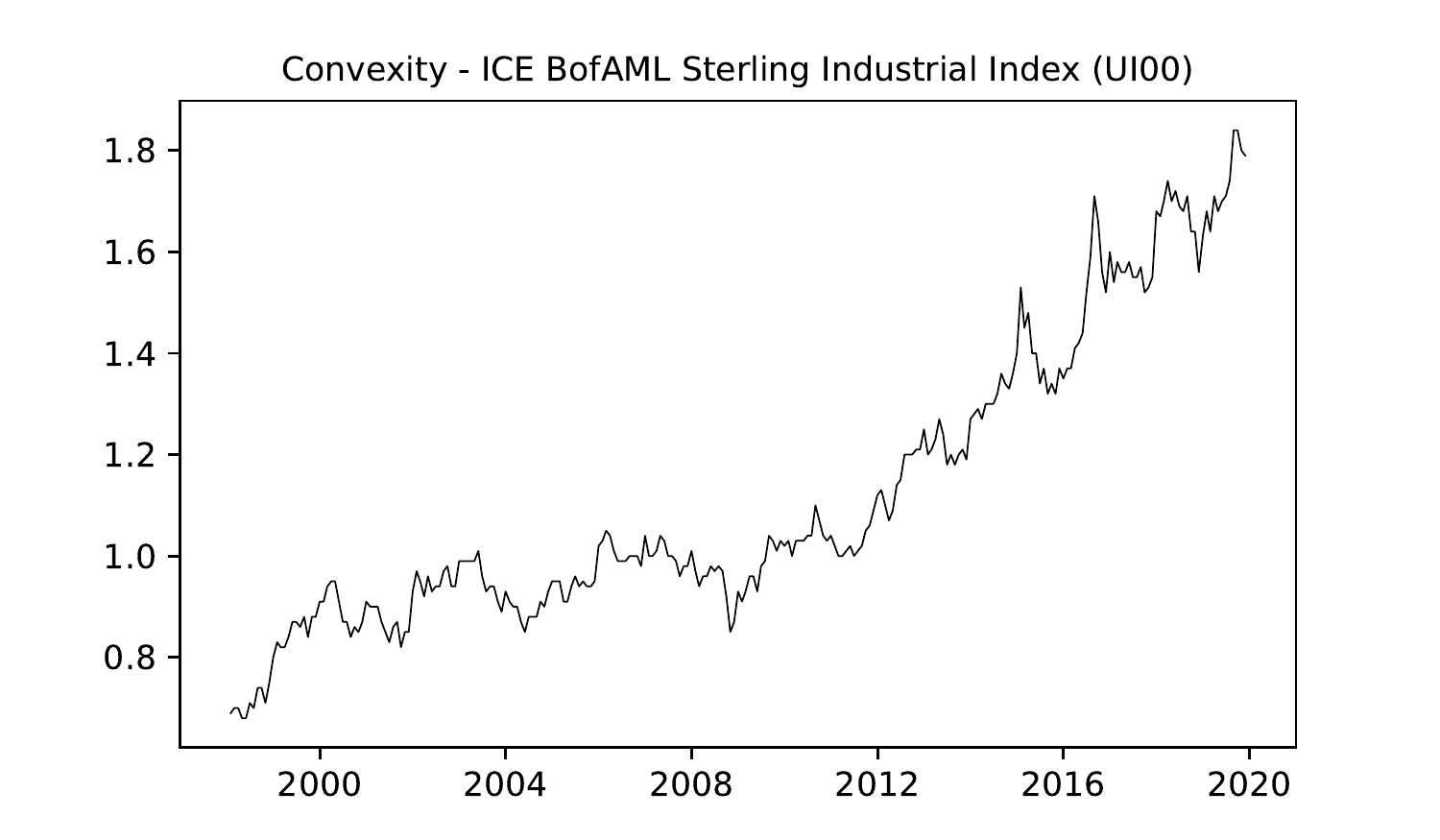}
}
\makebox[\textwidth][c]{
\includegraphics[scale=0.5]{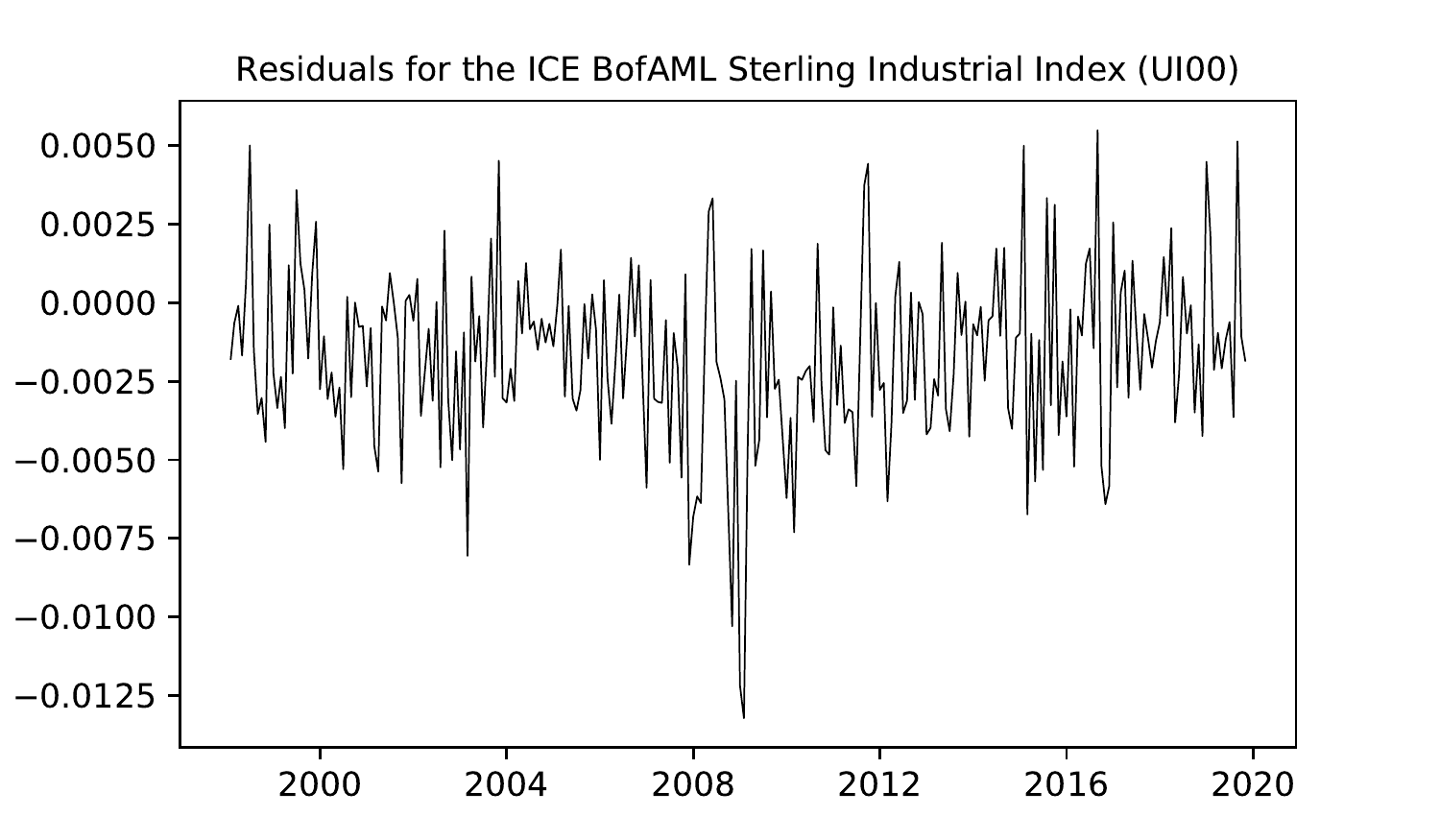}\includegraphics[scale=0.5]{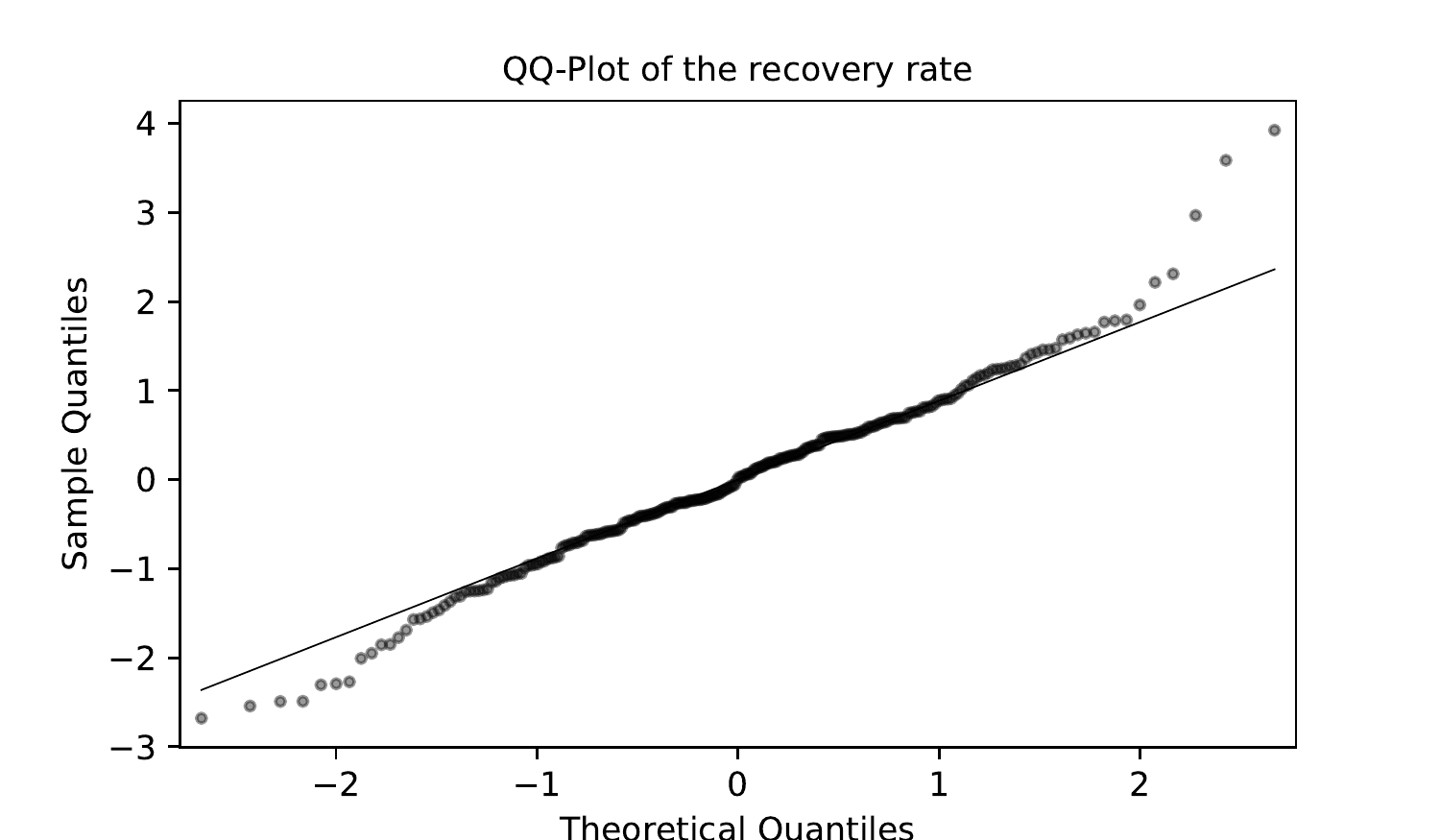}
}
\end{center}
\caption{The four topmost pictures illustrate the historical data used in the calibration of corporate bonds. The residuals used to estimate the recovery rate are illustrated in the bottom left picture. Next to it, the QQ-plot illustrating their good adherence to the lognormal model.}\label{fig:corporate}
\end{figure}

%% file: doc_model_uk_simulation.tex
\section{Simulation results}\label{simulation_results}

In this section we use the calibrated time series model in a simulation experiment. First, we use the time series model to generate scenarios for the risk factors $x$. After that, we apply the inverses of the transformations presented in Section~\ref{data_transformations}. Simulation results for 500k scenarios and a time horizon of 70 years are illustrated in Figures~\ref{fig:economic_financial_risk_factors} and~\ref{fig:survival_probabilities}. The figures plot the historical data and the median of the simulated scenarios with 95\% and 99\% confidence bands along a single simulated scenario. 

In the simulation results, one can observe the convergence of the medians to the values specified in Section~\ref{views}. In addition, the mean-reverting characteristic of the yields, spreads and the log-growths of CPI, AWE and GDP can be visually confirmed by the horizontal confidence bands. Finally, as described in Section~\ref{calibration_historical_data}, the model has been calibrated to reflect the recent behavior of inflation, which does not present the volatility levels observed in the 80s. In line with the results previously shown in~\cite{aro2011user} and~\cite{glover.mortality}, one notices that the survival probabilities for the 18-year-old cohorts has already reached an equilibrium level and now presents mean-reverting behavior. The probabilities for the older cohorts, however, are still improving. Figure~\ref{fig:risk_factors_2038} shows kernel density plots for the distribution of returns in each asset class in the year 2038. As one would expect, the short-term bonds present the smallest variance, long-term bonds the second largest, and stocks the largest. 

\subsection{Computation times}\label{computation_times}

Our simulator has been implemented in Python 3.7. The computation times obtained using an Intel(R) Core(TM) i7-7700HQ laptop with 32 GB of RAM (now a two year-old system) can be found in Table~\ref{timings}. As illustrated, the current implementation generates half a million scenarios in about one minute, and one million scenarios in less than three minutes. The figures show that about half of the time is spent on the simulation of the risk factors and the other half on the computation of the asset returns, where most of the time is spent with the returns on portfolios of corporate bonds. The software can be still be optimized with the introduction of antithetic sampling in the simulation of the risk factors and of a better algorithm for sampling the recovery rate for portfolios of corporate bonds. Even though the code is not fully optimized, it already shows the benefits of our approach, where a reduced number of risk factors is used for the survival probabilities and returns on bond portfolios.

\begin{table}[h]
\begin{center}
\begin{small}
\input{table_computational_time_small}
\end{small}
\caption{Computation times in seconds for the simulation of risk factors.}\label{timings}
\end{center}
\end{table}

\input{plots_simulations_risk_factors}

\input{plots_simulations_survival_probabilities_18_65_105}

\begin{figure}[ht]
\begin{center}
\makebox[\textwidth][c]{
\includegraphics[scale=0.5]{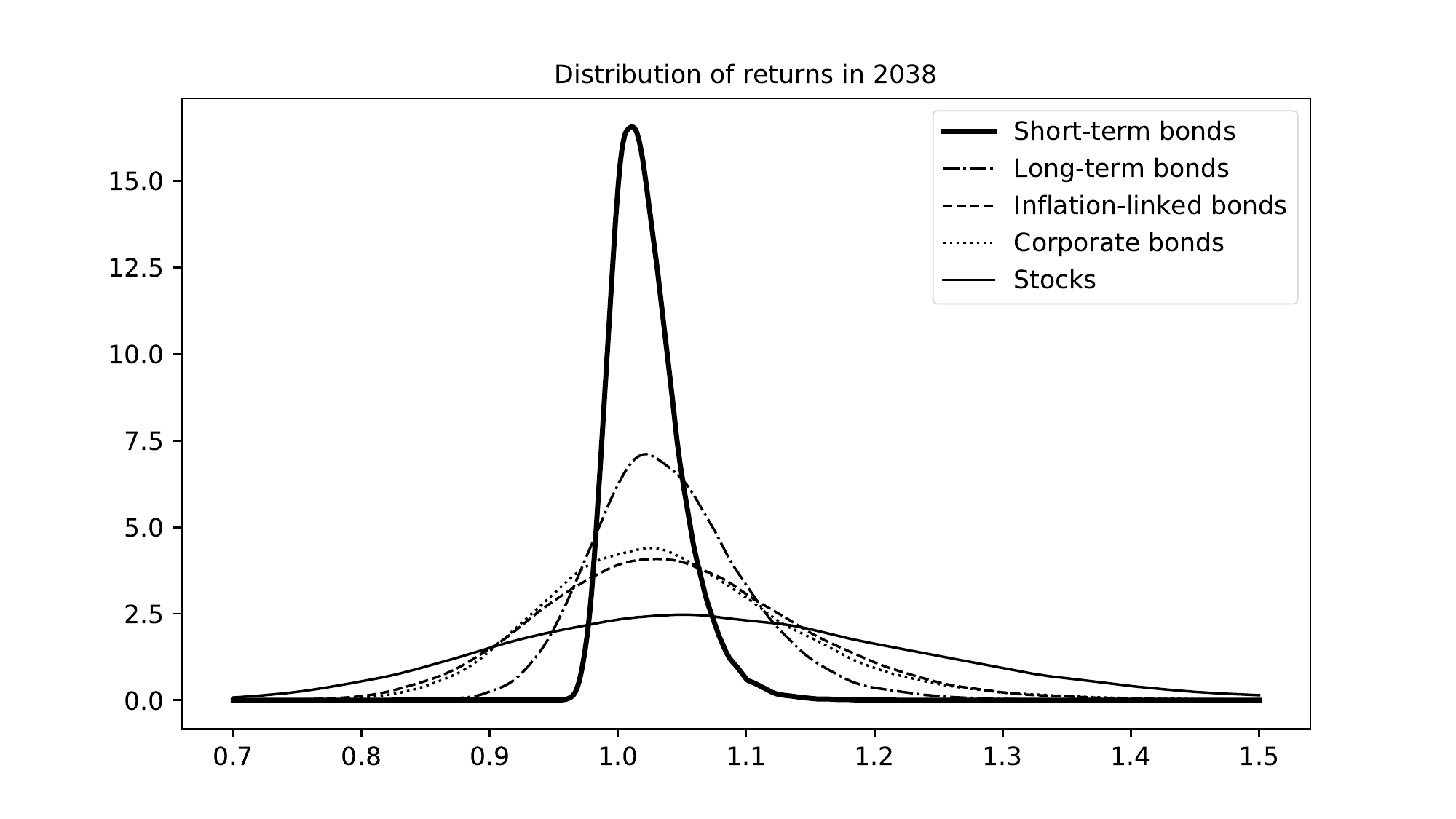}
}
\end{center}
%\caption{Distribution of returns and survival probabilities in a specific year.}\label{fig:risk_factors_2038}
\caption{The picture illustrates the distribution of the asset returns for a specific year using density plots. }\label{fig:risk_factors_2038}
\end{figure}

\begin{figure}[ht]
\begin{center}
\makebox[\textwidth][c]{
\includegraphics[scale=0.5]{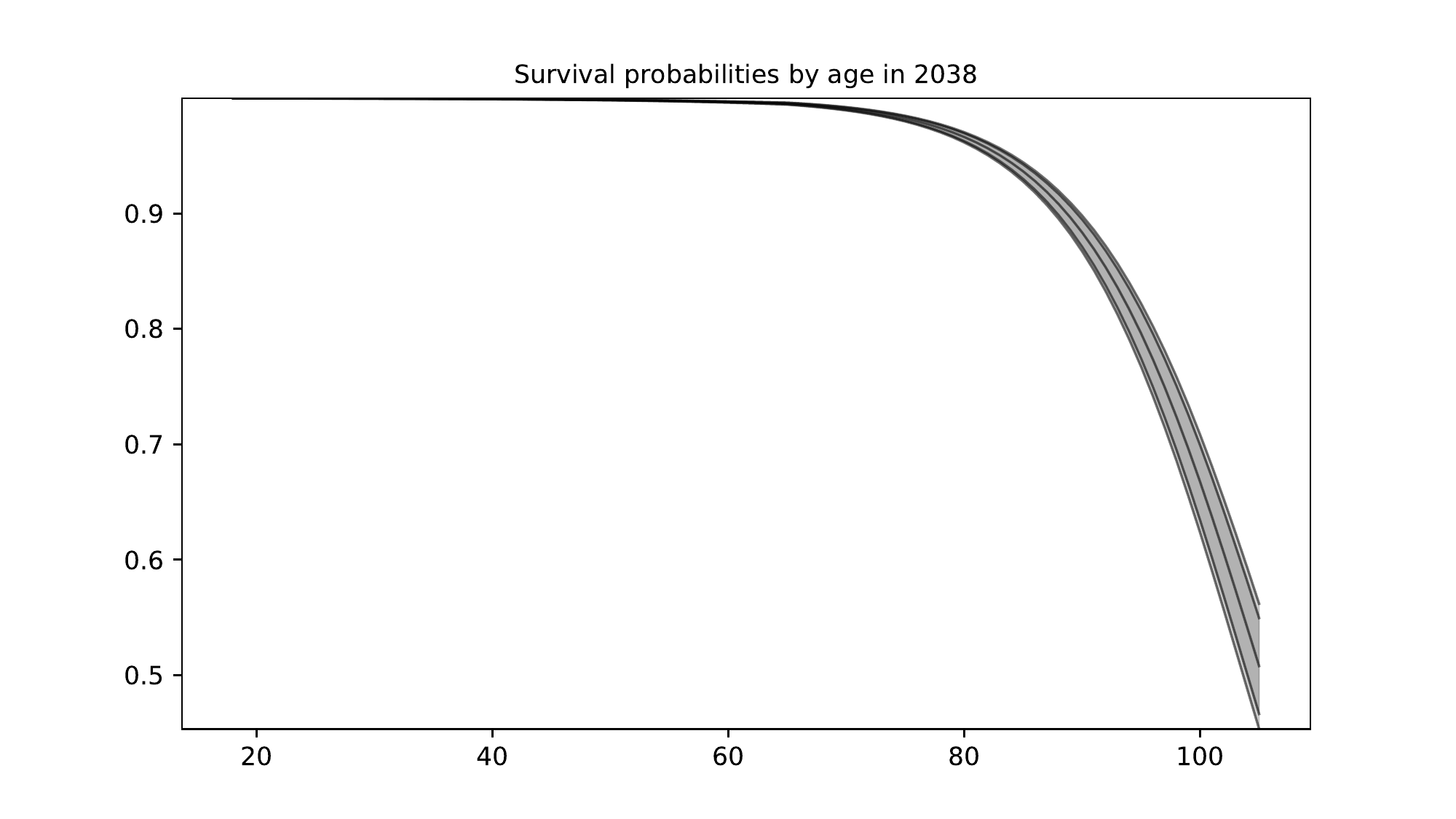}
}
\end{center}
%\caption{Distribution of returns and survival probabilities in a specific year.}\label{fig:risk_factors_2038}
\caption{The picture illustrates the median, the 95\% and 99\% confidence bands for the survival probabilities of females in a specific year.}\label{fig:risk_factors_2038}
\end{figure}

\subsection{Population sizes}\label{population_size}

As a first step to the computation of pension payments, we study the dynamics of a population using the models presented in Section~\ref{longevity}, in which Binomial random variables are used to track the total number of survivors from one year to the next, with survival probabilities that are based on the simulated mortality risk factors.
Simulation results for 100000 scenarios for cohorts of 18 and 85 year-old females of size 10, 100, and 1000 are illustrated in Figure~\ref{fig:population_size_simulations_size}.
In each plot, the median values and the 95\% and 99\% confidence bands for the population size are shown. In all experiments, simulations stopped when all individuals died or reached 105 years-old.

Inspecting the results, one can notice the different shapes of the curves obtained for the 18 and 85 year-old cohorts. For the younger cohort, the population size decreases slowly at first, and then accelerates its decrease as the population ages. In the simulations, the change happens around the year 2070, when the individuals reach their 70s.
For the older cohort, we find that the population size decreases rapidly. Unsurprisingly, these observations are understood when one looks into the survival probabilities for all ages at a specific point in time. As shown in  Figure~\ref{fig:risk_factors_2038}, survival probabilities remain fairly high until the age of 70 and rapidly decrease after that. 

The results also show a wide confidence interval for the times in which the population size becomes zero. These are particularly important for pension fund managers as they mark the end of all pension liabilities and also the investment horizon. For the smaller cohort of 85-year-old females, for example, in 95\% of the scenarios all individuals seem to die within a period of 11 years, between 2026 and 2037. For the larger cohorts, we expect the confidence intervals to be even larger.

Our results also show that larger populations will outlive smaller ones on average. This is particularly clear when comparing the medians of population sizes for the older cohort. For example, the median size for the smaller 85-year-old cohort is zero in the year 2032, but the same value is not reached before 2036 for the larger cohorts.
The result is easy to understand, as one of the factors used in the computation of medians for Binomial random variables is the number of trials, which corresponds here to the population size.

\input{plots_population_size}

\subsection{Pension payments}\label{pension_payments}

\begin{figure}[ht]
\makebox[\textwidth][c]{
\includegraphics[scale=0.55]{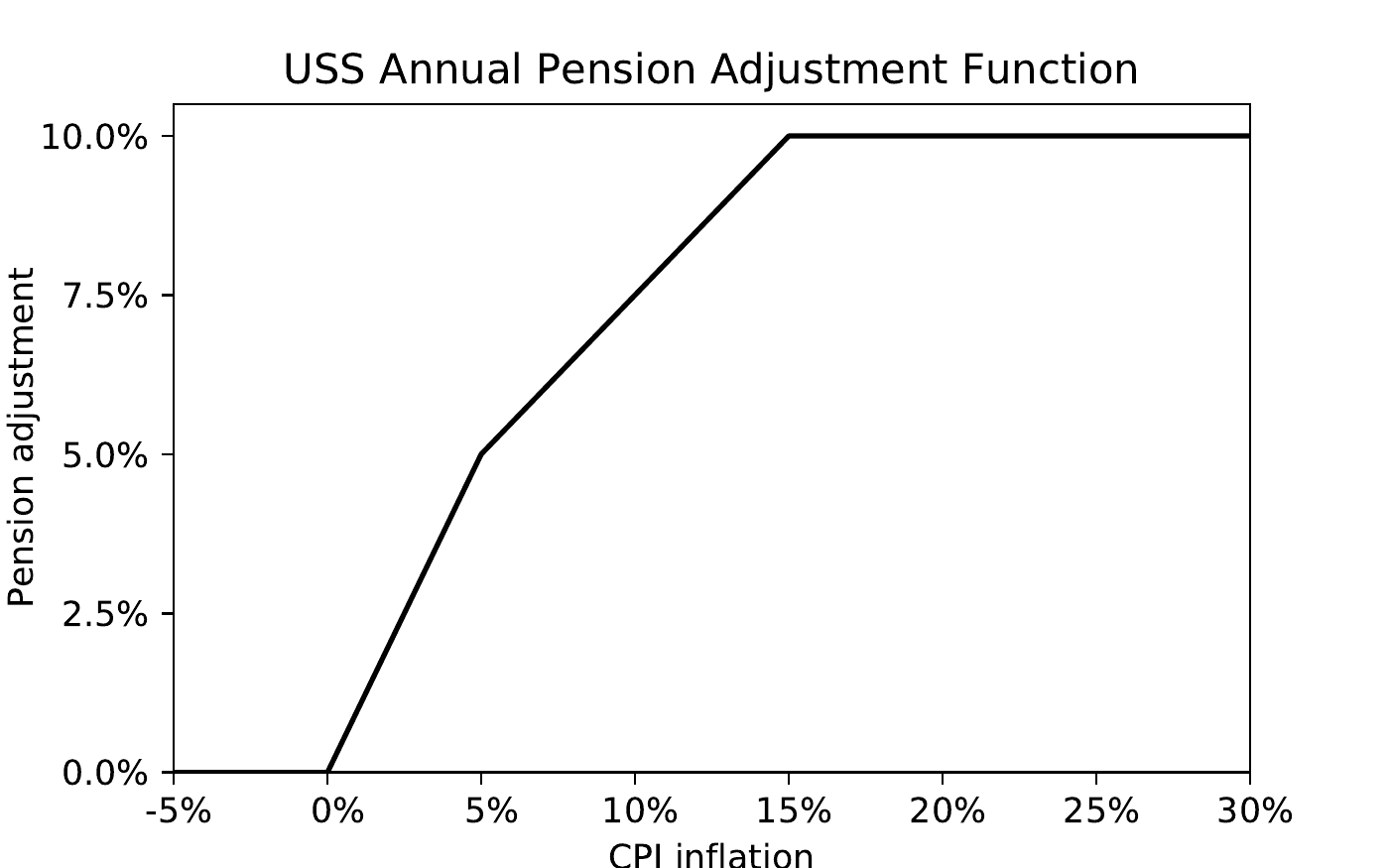}
}
\caption{Benefit adjustment function adopted by the Universities Superannuation Scheme (USS). As illustrated, no adjustments are made in the case of negative inflation, and only partial adjustments are offered for high inflation.}\label{fig:adj_func}
\end{figure}

We now study payments in the context of DB pension schemes, which are characterized by guaranteed benefits that are periodically adjusted in accordance to an index, typically inflation. 
The total pension payment to a population $B_t$ of pensioners is given by 
\begin{equation}
  c_t = \sum_{b \in B_t} F_t c_0^b,
\end{equation}
where $c_0^b$ is the initial benefit paid to a pensioner and $F_t$ an adjustment factor that is regularly updated. Considering annual adjustments that are based on inflation, as measured by the CPI, we assume that the adjustment factor is given by 
\begin{equation}
 F_t = \prod_{k=0}^t \lrs{1 + f_{adj}\lrp{\frac{\text{CPI}_{k} - \text{CPI}_{k-1}}{\text{CPI}_{k-1}}}}\label{eq:Ft},
\end{equation}
for an adjustment function $f_{adj}$ that is specific to each pension scheme. As an example, the adjustment function adopted by the Universities Superannuation Scheme (USS) is illustrated in Figure~\ref{fig:adj_func}.

Under the specifications above, one can easily identify the main sources of risk affecting pensions: indexation and longevity. In order to quantify the contribution of each risk factor to the overall financial risk, we recompute the payments removing the risk factors one at a time, by replacing them by their median values. This approach leads to four different computations, as illustrated in Figure~\ref{fig:payments_risk}, where the payments made to a group of 100 65-year-old female pensioners are shown. In the experiment, payments were adjusted for inflation. 

The top-left plot in Figure~\ref{fig:payments_risk} shows a deterministic projection of the future payments, computed with constant inflation at 2\% and median population sizes. In the top-right plot, the effects of the longevity risk included, as simulated survival probabilities are used to control the population sizes while maintaining inflation constant at 2\%. The bottom-left plot illustrates the inflation risk, as payments are adjusted in accordance to the USS regulations. Finally, the bottom-right plot shows the joint effects of the inflation and longevity risks, as no deterministic replacements are in place. 

The results show that inflation risk is dominant in the 
first years of the simulation, when the population sizes are usually larger, and that the longevity risk seems to play a larger role in the long run, as illustrated by the larger confidence bands found when it is present. Another interesting feature in the results is the growth of the real pension payments observed when in the inflation risk is present. Such growth corresponds to periods of negative inflation, and is a direct result of the USS pension adjustments. As shown in Figure~\ref{fig:adj_func}, payments are not decreased when inflation is negative. 

\input{plots_payment_risk_components}

\subsection{Longevity and economic growth}\label{gdp_longevity}

Evidence of the link between old-age longevity and economic growth has already been presented in~\cite{aro2014stochastic},~\cite{glover.mortality} and references therein. In our model, as in the second, the mortality risk factors associated with the longevity of the elderly are linked to economic growth through the real AWE. This link is tested here by increasing the long-term median of the real GDP log-growth in our time series model from 2\% to 8\% and examining the impact of this change for a cohort of 1000 85-year-old female pensioners. 
Simulation results, illustrated in Figure~\ref{fig:link_gdp_mortality}, show an increase in the survival probabilities of the elderly (represented by the risk factor $v^{3,f}$) that seems to cause an increase in the population sizes and, consequently, on pension payments as well.

\input{plots_gdp_mortality_link}

\subsection{ALM study of a closed DB fund}

We now focus on the asset-liability management of a closed DB fund, that is, a scheme that is closed to new members and to accrual of new benefits. According to the estimates provided in the ``Purple Book''~\cite{purplebook2017short}, closed funds correspond to 41\% of the 5,436 DB-funds in the UK that are eligible for protection from the Pension Protection Fund (PPF). 

More specifically, we consider a pension scheme with the liabilities described in Section~\ref{pension_payments} and £200k in assets. In each scenario, we simulate the fund 40 years into the future computing the remaining wealth at the end of each year after collecting the investment returns and paying out the pension benefits. At the beginning of each year, the remaining wealth is reallocated among the available asset classes in fixed proportions. When the fund goes into deficit, we assume that the remaining pension payments are financed by borrowing money at the short-term interest rate.

The proportions used in rebalancing are based on the 2008 and 2019 average asset allocations in Tables~\ref{tab:purplebookallocation} and~\ref{tab:purplebookallocationbonds}, which have been adjusted to compensate for the asset classes that are not available in our models (e.g. property, hedge funds, etc.). The 2008 and 2019 allocations have been chosen for illustrating how the investment strategies adopted by pension funds in the UK have changed after the financial crisis of 2008. Before the crisis, funds used to invest 53.6\% of its wealth in equities and 32.9\% in bonds. At the moment, average investments are at 24.0\% in equities and 62.8\% in bonds. 

As one would expect, the higher proportion of stocks in the 2008 allocation results in wider distribution for the net wealth. Interestingly though, the uncertainty is mostly on the upside as the downside risk seems almost the same for both strategies. It would be natural to try to optimize the investment strategy by minimizing a give performance criterion over all feasible strategies that are adapted to the information available to the fund managers over time. Preliminary results in that direction were obtained in \cite{hilli2011optimal} for the case without longevity risk. Extensions to the general setting will be explored in a subsequent paper.

\input{plots_wealth}

%% file: table_computational_time_small.tex
\begin{tabular}{rrrrr}
%\begin{tabular}{ccccc}

\toprule
Scenarios  &  Total &  \thead{Simulation \\of risk factors}  &  Returns &  \thead{Returns on\\ corporate bonds}\\
\midrule
 100k &  12.19 &                    5.61 &     6.15 &                     5.66 \\
 \rowcolor{black!5}200k &  24.23 &                   11.13 &    12.26 &                    11.27 \\
 300k &  36.65 &                   16.80 &    18.57 &                    17.08 \\
 \rowcolor{black!5}400k &  49.70 &                   23.51 &    24.50 &                    22.50 \\
 500k &  62.64 &                   29.05 &    31.49 &                    29.01 \\
 \rowcolor{black!5}600k &  78.83 &                   34.79 &    41.50 &                    38.54 \\
 700k &  93.33 &                   41.01 &    49.34 &                    45.69 \\
 \rowcolor{black!5}800k & 113.35 &                   50.20 &    57.75 &                    51.93 \\
 900k & 130.38 &                   57.63 &    65.41 &                    58.62 \\
 \rowcolor{black!5}1000k & 146.53 &                   64.56 &    74.08 &                    66.39 \\
\bottomrule

\end{tabular}

%% file: plots_simulations_risk_factors.tex
\begin{figure}[ht]
\begin{center}
\makebox[\textwidth][c]{
\includegraphics[scale=0.5]{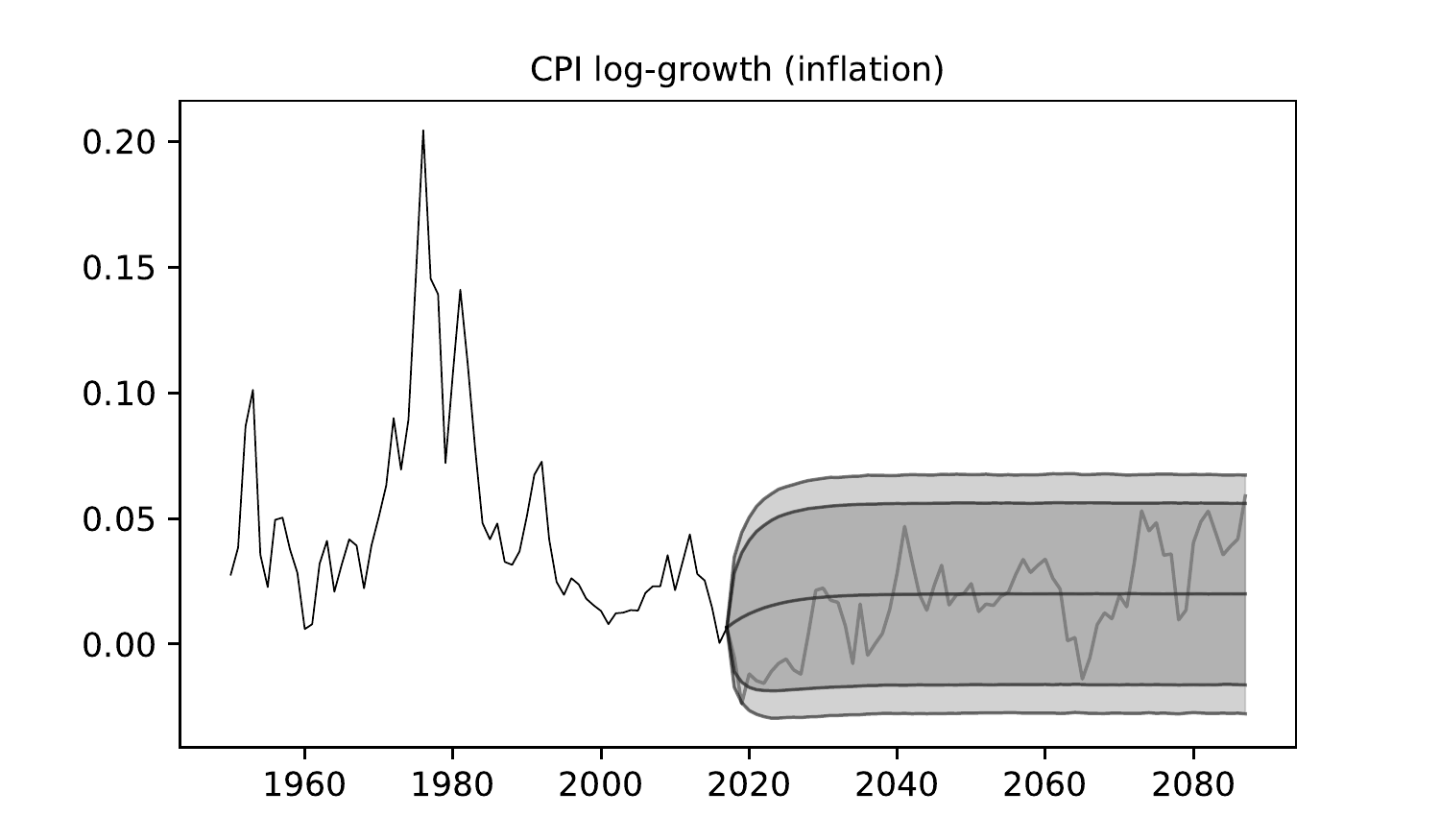}\includegraphics[scale=0.5]{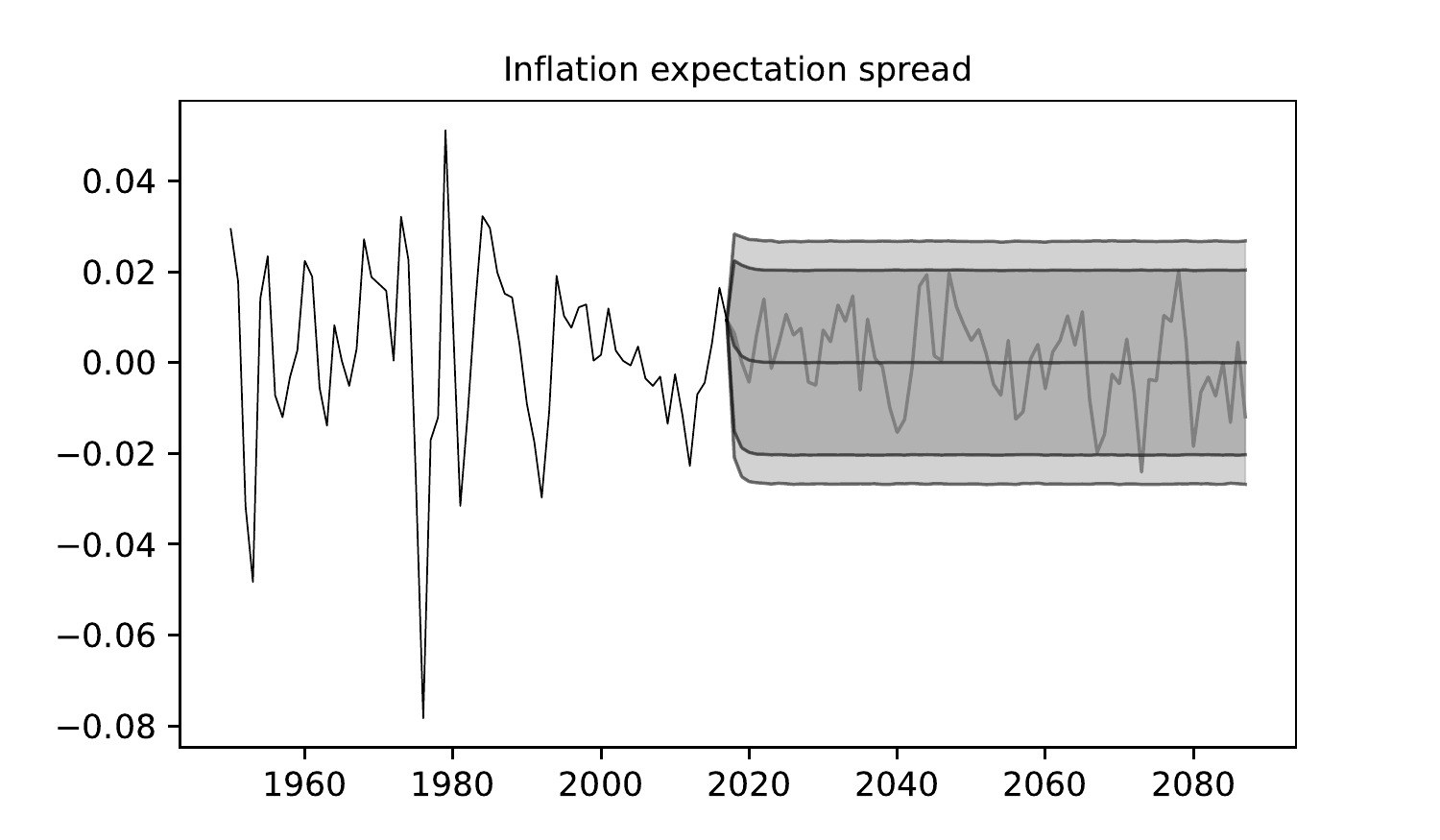}
}
\makebox[\textwidth][c]{
\includegraphics[scale=0.5]{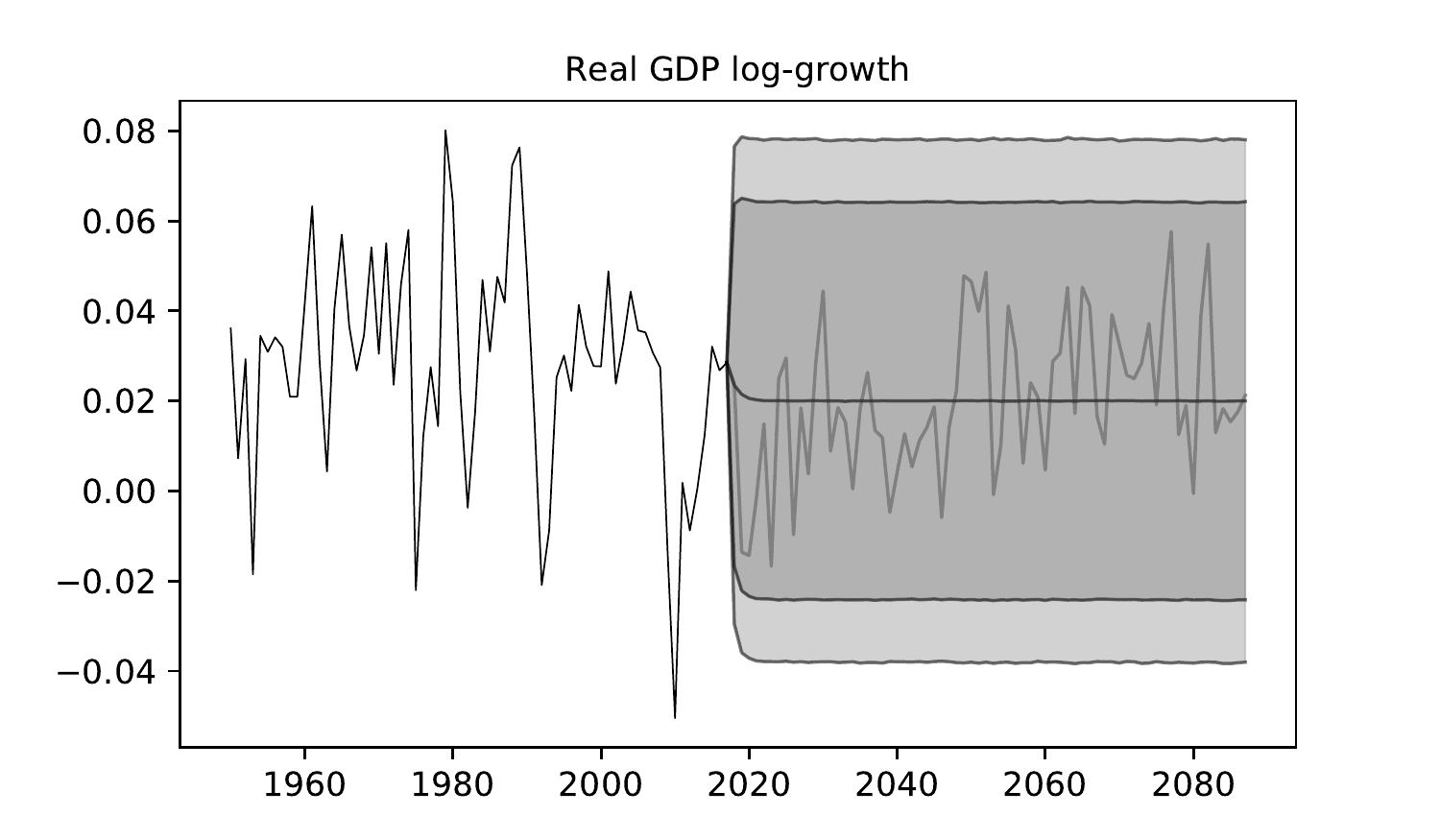}\includegraphics[scale=0.5]{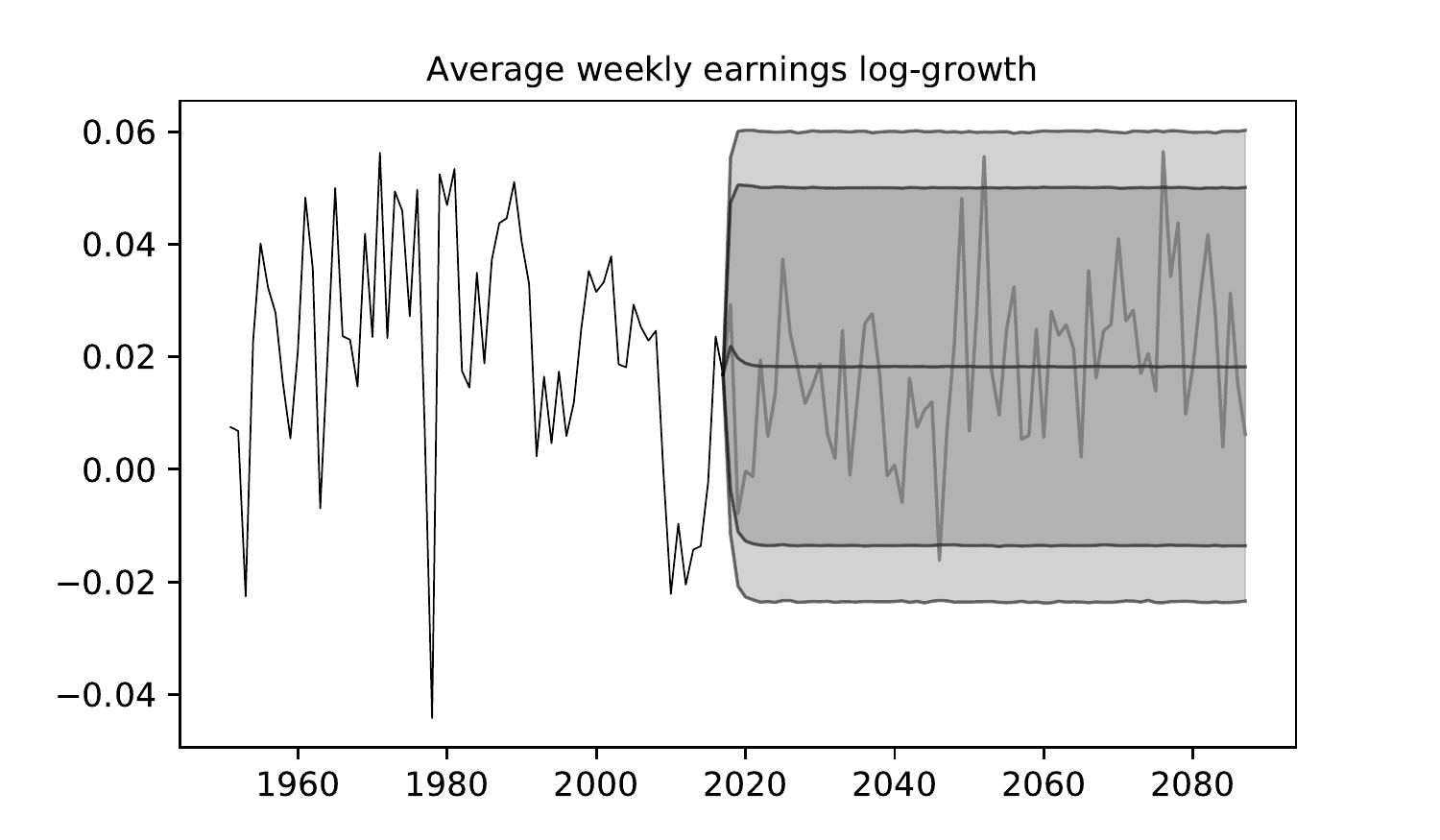}
}
\makebox[\textwidth][c]{
\includegraphics[scale=0.5]{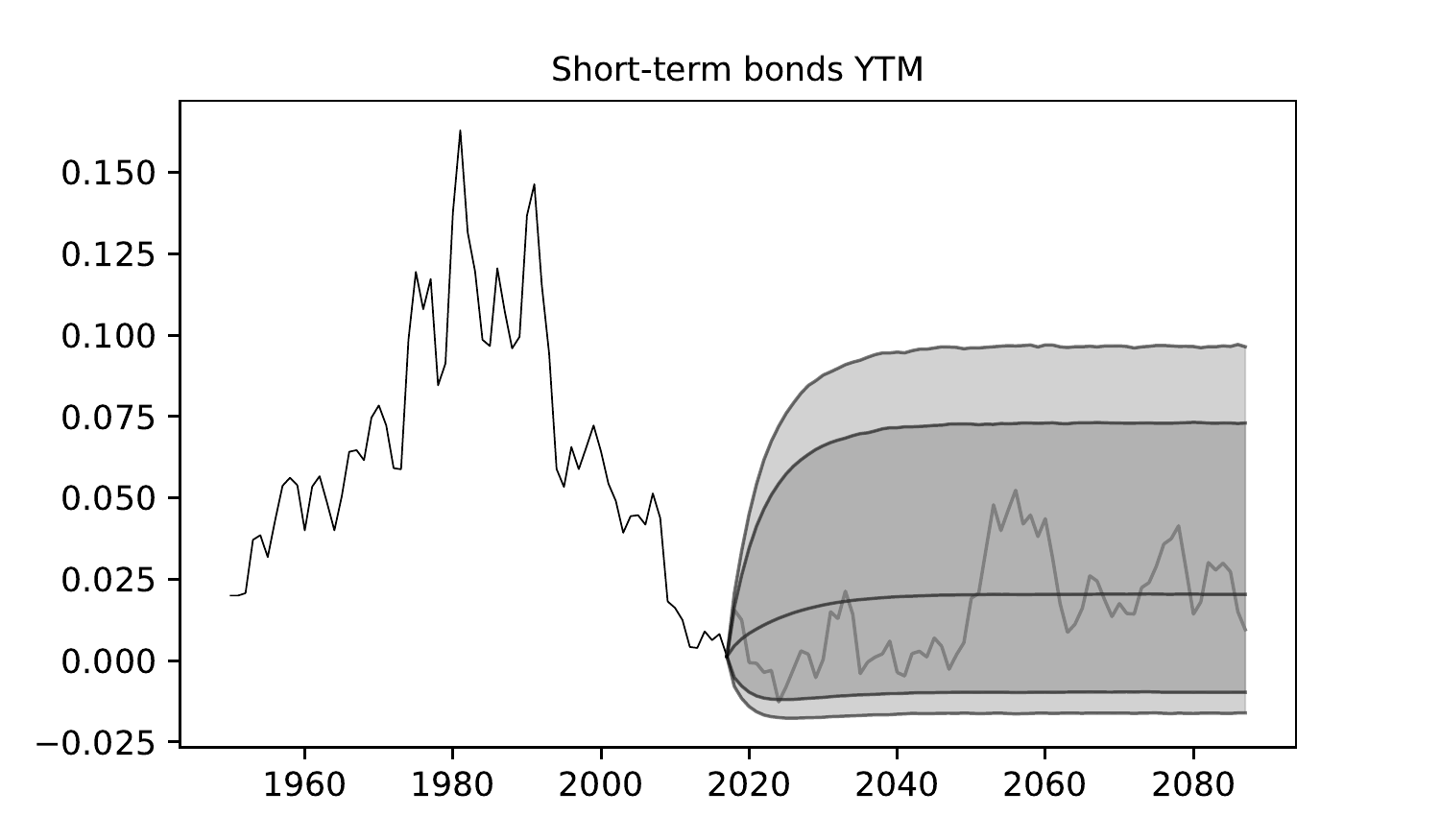}\includegraphics[scale=0.5]{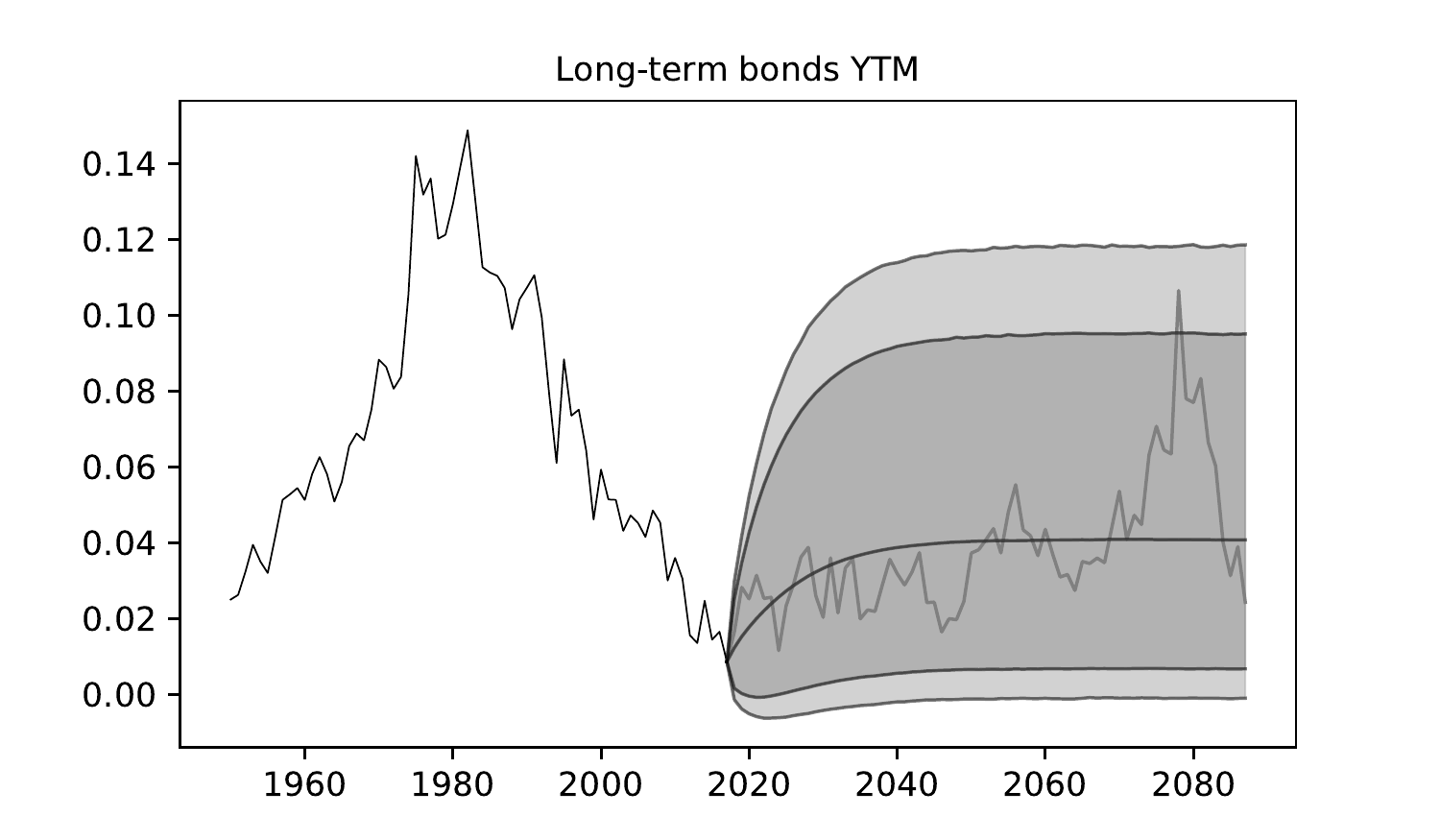}
}
\makebox[\textwidth][c]{
\includegraphics[scale=0.5]{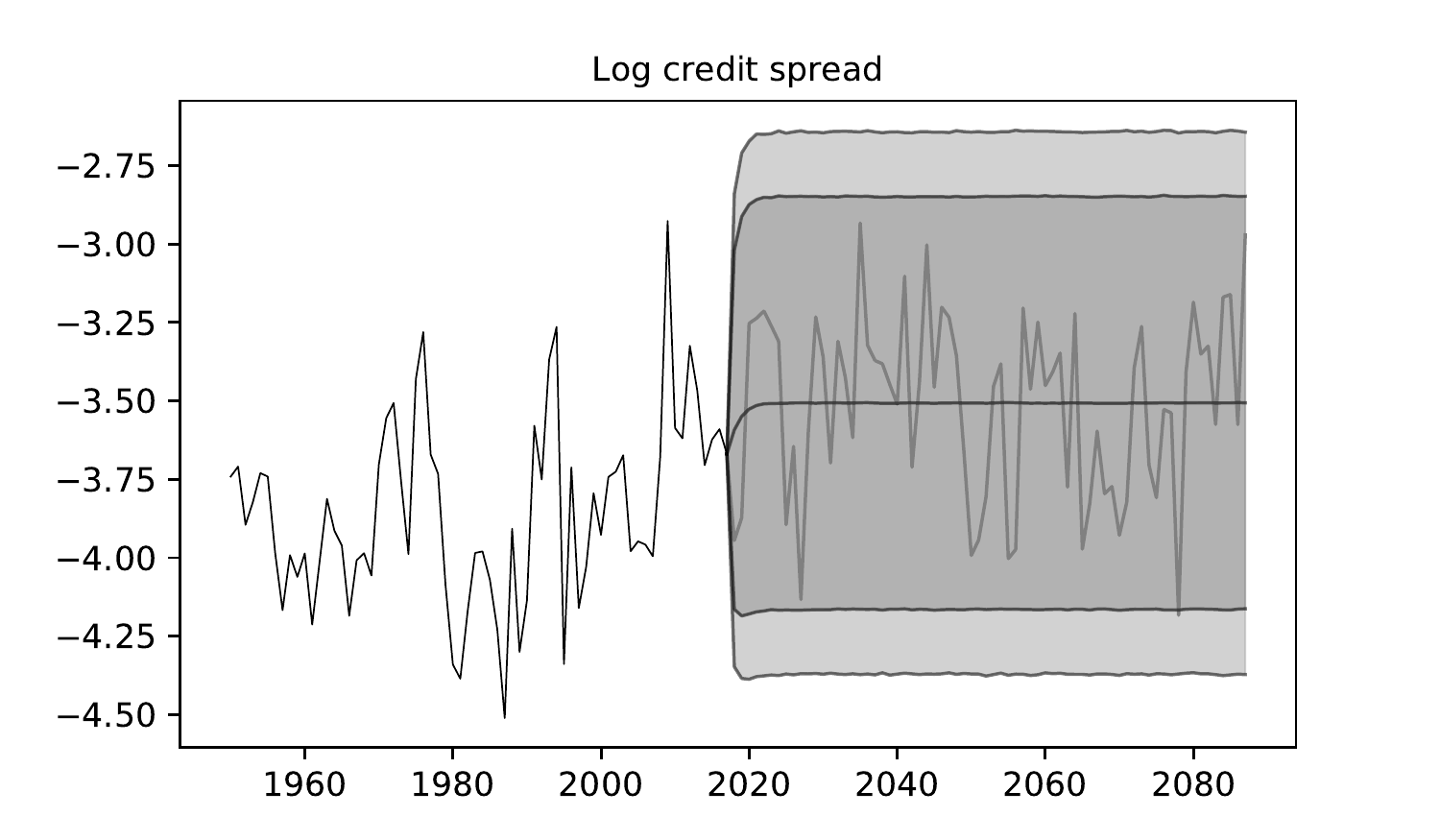}\includegraphics[scale=0.5]{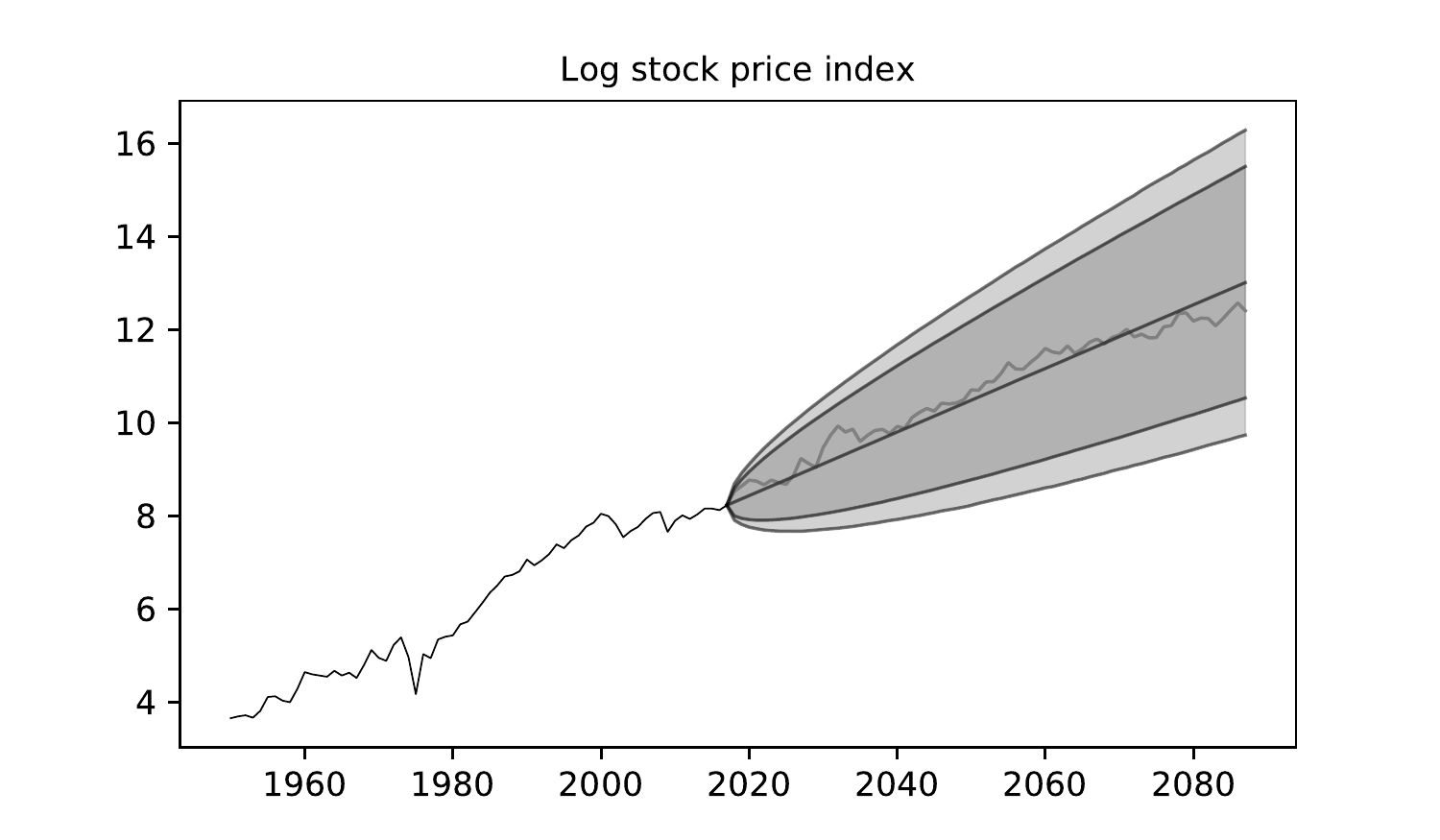}
}
\end{center}
\caption{Simulated scenarios for the economic and financial risk factors. The plots show the historical risk factors with a single simulated scenario and the 95\% and 99\% confidence bands.}\label{fig:economic_financial_risk_factors}
\end{figure}

%% file: plots_simulations_survival_probabilities_18_65_105.tex
\begin{figure}[ht]
\begin{center}
\makebox[\textwidth][c]{
\includegraphics[scale=0.5]{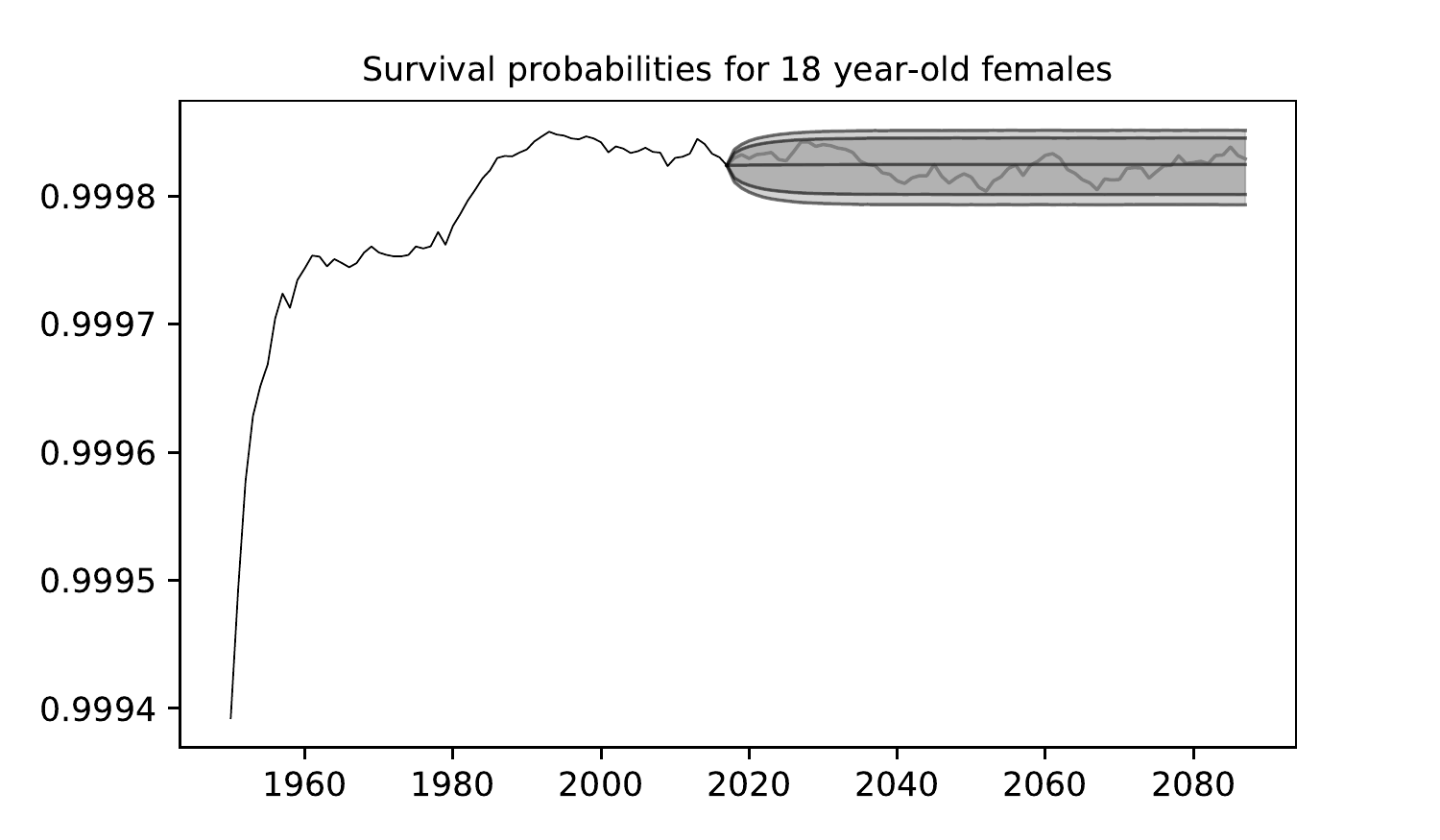}\includegraphics[scale=0.5]{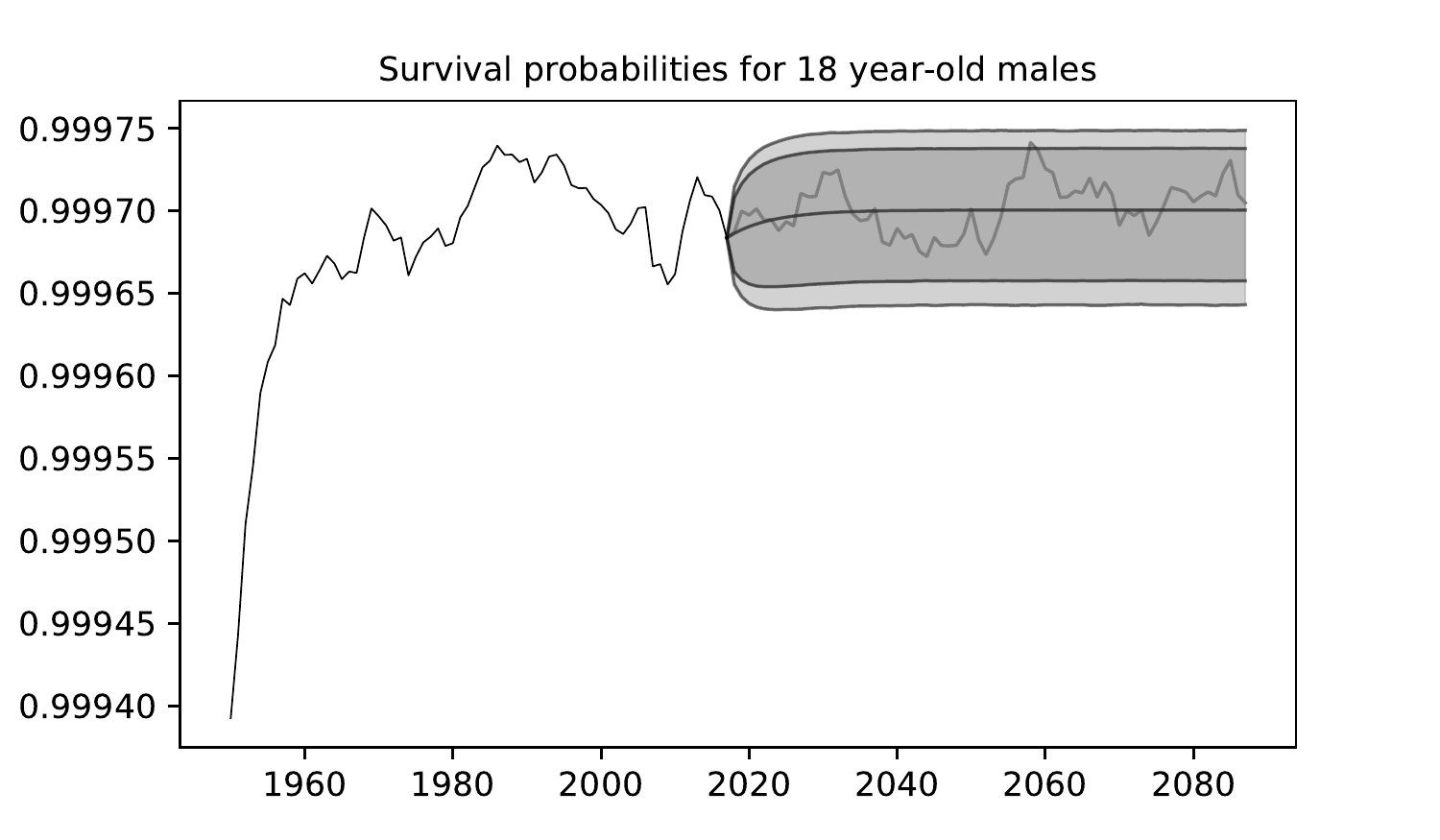}
}

\makebox[\textwidth][c]{
\includegraphics[scale=0.5]{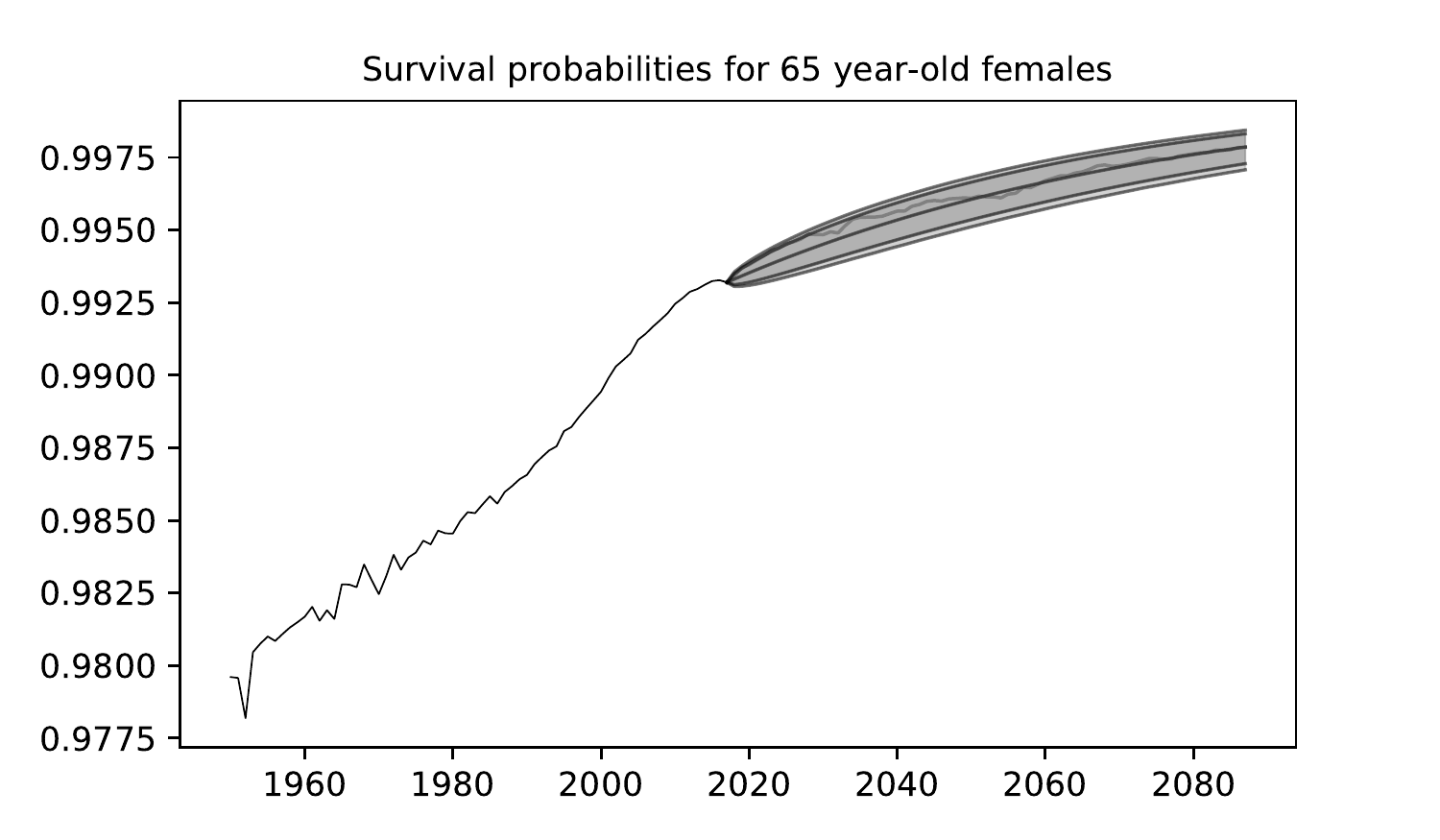}\includegraphics[scale=0.5]{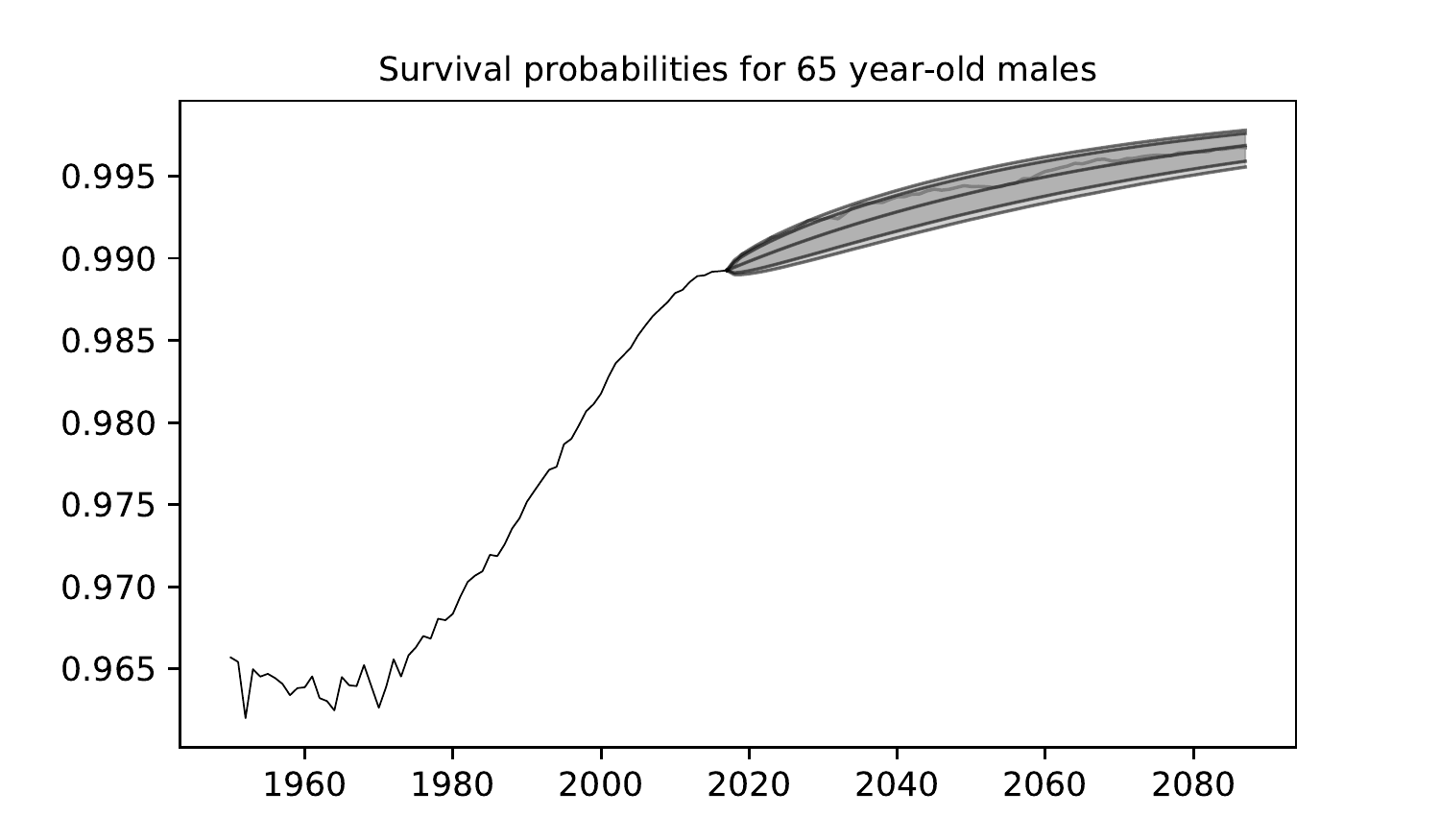}
}

\makebox[\textwidth][c]{
\includegraphics[scale=0.5]{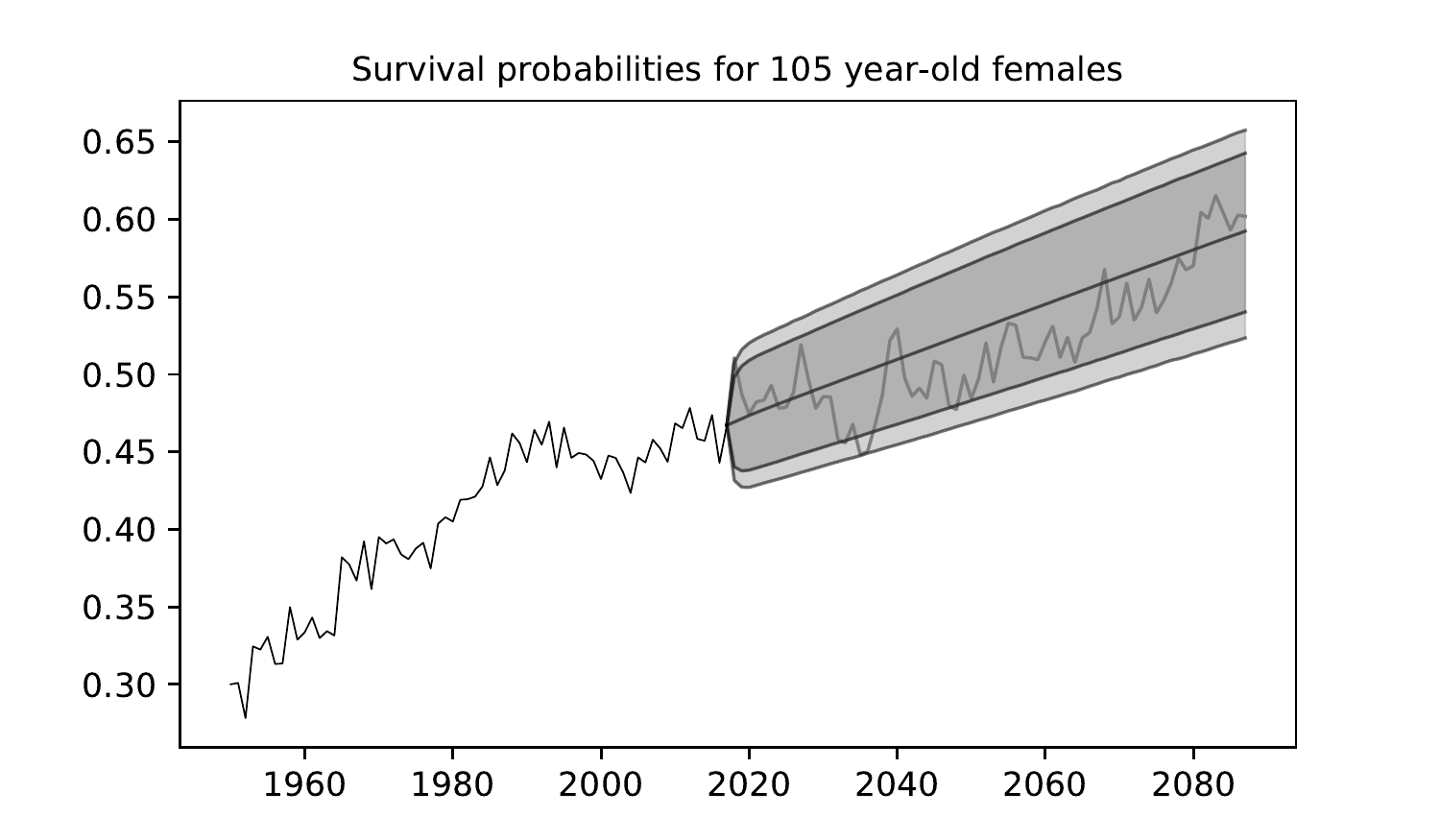}\includegraphics[scale=0.5]{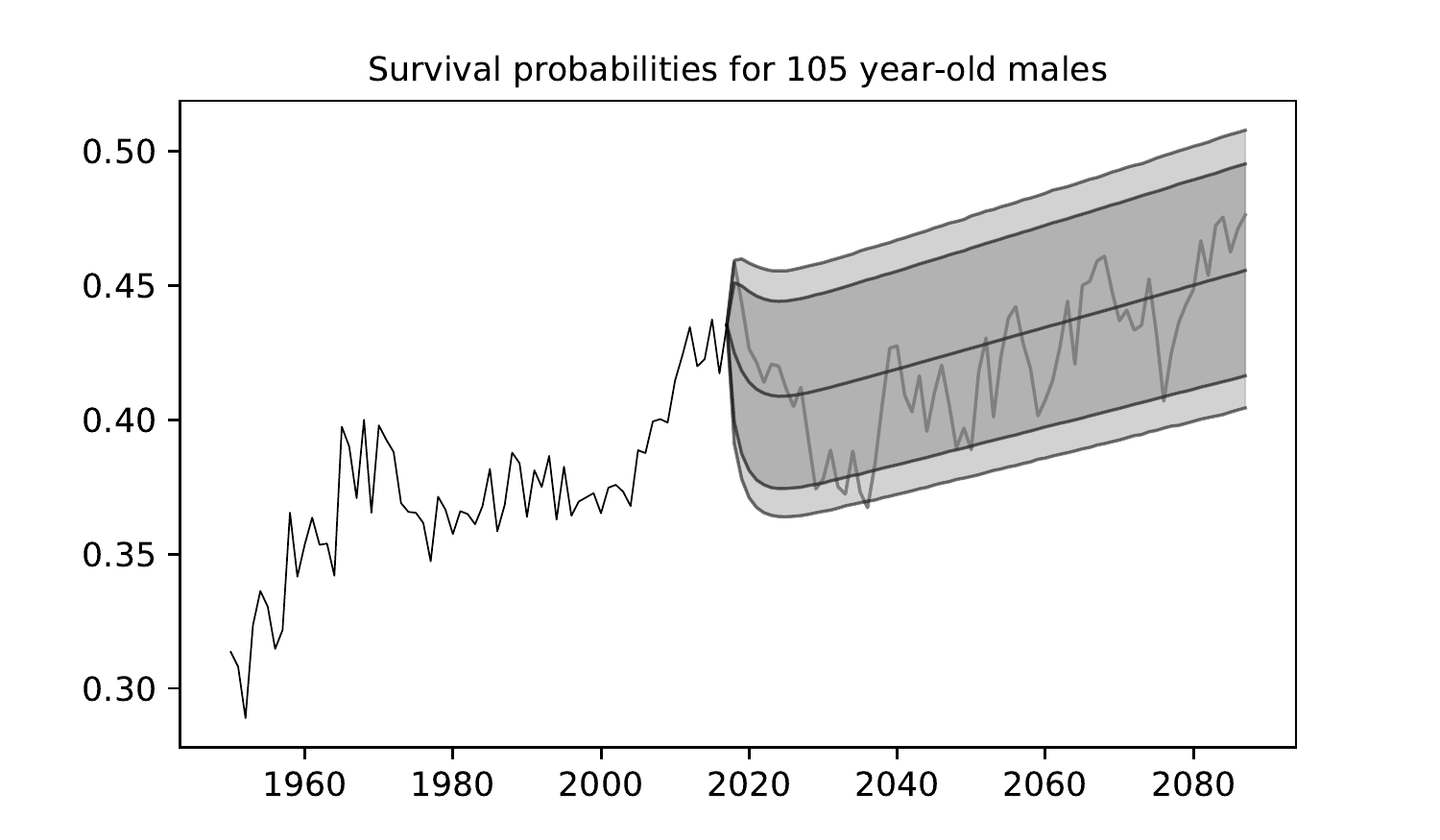}
}
\end{center}
\caption{Simulated scenarios for the survival probabilities of the 18, 65 and 105 year-old cohorts. The plots show the historical probabilities with a single simulated scenario and the 95\% and 99\% confidence bands.}\label{fig:survival_probabilities}

\end{figure}

%% file: plots_population_size.tex
\begin{figure}[ht]
\begin{center}
\makebox[\textwidth][c]{
\includegraphics[scale=0.5]{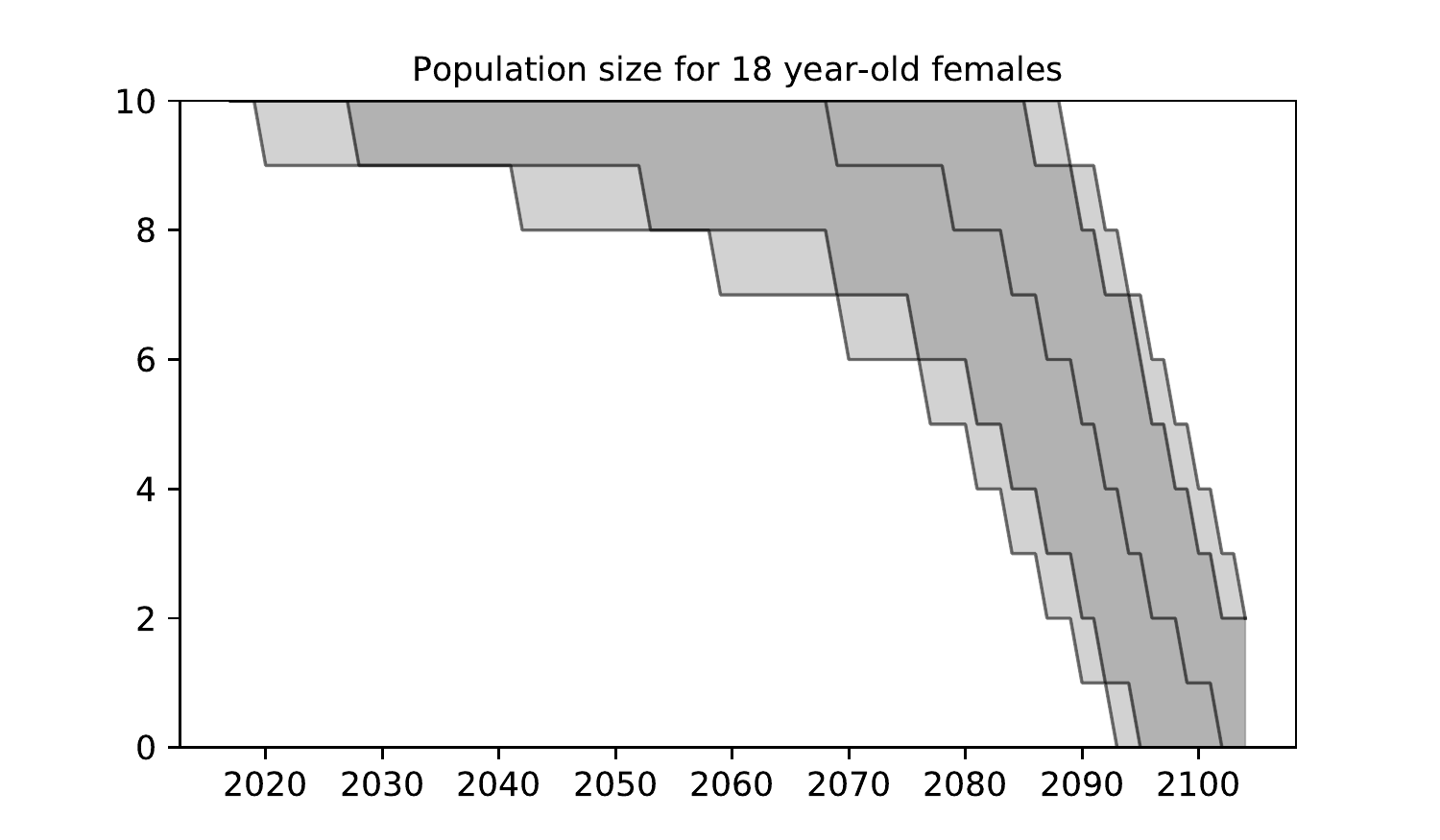}\includegraphics[scale=0.5]{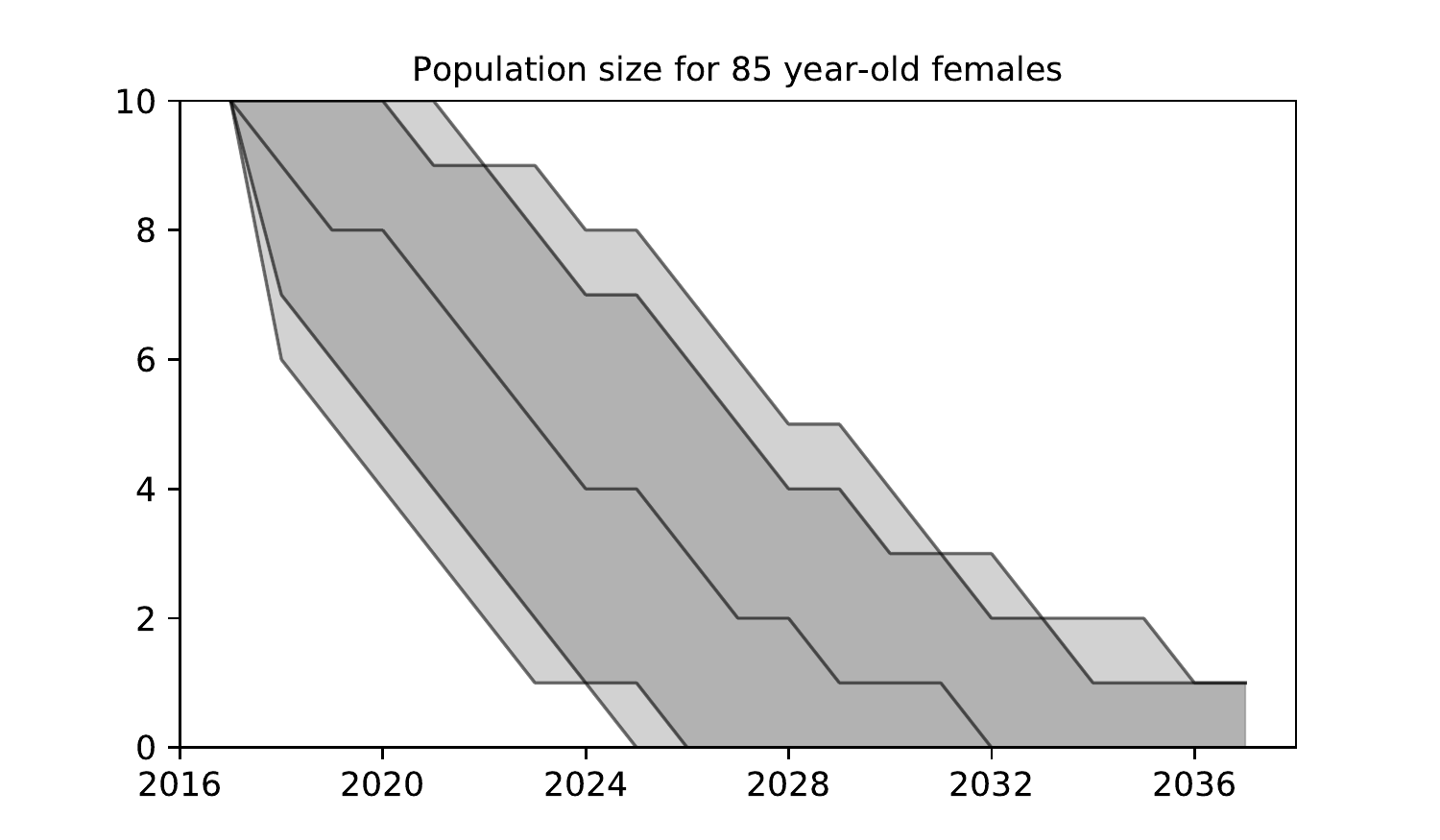} 
}
\makebox[\textwidth][c]{
\includegraphics[scale=0.5]{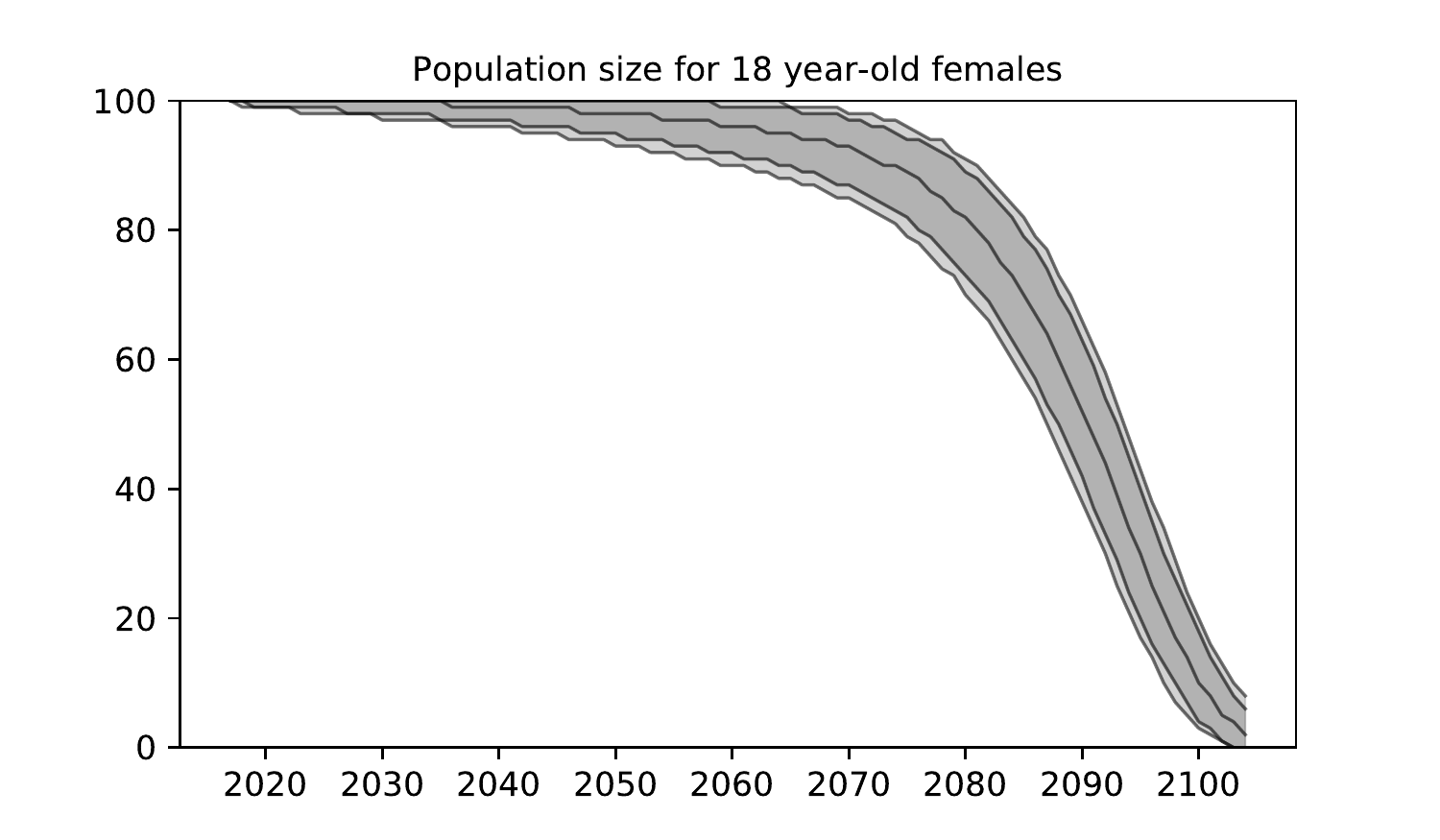}\includegraphics[scale=0.5]{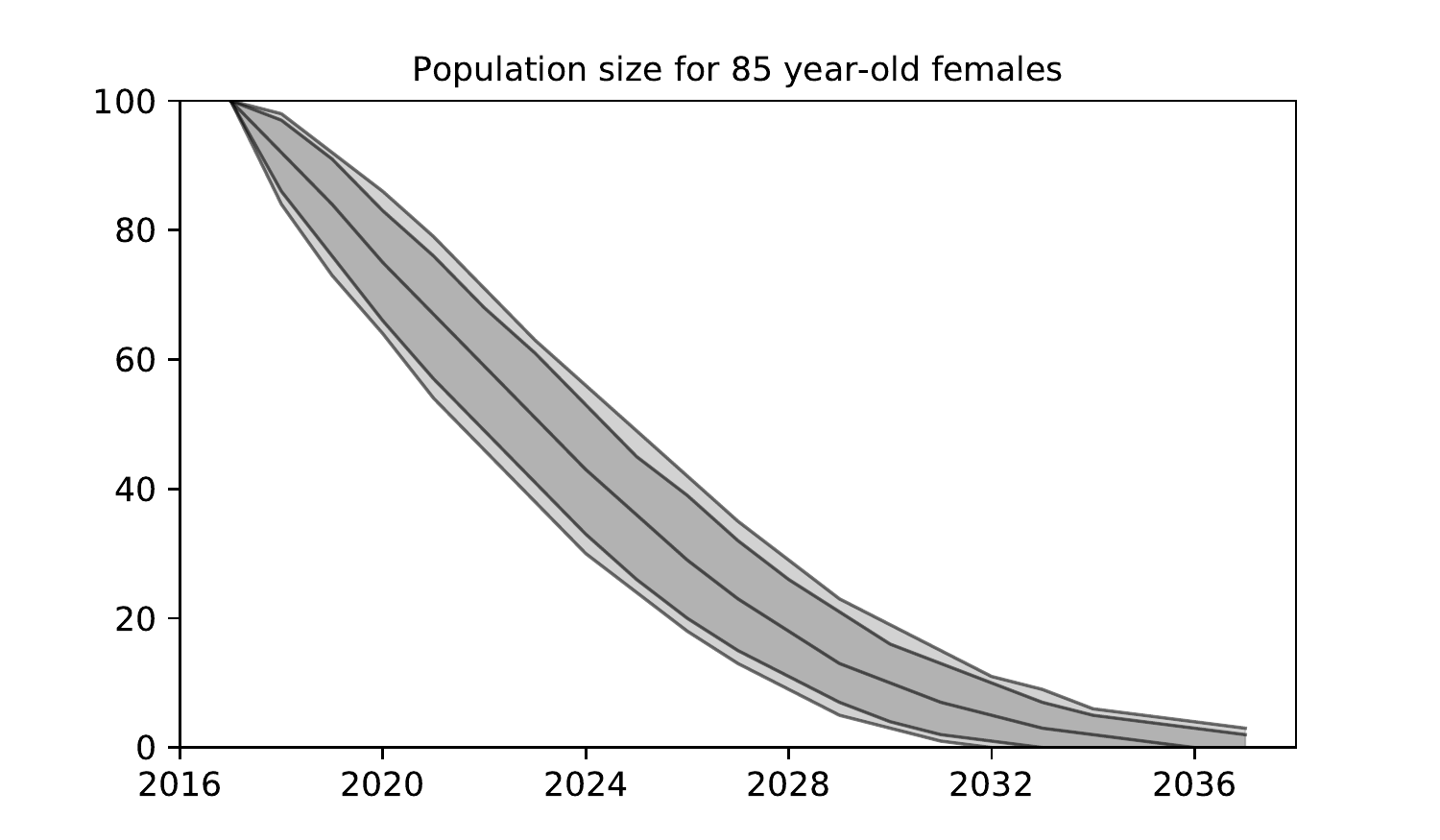} 
}
\makebox[\textwidth][c]{
\includegraphics[scale=0.5]{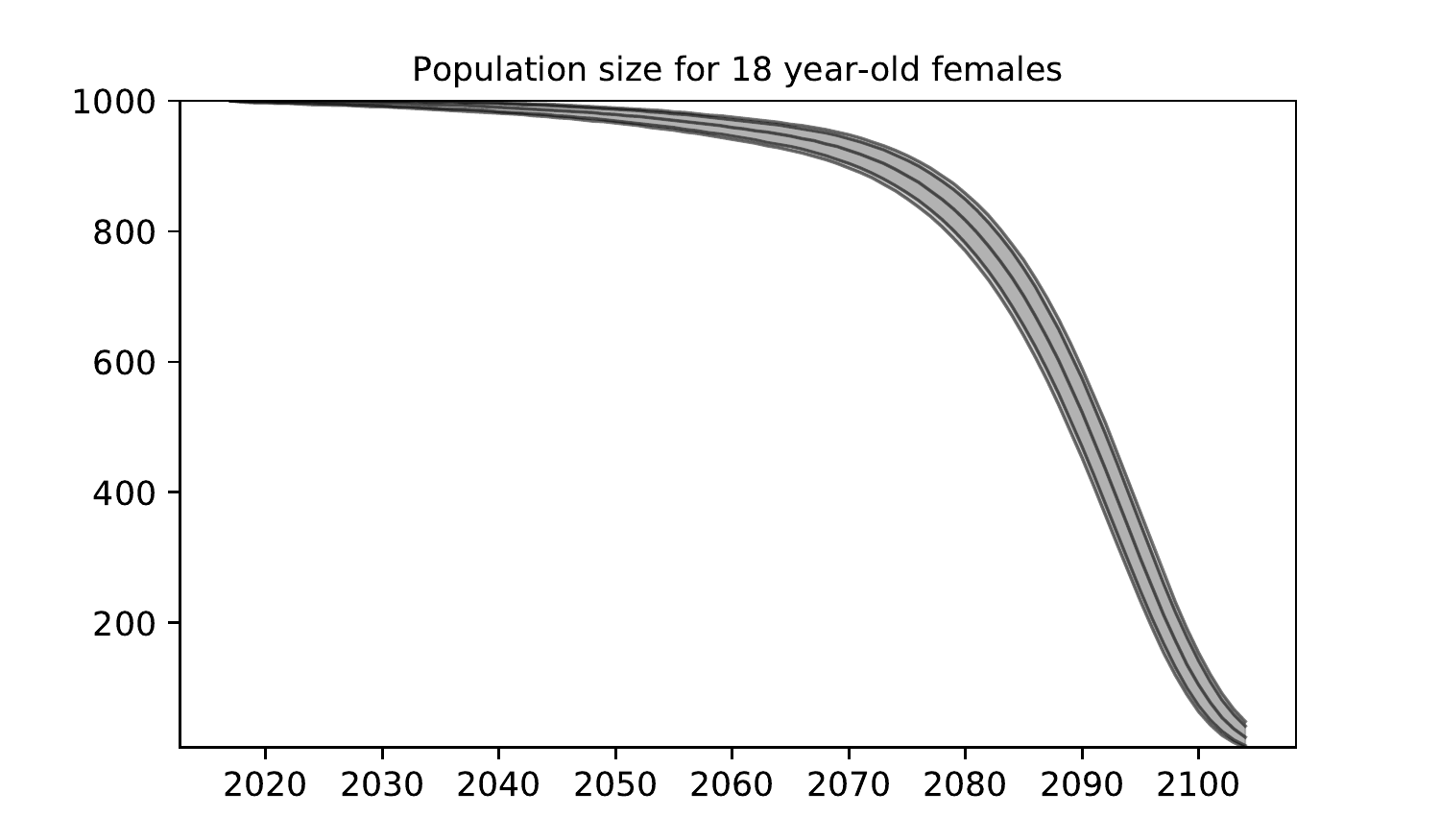}\includegraphics[scale=0.5]{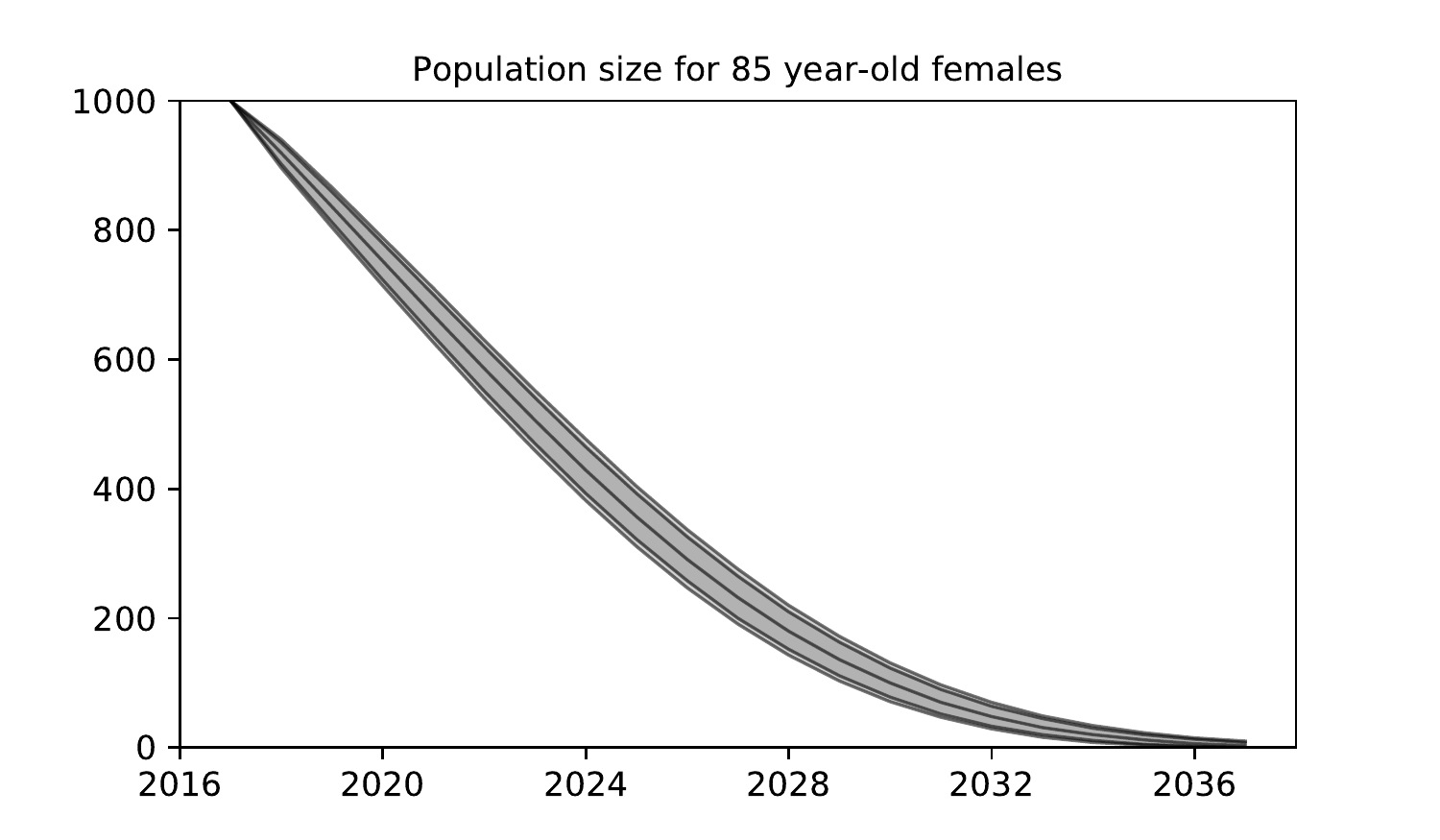} 
}
\end{center}
\caption{Simulation of population sizes with survival probabilities from the UK model. Plots on the left show the median values of the population size with the 95\% and 99\% confidence bands for cohorts of 18-year-old females with sizes 10, 100, and 1000. Results for 85-year-old females are shown on the right. The top plots clearly illustrate the discrete nature of the Binomial random variables used to model the population sizes.}\label{fig:population_size_simulations_size}
\end{figure}

%% file: plots_payment_risk_components.tex
\begin{figure}[ht]
\begin{center}
\makebox[\textwidth][c]{
\includegraphics[scale=0.5]{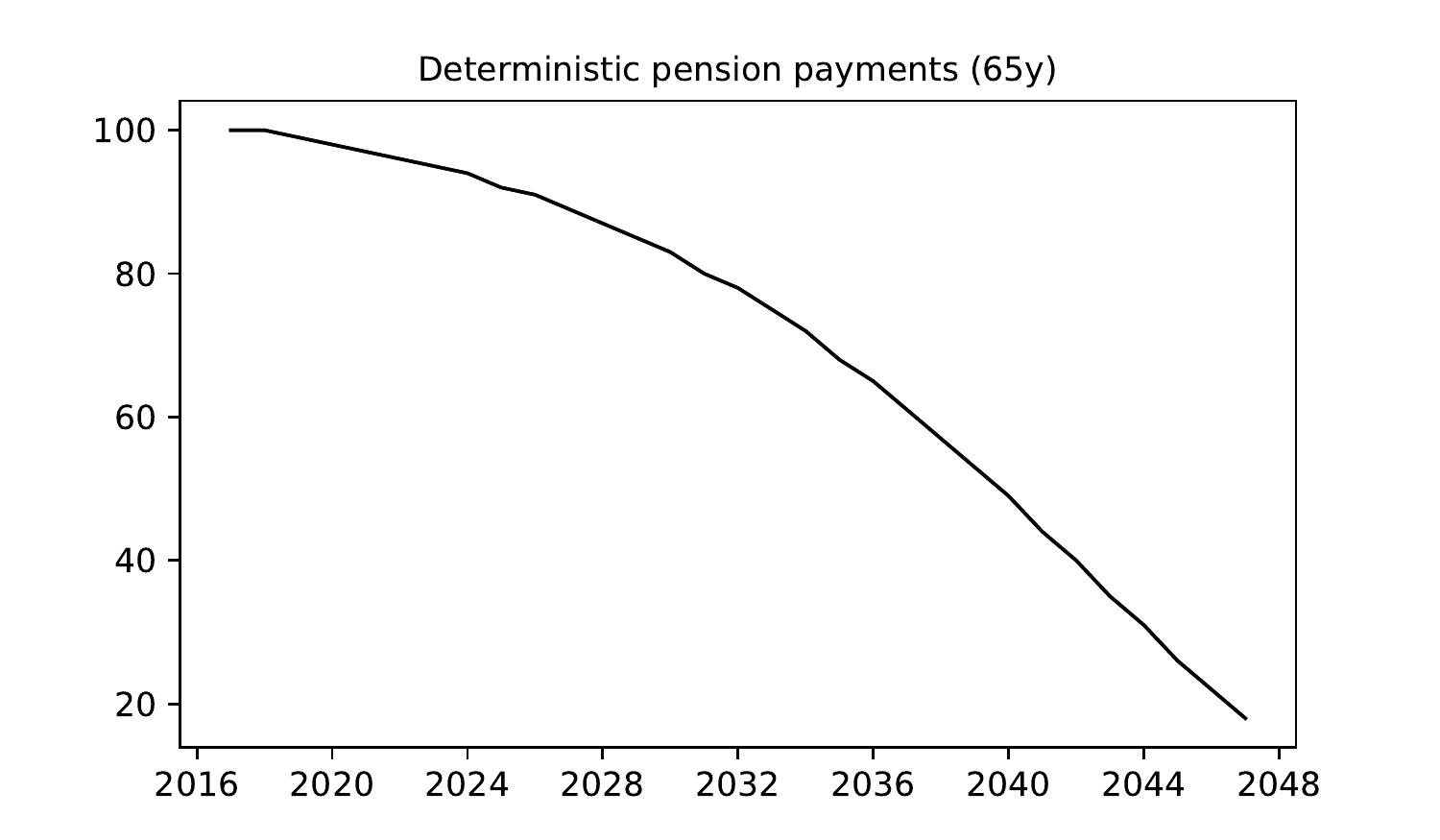}\includegraphics[scale=0.5]{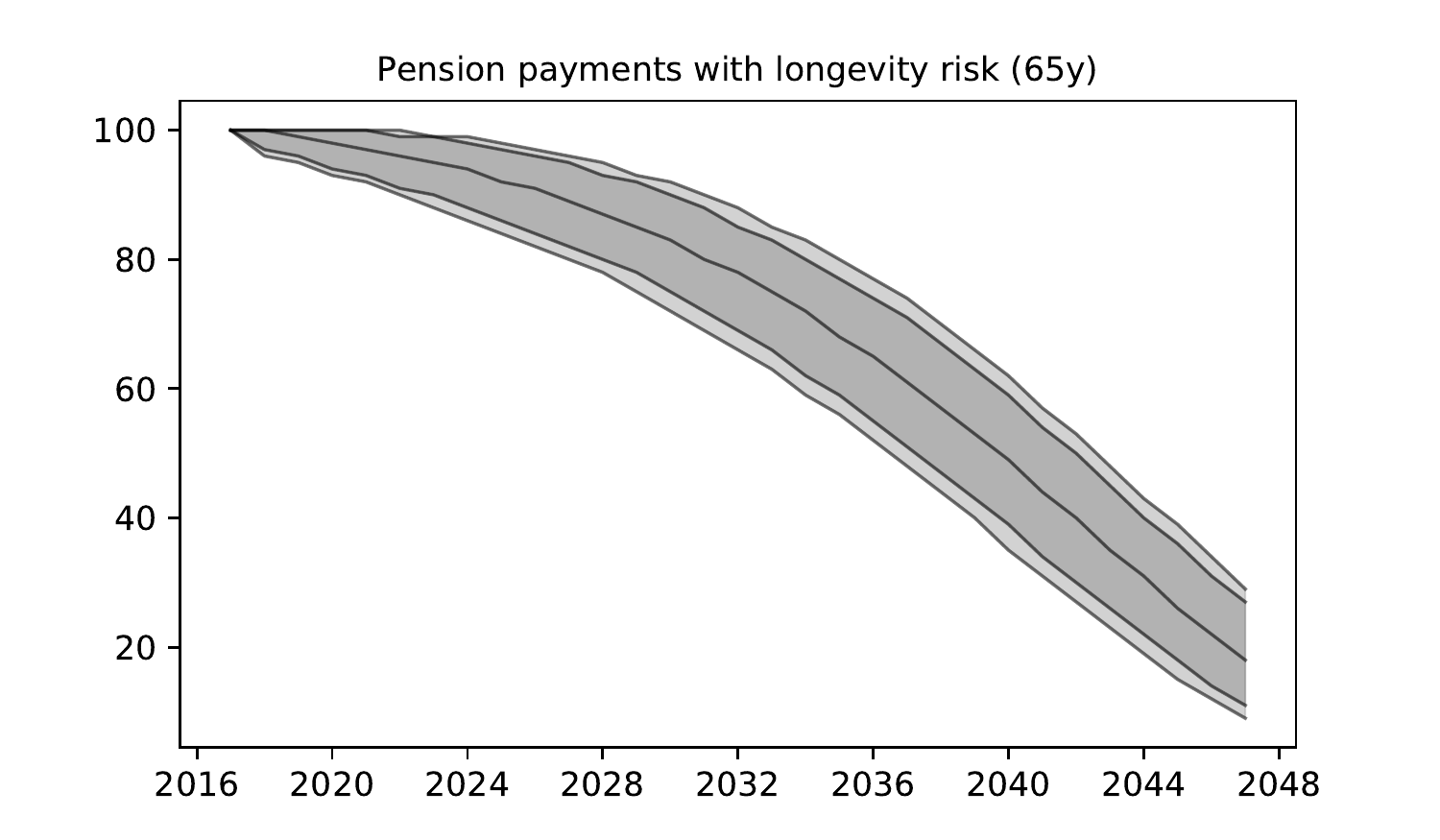}
}
\makebox[\textwidth][c]{
\includegraphics[scale=0.5]{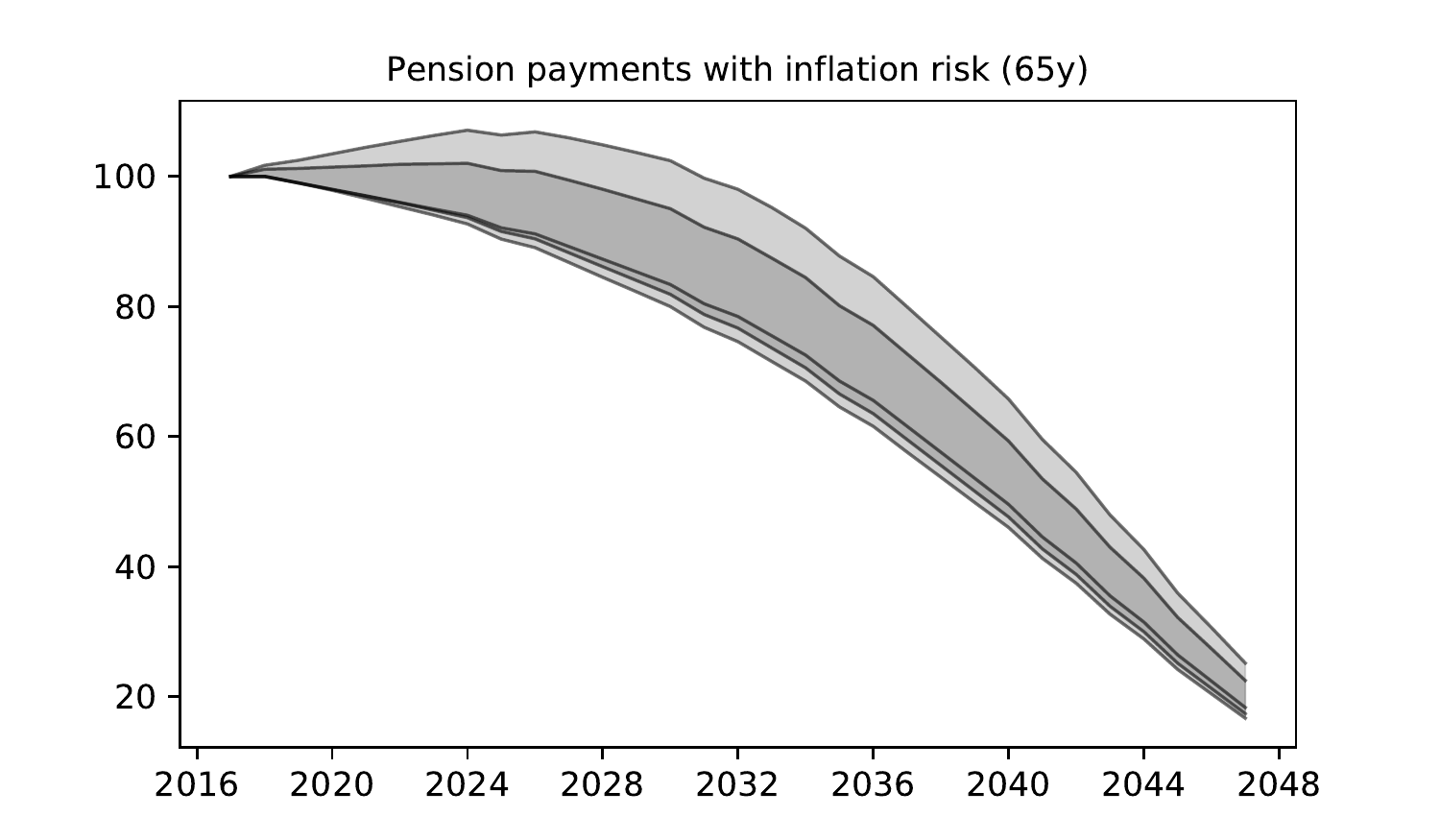}\includegraphics[scale=0.5]{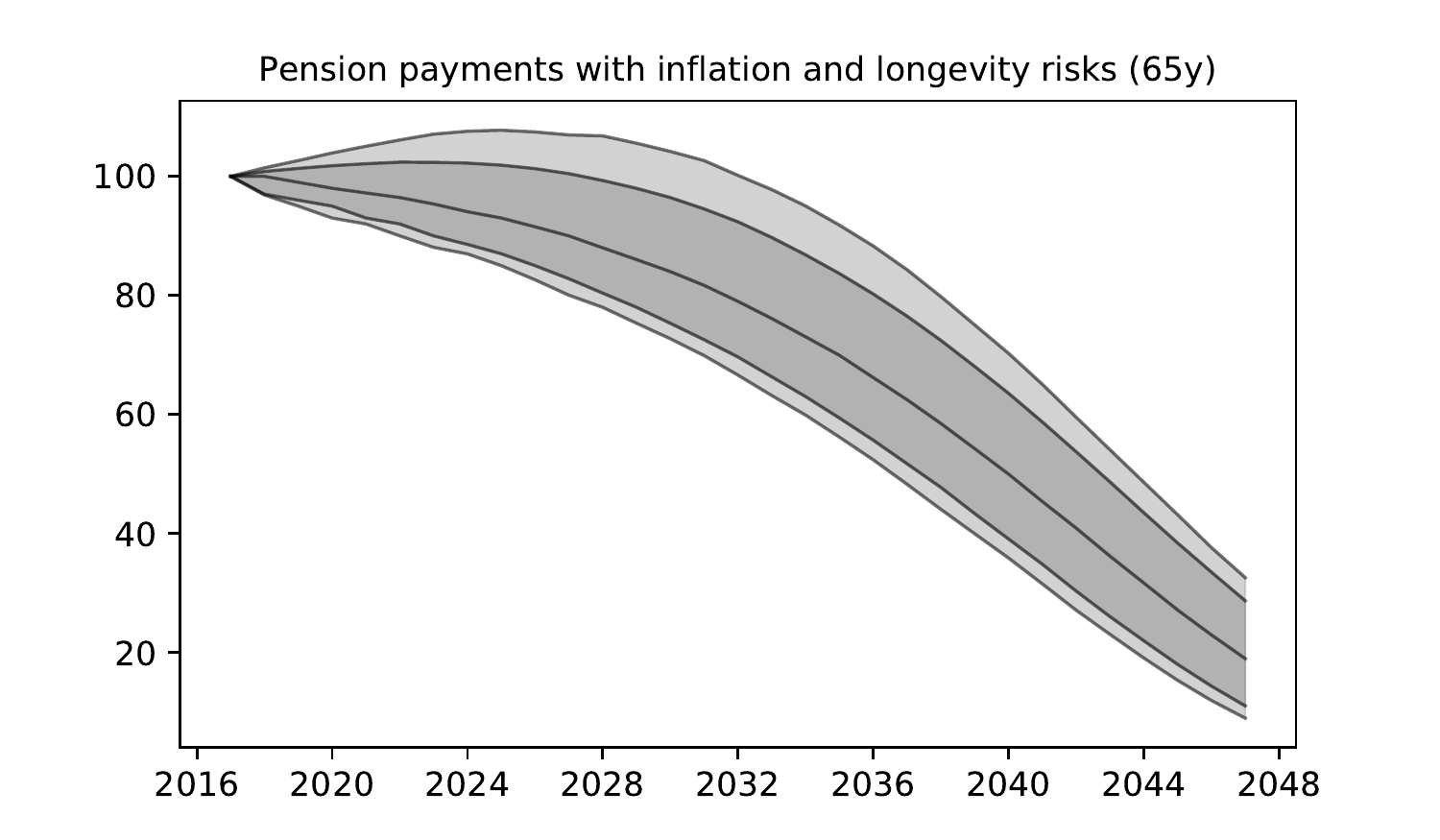} 
}
\end{center}
\caption{
Real values of pension payments obtained for different risk factors. The top-right plot shows the 95\% and 99\% confidence bands obtained for the longevity risk. In the bottom-left plot, the same bands and shown for a constant inflation of 2\%. Finally, the bottom-right plot shows the result obtained when both longevity and inflation risk are taken into account.
}\label{fig:payments_risk}
\end{figure}

%% file: plots_gdp_mortality_link.tex
\begin{figure}[h]
\begin{center}
\makebox[\textwidth][c]{
\includegraphics[scale=0.5]{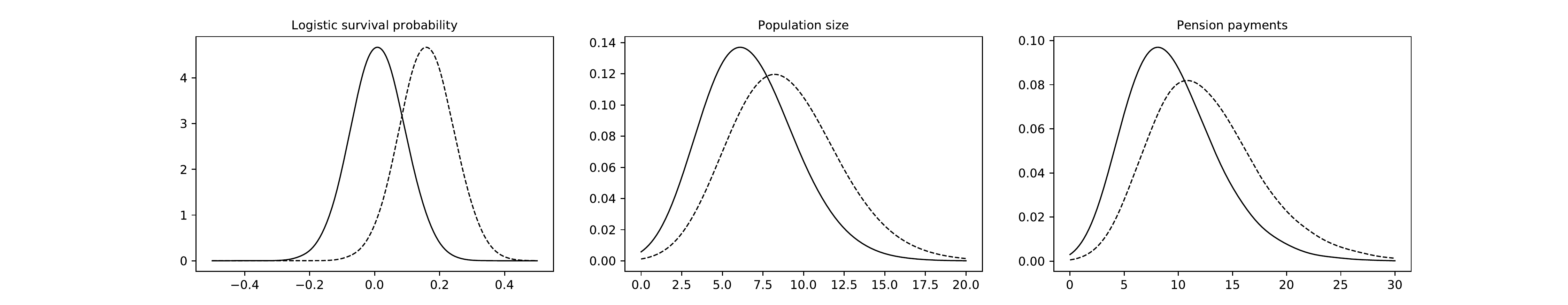}
}
\end{center}
\caption{The leftmost picture shows the distribution of the risk factor $v^{3,f}$ in 2035. The center plot shows the distribution of population sizes at the end of the same year for a cohort that had one thousand 85 year-old females in 2016. The corresponding pension payments for 2035 are shown in the rightmost plot. Dashed lines show the effect of increasing the real GDP log-growth long-term median to 8\%.
}\label{fig:link_gdp_mortality}
\end{figure}

%% file: plots_wealth.tex
\begin{figure}[ht]
\begin{center}
\makebox[\textwidth][c]{
\includegraphics[scale=0.5]{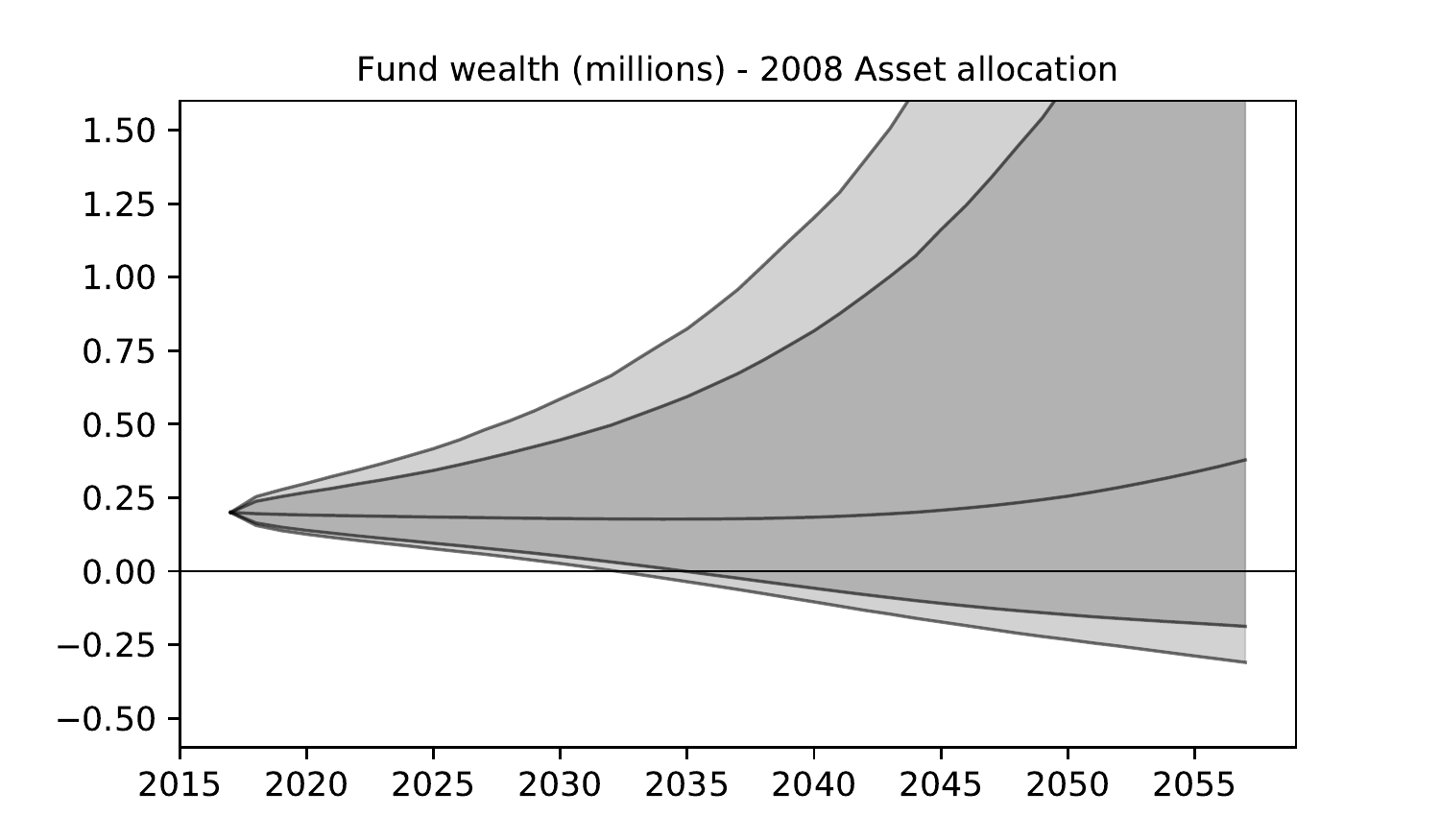}\includegraphics[scale=0.5]{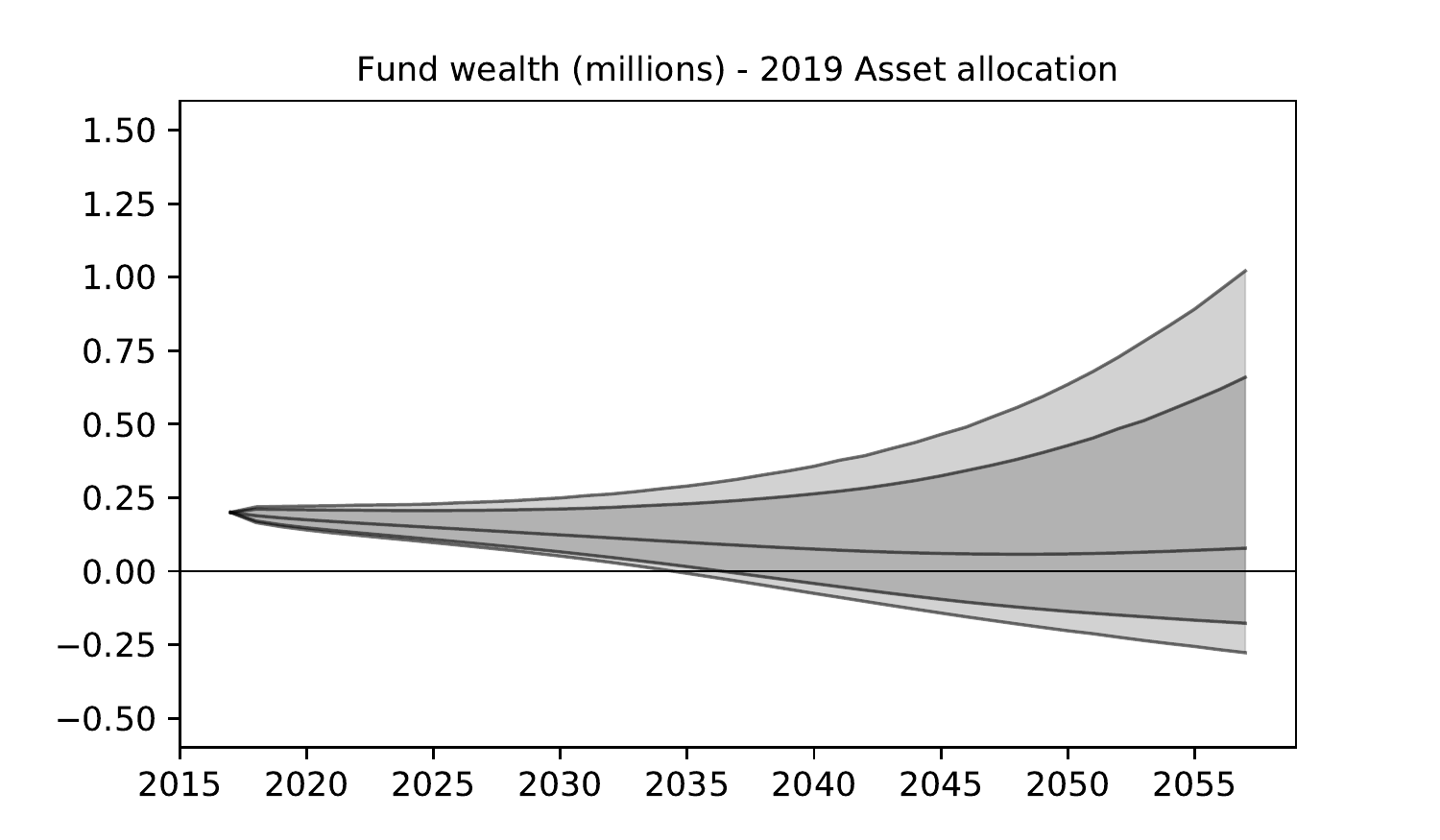}
}
\end{center}
\caption{The plots compare simulations for the wealth of a pension fund using the 2008 and 2019 average asset allocations from tables~\ref{tab:purplebookallocation} and~\ref{tab:purplebookallocationbonds}. In the most recent allocation, over 60\% of the funds are invested in bonds, resulting in lower risks, but also in lower gains on average, as illustrated by the median of the wealth in the plots. The 95\% and 99\% quantiles are also illustrated in the plots.}\label{fig:wealth_alloc}
\end{figure}